\documentclass[twocolumn]{aastex631}

\newcommand{\tI}{t_{\mathrm{I}}}
\newcommand{\tII}{t_{\mathrm{I}\hspace{-1pt}\mathrm{I}}}
\newcommand{\tIII}{t_{\mathrm{I}\hspace{-1pt}\mathrm{I}\hspace{-1pt}\mathrm{I}}}
\newcommand{\tIV}{t_{\mathrm{I}\hspace{-1pt}\mathrm{V}}}
\newcommand{\Ttot}{T_{\mathrm{tot}}}
\newcommand{\Tfull}{T_{\mathrm{full}}}

\newcommand{\ti}{t_{\mathrm{i}}}
\newcommand{\te}{t_{\mathrm{e}}}
\newcommand{\taui}{\tau_{\mathrm{i}}}
\newcommand{\taue}{\tau_{\mathrm{e}}}

\newcommand{\xii}{x_{\mathrm{i}}}
\newcommand{\xe}{x_{\mathrm{e}}}
\newcommand{\yi}{y_{\mathrm{i}}}
\newcommand{\ye}{y_{\mathrm{e}}}

\newcommand{\ki}{k_{\mathrm{i}}}
\newcommand{\ke}{k_{\mathrm{e}}}

\newcommand{\Rp}{R_{\mathrm{p}}}
\newcommand{\Rpxip}{R_{\mathrm{p}}^{x\mathrm{i}+}}
\newcommand{\Rpxin}{R_{\mathrm{p}}^{x\mathrm{i}-}}
\newcommand{\Rpxep}{R_{\mathrm{p}}^{x\mathrm{e}+}}
\newcommand{\Rpxen}{R_{\mathrm{p}}^{x\mathrm{e}-}}
\newcommand{\Rs}{R_{\mathrm{s}}}

\usepackage{amsmath}
\usepackage{amsfonts}
\usepackage{mathcomp}
\usepackage{bm}

\begin{document}

\title{Probing Two-dimensional Asymmetries of an Exoplanet Atmosphere from Chromatic Transit Variation}


\author[0000-0002-4093-8736]{Shotaro Tada}
\email{shotarotada1@gmail.com}
\affiliation{Graduate University for Advanced Studies, SOKENDAI, 2-21-1 Osawa, Mitaka, Tokyo 181-8588, Japan}

\author[0000-0003-3309-9134]{Hajime Kawahara}
\affiliation{Institute of Space and Astronautical Science, Japan Aerospace Exploration Agency, 3-1-1 Yoshinodai, Chuo-ku, Sagamihara, Kanagawa 252-5210, Japan}
\affiliation{Department of Astronomy, Graduate School of Science, The University of Tokyo, 7-3-1 Hongo, Bunkyo-ku, Tokyo 113-0033, Japan}

\author[0000-0003-3800-7518]{Yui Kawashima}
\affiliation{Department of Astronomy, Graduate School of Science, Kyoto University, Kitashirakawa Oiwake-cho, Sakyo-ku, Kyoto 606-8502, Japan}
\affiliation{Frontier Research Institute for Interdisciplinary Sciences, Tohoku University, 6-3 Aramaki aza Aoba, Aoba-ku, Sendai, Miyagi 980-8578, Japan}
\affiliation{Department of Geophysics, Graduate School of Science, Tohoku University, 6-3 Aramaki aza Aoba, Aoba-ku, Sendai, Miyagi 980-8578, Japan}
\affiliation{Institute of Space and Astronautical Science, Japan Aerospace Exploration Agency, 3-1-1 Yoshinodai, Chuo-ku, Sagamihara, Kanagawa 252-5210, Japan}
\affiliation{Cluster for Pioneering Research, RIKEN, 2-1 Hirosawa, Wako, Saitama 351-0198, Japan}

\author[0000-0001-6181-3142]{Takayuki Kotani}
\affiliation{Astrobiology Center, 2-21-1 Osawa, Mitaka, Tokyo 181-8588, Japan}
\affiliation{Graduate University for Advanced Studies, SOKENDAI, 2-21-1 Osawa, Mitaka, Tokyo 181-8588, Japan}
\affiliation{National Astronomical Observatory of Japan, 2-21-1 Osawa, Mitaka, Tokyo 181-8588, Japan}

\author[0000-0003-1298-9699]{Kento Masuda}
\affiliation{Department of Earth and Space Science, Graduate School of Science, Osaka University, 1-1 Machikaneyama-cho, Toyonaka, Osaka 560-0043, Japan}




\begin{abstract}
We propose a new method for investigating atmospheric inhomogeneities in exoplanets through transmission spectroscopy.
Our approach links chromatic variations in conventional transit model parameters—central transit time, total and full durations, and transit depth—to atmospheric asymmetries. 
By separately analyzing atmospheric asymmetries during ingress and egress, we can derive clear connections between these variations and the underlying asymmetries of the planetary limbs. Additionally, this approach enables us to investigate differences between the limbs slightly offset from the terminator on the dayside and the nightside.
We applied this method to JWST's NIRSpec/G395H observations of the hot Saturn exoplanet WASP-39 b. 
Our analysis suggests a higher abundance of $\mathrm{CO_2}$ on the evening limb compared to the morning limb and indicates a greater probability of $\mathrm{SO_2}$ on the limb slightly offset from the terminator on the dayside relative to the nightside. These findings highlight the potential of our method to enhance the understanding of photochemical processes in exoplanetary atmospheres.
\end{abstract}

\keywords{Exoplanet atmospheres (487), Exoplanets (498), Transits (1711), Transmission spectroscopy (2133)}

\section{Introduction} \label{sec:introduction}

The unprecedentedly precise transmission spectra obtained by the James Webb Space Telescope (JWST) have significantly advanced the study of close-in exoplanet atmospheres \citep[e.g.,][]{2023Natur.614..659R}. For instance, the detection of $\mathrm{SO_2}$ in the atmosphere of the hot Saturn WASP-39 b marked the first unambiguous identification of a photochemically produced molecule in an exoplanetary atmosphere \citep{2023Natur.614..664A, 2024Natur.626..979P, 2023Natur.617..483T}, showcasing JWST's capability to probe photochemical processes.
Simulations using a 3D Global Circulation Model (GCM) and a two-dimensional photochemical model have further suggested that $\mathrm{SO_2}$ may accumulate on the planet’s nightside \citep{2023ApJ...959L..30T}. 
While ultraviolet radiation from the host star drives photochemical reactions on the dayside, these reactions, coupled with atmospheric circulation, affect the global distribution of chemical species.
These findings highlight the importance of studying atmospheric inhomogeneities to understand the interplay between photochemical processes and atmospheric circulation.

Studies on atmospheric inhomogeneities are progressing rapidly.
Recently, \cite{2023Natur.614..659R} discovered that the central transit time of WASP-39 b observed with JWST's NIRSpec/PRISM varies by seconds depending on the wavelength, suggesting potential morning-evening asymmetries in the planet's limb.  
This finding aligns with predictions from studies using GCMs, which suggest that atmospheric asymmetries could cause the wavelength dependence of the central transit time \citep[][]{2012ApJ...751...87D}.
Moreover, several studies using GCMs have investigated the effects of atmospheric inhomogeneities on transmission spectra \citep[e.g.,][]{2010ApJ...709.1396F, 2017ApJ...845L..20K} and the shape of transit light curves \citep[e.g.,][]{2016ApJ...820...78L, 2024A&A...685A.125F}.

One approach to characterizing atmospheric asymmetries from such distortions in transit light curves is to analyze the data using a dedicated aspherical planet model \citep{2016A&A...589A..52V, 2021AJ....162..165E, 2023MNRAS.519.5114G} instead of conventional transit modeling\footnote{We refer to the transit model that assumes a circular planetary shadow and Keplerian motion as the conventional transit model.}. The advantage of this approach is that it allows for the direct prediction of transit light curves from the asymmetrical model. 
\citet{2024Natur.632.1017E} applied one such model to the previously mentioned JWST data of WASP-39 b, deriving for the first time separate transmission spectra for the morning and evening limbs. Similarly, \citet{2024NatAs.tmp..231M} used the same model to analyze JWST data of the warm Neptune WASP-107 b and derived separate transmission spectra for its morning and evening limbs.

However, these studies have predominantly focused on asymmetries in the morning-evening direction, ignoring potential asymmetries in the day-night or north-south directions.
Day-night asymmetries, in particular, are closely linked to photochemical processes, making them critical for understanding photochemical processes on exoplanets.
While high-resolution transmission spectroscopy has demonstrated the ability to separately investigate the atmospheric properties of the dayside and nightside of the terminator \citep{2022MNRAS.515..749G}, it remains unclear how such day-night asymmetries can be effectively studied using low- to mid-resolution transmission spectroscopy.
\citet{2024A&A...685A.125F} demonstrated that the effect of planetary rotation can modify the shape of transit light curves, as the slightly visible dayside or nightside influences the apparent radii of the limbs. This finding suggests the potential to explore day-night asymmetries through detailed analyses of transit light curves.

In this paper, we explore atmospheric asymmetries beyond the morning-evening direction by linking the color dependence of conventional transit model parameters to atmospheric asymmetries on the planetary limbs. 
We analyze not only the color dependence of the central transit time but also the duration for the first time.
By separately analyzing asymmetries during ingress and egress, we can derive clear connections between these chromatic variations and the underlying asymmetries of the planetary limbs. This approach also enables us to investigate differences between the limbs slightly offset from the terminator on the dayside and the nightside.

We refer to the variations in transit parameters with wavelength as chromatic transit variation (CTV) and aim to establish a general framework that links CTV to atmospheric asymmetries in planetary atmospheres.
The structure of this paper is as follows: In \S 2, we formulate the relationship between CTV and atmospheric asymmetries. In \S 3, we validate the proposed method using synthetic data. In \S 4, we apply this method to JWST observations of WASP-39 b. \S 5 discusses the implications for planetary atmospheres, and \S 6 outlines future prospects.

\section{Impact of Atmospheric Asymmetry on Conventional Transit Modeling} \label{sec:formulation}

\subsection{How Does Atmospheric Asymmetry Affect Transit Parameters?}\label{subsec:concept}

In transmission spectroscopy, chromatic variations in the depth of the transit light curve are observed due to differences in the apparent radius of a planet at different wavelengths. These variations arise from the absorption and scattering of light by atmospheric molecules, clouds, and haze in the planet's limb. 
Spatial asymmetries in the atmospheric structure of a planet's limb, such as temperature and molecular abundance, also affect the transit light curve. How would these spatial asymmetries alter the parameters in the transit light curve?

\begin{figure*}[t]
\centering
\includegraphics[width=\linewidth]{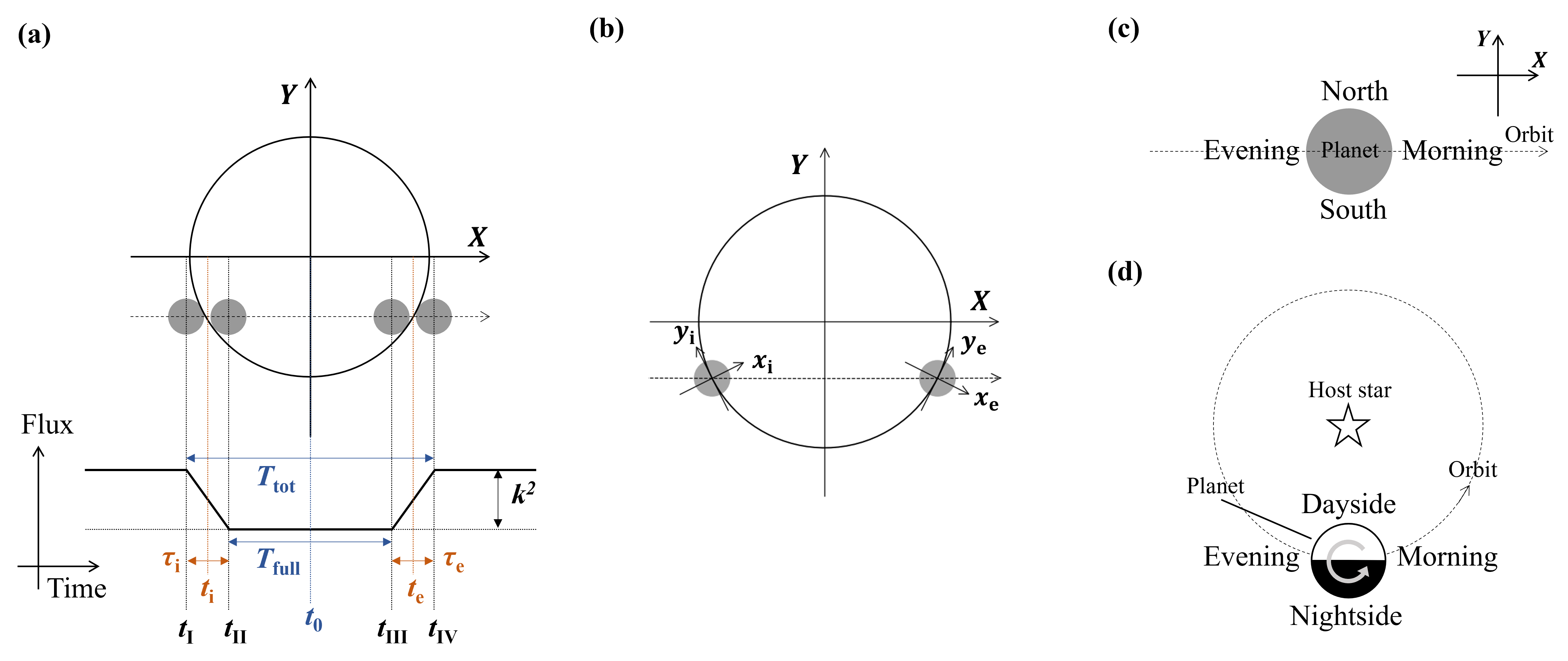}
\caption{Definitions of parameters, coordinate systems, and directional terms used in this paper. (a) The definitions of parameters in the conventional transit model. $k^2$ represents the depth of the light curve. The four contact times, $\tI$, $\tII$, $\tIII$, and $\tIV$, are defined as the times when the planet is externally and internally tangent to the stellar disk. These contact times can be converted as follows: $\ti = (\tI + \tII)/2$, $\taui = \tII - \tI$, $\te = (\tIII + \tIV)/2$, and $\taue = \tIV - \tIII$. For a circular orbit, there is a constraint $\taui = \taue$, leading to the following useful conversions: $\Ttot = \tIV - \tI$, $\Tfull = \tIII - \tII$, and $t_0 = (\tI + \tIV)/2 = (\tII + \tIII)/2$. 
(b) The coordinate systems used in this paper. The $X$--$Y$ coordinate system is fixed relative to the celestial sphere. The $\yi$--axis and $\ye$--axis are aligned with the tangent lines of the stellar disk at the points where the planetary orbit intersects the edge of the host star. The $\xii$--axis and $\xe$--axis are oriented perpendicularly to the $\yi$--axis and $\ye$--axis, respectively. The $\xii$--$\yi$ and $\xe$--$\ye$ coordinate systems are fixed relative to the planet's center of mass and rotate synchronously with the planet's orbital motion, regardless of whether the planet itself is tidally locked.
(c) The definitions of the north, south, morning, and evening.
(d) The definitions of the morning, evening, dayside, and nightside. The planet's rotation is in the same direction as its orbital motion. 
\label{fig:definition}}
\end{figure*}

Panel (a) of Figure \ref{fig:definition} shows the parameters of the transit light curve. The key characteristics of the light curve are described by five parameters: the transit depth $k^2$ and the four contact times, $\tI$, $\tII$, $\tIII$, and $\tIV$. Here, $k$ is the ratio of the planetary radius to the host star's radius. These contact times can be converted into the timing of ingress $\ti = (\tI + \tII)/2$, the duration of ingress $\taui = \tII - \tI$, the timing of egress $\te = (\tIII + \tIV)/2$, and the duration of egress $\taue = \tIV - \tIII$. For a circular orbit, there is a constraint $\taui = \taue$, leading to the useful conversions $t_0 = (\tI + \tIV)/2 = (\tII + \tIII)/2$, $\Ttot = \tIV - \tI$, and $\Tfull = \tIII - \tII$. Here, $t_0$ is the central transit time, $ \Ttot $ refers to the total duration during which any part of the planet is transiting, and $ \Tfull $ refers to the full duration during which the entire planet is transiting.

The coordinate systems used in this paper are shown in Panel (b) of Figure \ref{fig:definition}. The $X$--$Y$ coordinate system follows \citet{2010exop.book...55W}. The host star is centered in this coordinate system. The dashed line represents the orbit of the planet's center of mass. To clarify the discussion, we also define the $\xii$--$\yi$ and $\xe$--$\ye$ coordinate systems. The $\yi$--axis and $\ye$--axis are aligned with the tangent lines of the stellar disk where the planet's orbit intersects the edge of the star, while the $\xii$--axis and $\xe$--axis are perpendicular to the $\yi$--axis and $\ye$--axis, respectively. The $X$--$Y$ coordinate system is fixed relative to the celestial sphere, while the $\xii$--$\yi$ and $\xe$--$\ye$ coordinate systems are fixed relative to the planet's center of mass and rotate synchronously with the planet's orbital motion, regardless of whether the planet itself is tidally locked. This means that the $\xii$--$\yi$ and $\xe$--$\ye$ coordinate systems rotate in the same way as the planet's day-night terminator (see Figure \ref{fig:3d_wasp39b}).

In this paper, we assume that the planet’s rotation is in the same direction as its orbital motion and refer to the planet’s leading limb as the morning limb and the trailing limb as the evening limb. This assumption includes the case of synchronous rotation due to tidal locking.
Additionally, we define the positive $Y$ direction as north, and the negative $Y$ direction as south.
Panel (c) and (d) of Figure \ref{fig:definition} illustrate the definitions of these directional terms used in this paper. 

\begin{figure*}[t]
\centering
\includegraphics[width=0.9\linewidth]{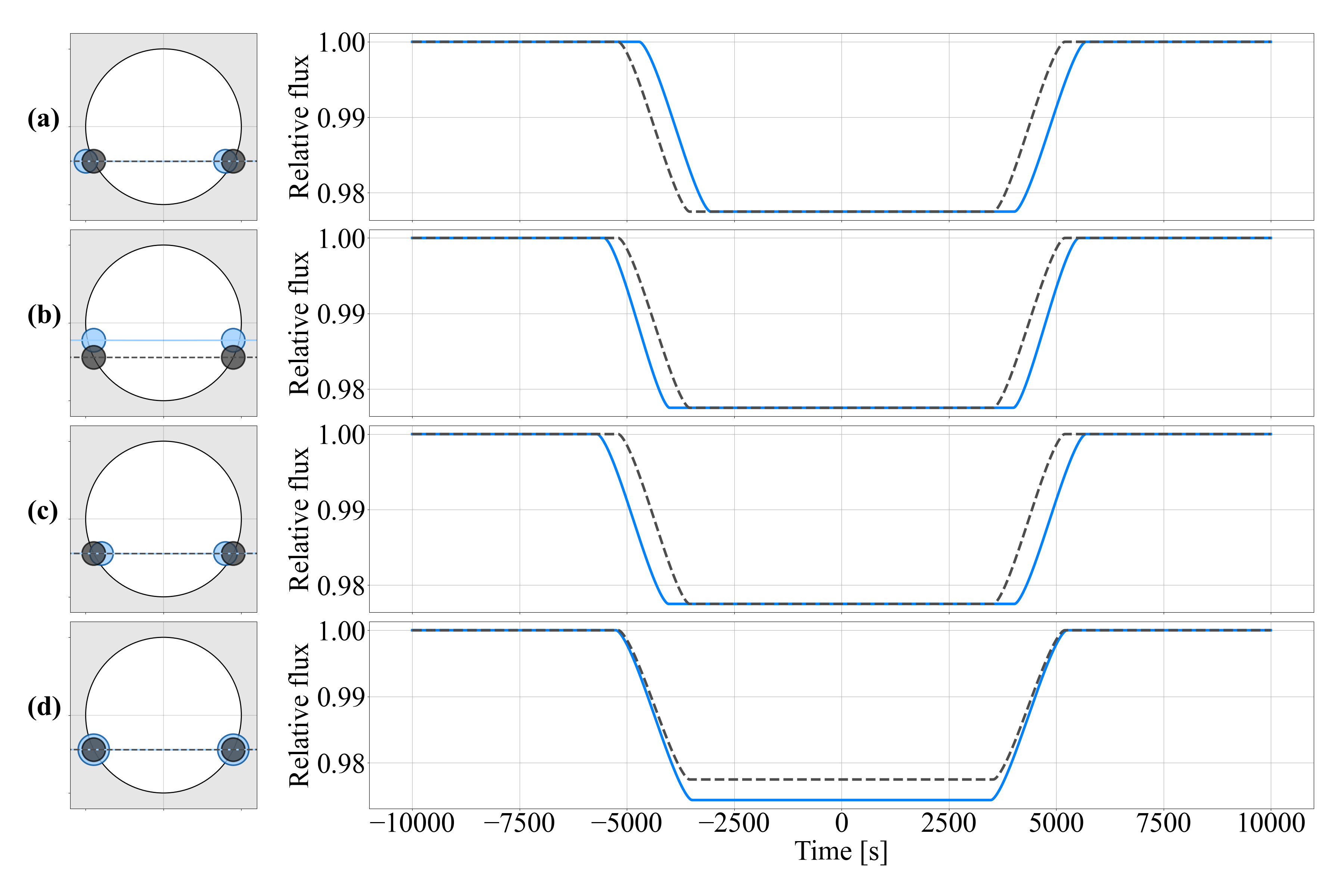}
\caption{Patterns of atmospheric asymmetry and their different impacts on light curves.
The blue solid lines and grey dashed lines in the figures on the right of each panel correspond to the light curves of the blue and grey disks of the planet's shadow in the schematics on the left, respectively. 
(a) A displacement of the planet's shadow along the $X$--axis (morning-evening asymmetry) causes a shift of the central transit time $t_0$. 
(b) A displacement along the $Y$--axis (north-south asymmetry) causes a change in the transit duration. 
(c) Opposite displacements along the $X$--axis between ingress and egress also alter the transit duration. 
(d) Changes in the apparent radius result in variations in both the transit depth and duration. In Panel (d), the radius of the blue disk is exaggerated by a factor of 5 compared to the disk producing the blue light curve, to highlight the differences between the grey and blue disks.
\label{fig:ponchi}}
\end{figure*}

In our model, atmospheric asymmetries are represented by slight displacements of the center of the planet's circular shadow disk from the center of mass. 
Figure \ref{fig:ponchi} illustrates four different patterns of light curve changes. We can associate changes in the parameters of the conventional transit model with combinations of these patterns. 

As shown in Panel (a) of Figure \ref{fig:ponchi}, a displacement of the planet's shadow toward the negative $X$ direction at certain wavelengths (morning-evening asymmetry) causes an overall delay in the transit timing, shifting the central transit time, $t_0$, to a later time \citep[e.g.][]{2013MNRAS.435.3159D, 2024Natur.632.1017E}. 
On the other hand, a displacement along the $Y$--axis (north-south asymmetry) affects the transit duration, as shown in Panel (b). The transit duration is also altered when the displacements along the $X$--axis are in opposite directions during ingress and egress (Panel (c)). This pattern reflects the difference in the light conditions of the limb between ingress and egress, which we further discuss in \S \ref{subsec:lightcondition}.
Changes in the planet's radius affect the transit depth $k^2$ and duration, without altering the central transit time, as shown in Panel (d).

To account for any asymmetries, it is necessary to introduce an additional pattern where the displacement of the planet's shadow along the $Y$--axis differs between ingress and egress. However, this pattern is difficult to reconcile with the conventional transit model, as it results in different durations for ingress and egress, while the conventional model typically allows only slight differences between them \citep{2010exop.book...55W}.

However, classifying asymmetries into four patterns in Figure \ref{fig:ponchi}, and considering their impact on the transit parameters, especially the durations $\Ttot$ and $\Tfull$,  makes the analysis complex and difficult to interpret.
To improve the clarity of the relationship between the light curve and asymmetries, we consider asymmetries in ingress and egress separately, focusing on the timing of ingress $\ti$, the duration of ingress $\taui$, the timing of egress $\te$, and the duration of egress $\taue$. 
Note that when assuming a circular orbit, there is a constraint of $\taui = \taue$, which we will discuss further in \S \ref{subsec:taui=taue}.

\begin{figure*}[ht!]
\centering
\includegraphics[width=0.9\linewidth]{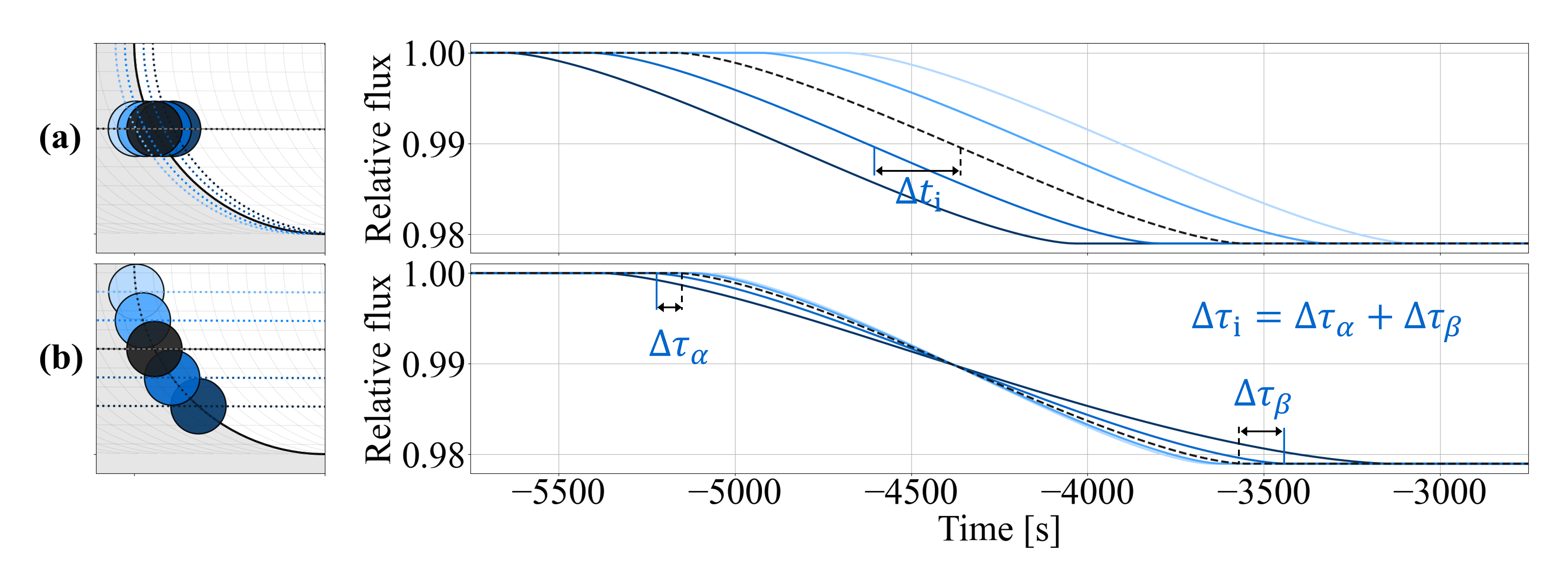}
\caption{Changes in timing and duration of ingress caused by atmospheric asymmetries.
The colored solid lines and grey dashed lines in the graphs on the right correspond to the light curves of the planet's shadow disks with the same color in the schematics on the left, respectively. Thin grey lines in the schematics on the left show the lines of constant $\ti' (\sim \ti)$ or $\taui$. (a) Asymmetry in the direction of orbital motion preserves the duration of ingress $\taui$. The colored dotted lines in the schematics on the left show the lines of constant $\ti'$ for each planet's shadow disk. (b) Asymmetry along the host star's edge preserves the timing of ingress $\ti'$. The colored dotted lines in the schematics on the left show the lines of constant $\taui$ for each planet's shadow disk.
\label{fig:ponchi_ingress}}
\end{figure*}

We find that $\taui$ remains constant with asymmetries in the direction of orbital motion, which is nearly aligned with the $X$ direction, assuming the planet moves in uniform linear motion on the celestial sphere during ingress (Panel (a) of Figure \ref{fig:ponchi_ingress}).
In contrast, asymmetries perpendicular to the orbital motion, which is nearly aligned with the $Y$ direction, cause changes in $\taui$. 

On the other hand, $\ti$ remains constant with asymmetries along the host star's edge, which is nearly aligned with the $\yi$ direction (Panel (b) of Figure \ref{fig:ponchi_ingress}). 
Strictly speaking, these asymmetries preserve the timing when the center of the planet's shadow disk intersects the host star's edge $\ti'$, which is slightly different from $\ti$ (see Appendix \ref{ap:duration} for details).
Asymmetries perpendicular to the host star's edge, which is nearly aligned with the $\xii$ direction, cause changes in $\ti'$.

Asymmetries in these directions do not strictly preserve the timing or duration because the projected velocity of the planet on the celestial sphere is not constant during ingress or egress. However, even considering this, the changes in those timings and durations are slight.

These patterns hold for egress as well: asymmetries in the direction perpendicular to the orbital motion, which is almost the same as the $Y$ direction, cause changes in $\taue$, while asymmetries in the direction perpendicular to the host star's edge, which is almost the same as the $\xe$ direction, cause changes in $\te$.

We can determine the displacement vector of the center of the planet's shadow disk from the center of mass for both ingress and egress separately.
To first order, the $\xii$ component of the vector is determined using $\ti$, and the $Y$ component using $\taui$ for ingress. For egress, the $\xe$ component is determined using $\te$, and the $Y$ component using $\taue$.

\begin{figure}[t!]
\centering
\includegraphics[width=\linewidth]{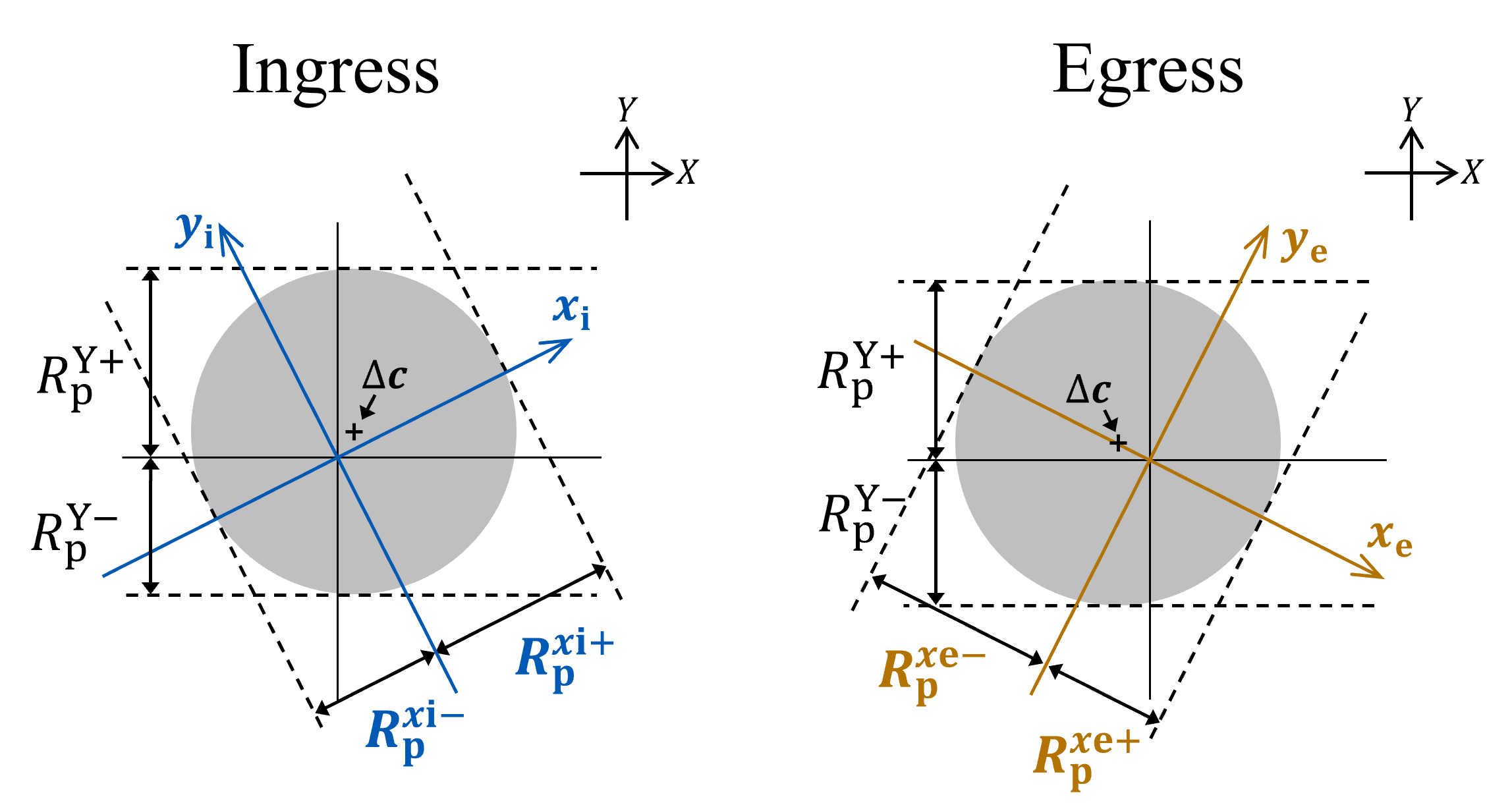}
\caption{Definitions of $\Rpxip$, $\Rpxin$, $\Rpxep$, $\Rpxen$, $\Rp^{Y+}$, and $\Rp^{Y-}$.  
The origin of the $\xii$--$\yi$ and $\xe$--$\ye$ coordinate systems corresponds to the planet's center of mass. 
Grey shadows represent the planet’s shadow at a certain wavelength, with two-dimensional displacement vectors $\Delta \bm{c}$.
$\Rp^{Y+}$ and $\Rp^{Y-}$ are the lengths of the projected planetary shadow onto the $Y$--axis, measured from the planet's center of mass in the positive $Y$ direction and negative $Y$ direction, respectively.
Similarly, $\Rpxip$ and $\Rpxin$ are the lengths of the projected planetary shadow onto the $\xii$--axis, measured from the planet's center of mass in the positive and negative $\xii$ direction, respectively. 
The $\xii$--$\yi$ plane corresponds to the $X$--$Y$ plane at ingress (left).
$\Rpxep$ and $\Rpxen$ are the lengths of the projected planetary shadow onto the $\xe$--axis, measured from the planet's center of mass in the positive and negative $\xe$ direction, respectively.
The $\xe$--$\ye$ plane corresponds to the $X$--$Y$ plane at egress (right).
\label{fig:def_Rp}}
\end{figure}

We perform an order-of-magnitude estimation of the differences in the ingress timing $\Delta \ti$ and duration $\Delta \taui$ from the reference. $\Delta \ti$ arises from the difference in the apparent planetary radius in the direction perpendicular to the host star's edge (the $\xii$ direction), $\Delta \Rp^{x\mathrm{i}} = \Rpxip - \Rpxin $. 
Here, $\Rpxip$ and $\Rpxin$ are the lengths of the projected planetary shadow onto the $\xii$--axis, measured from the planet's center of mass in the positive $\xii$ direction and negative $\xii$ direction, respectively (Figure \ref{fig:def_Rp}).
Similarly, $\Delta \taui$ arises from the difference in the apparent planetary radius in the direction perpendicular to the orbital motion (the $Y$ direction), $\Delta \Rp^{Y} = \Rp^{Y+} - \Rp^{Y-}$. $\Rp^{Y+}$ and $\Rp^{Y-}$ are the length of the projected planetary shadow onto the $Y$--axis, measured from the planet's center of mass in the positive $Y$ direction and negative $Y$ direction, respectively (Figure \ref{fig:def_Rp}).
$\Delta \ti$ and $\Delta \taui$ are approximated as
\begin{align}
\Delta \ti &\sim -\frac{\Delta \Rp^{x\mathrm{i}}}{2v\sqrt{1-b^2}},  \\
\Delta \taui &\sim \frac{\Delta \Rp^{Y}}{2v} \left(\frac{b}{\sqrt{(1+k)^2-b^2}} - \frac{b}{\sqrt{(1-k)^2-b^2}}\right) \label{eq:dtaui}.
\end{align}
Here, $v \sim 2\pi a/P$ is the velocity of planetary orbital motion, where $a$ is the semi-major axis and $P$ is the orbital period. $k^2$ represents the depth of the light curve, and $b$ is the impact parameter. $\Delta \taui$ can be derived by differentiating $\taui \sim \Rs /v \left(\sqrt{(1+k)^2-b^2}-\sqrt{(1-k)^2-b^2}\right)$ with respect to $b$ and using the conversion $\Delta b = -\Delta \Rp^{Y} / (2\Rs)$. 

Assuming WASP-39 b with $P = 4\ \mathrm{days}$, $a = 11.4\ \Rs$, $k = 0.145$, $b = 0.45$, $\Rs = 0.9\ \mathrm{R_{\odot}} = 6.3\times 10^{8}\ \mathrm{m}$ \citep[]{2011A&A...531A..40F}, and $\Delta \Rp^{x\mathrm{i}}, \Delta \Rp^{Y} = 1000\ \mathrm{km}$ ($\sim$ scale height), we obtain $\Delta \ti \sim -4\ \mathrm{s}$ and $\Delta \taui \sim -0.7\ \mathrm{s}$.
This magnitude of differences is measurable given the precision of JWST.

From these estimates, we find that the uncertainty in the atmospheric asymmetries in the $Y$ direction is greater than that in the $\xii$ direction. For WASP-39 b, the uncertainty of $\Delta \Rp^{Y}$ is approximately $4 \div 0.7 \times 2 \sim 10$ times greater than that of $\Delta \Rp^{x\mathrm{i}}$, considering the factor of 2 in the conversions $\ti = (\tI + \tII)/2$ and $\taui = \tII - \tI$. Similar considerations apply to egress. Therefore, for both ingress and egress, the direction in which asymmetries can be measured with the highest precision is perpendicular to the host star's edges (the $\xii$ and $\xe$ directions). This motivated our use of the $\xii$--$\yi$ and $\xe$--$\ye$ coordinate systems in the formulation in \S \ref{subsec:formulation}. In the case where $b \sim 0$, $\Delta \taui$ is nearly zero, meaning that almost no information can be obtained about asymmetries along the $Y$ axis.

\subsection{Light Conditions of Limb During Transit}\label{subsec:lightcondition}

\begin{figure*}[t]
\centering
\includegraphics[width=\linewidth]{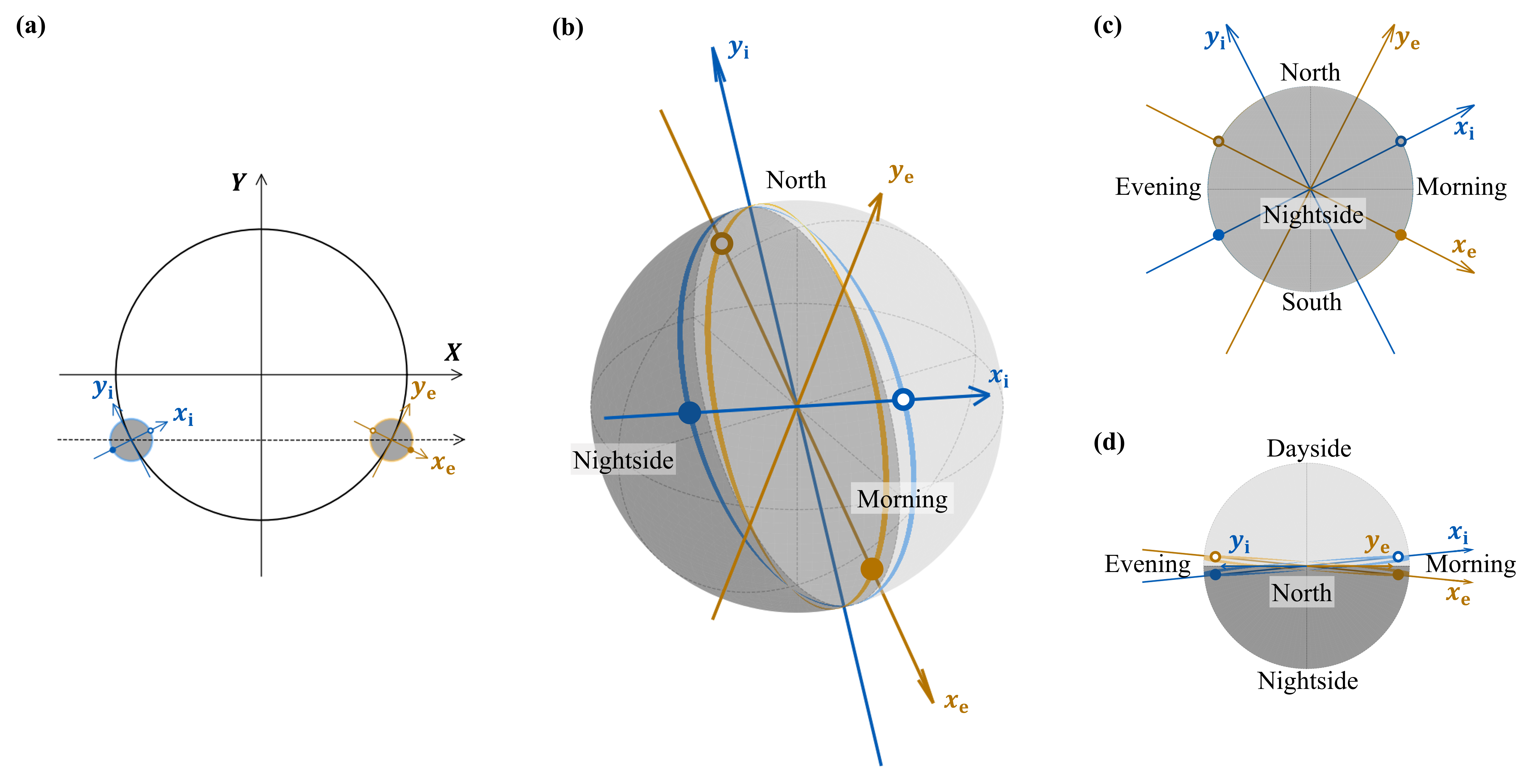}
\caption{Positions of limbs observable during ingress and egress on the planet, as expected from the orbital parameters of WASP-39 b. Panel (a) is similar to Panel (b) of Figure \ref{fig:definition}, but the $\xii$--axis and $\yi$--axis are colored blue, and the $\xe$--axis and $\ye$--axis are colored orange. The blue and orange points indicate where the limbs intersect the $\xii$--axis or $\xe$--axis. The white-filled points represent those on the dayside of the northern hemisphere, while the filled points represent those on the nightside of the southern hemisphere.
The limb observable during ingress is colored blue and the limb observable during egress is colored orange.
Panels (b), (c), and (d) show the positions of the $\xii$--axis, $\yi$--axis (blue), $\xe$--axis, and $\ye$--axis (orange), and the limbs observable during ingress (blue belt) and egress (orange belt) on the planet. The finite width of each belt reflects the shift caused by the planet's rotation during ingress or egress. 
The white and gray hemispheres represent the dayside and nightside of the planet, respectively. Panel (c) shows the view from the nightside, and Panel (d) shows the view from the north. The limbs observable during ingress and egress are shifted slightly away from the terminator. The $\yi$--axis and $\ye$--axis intersect the terminator.
\label{fig:3d_wasp39b}}
\end{figure*}

Displacements of the planet's shadow along the $Y$--axis can arise not only from north-south asymmetries in the planet's day-night terminator itself but also from the effect of an orbital inclination that is not exactly 90 degrees, which allows a slight view of the dayside (Figure \ref{fig:3d_wasp39b}). 
Therefore, displacements along the $Y$--axis can also be interpreted as differences between the limbs slightly offset from the terminator on the dayside versus the nightside.
This creates different light conditions between the north and south limbs. These differences in light conditions could cause variations in photochemical processes, resulting in different atmospheric properties.

\begin{figure}[ht]
\centering
\includegraphics[width=\linewidth]{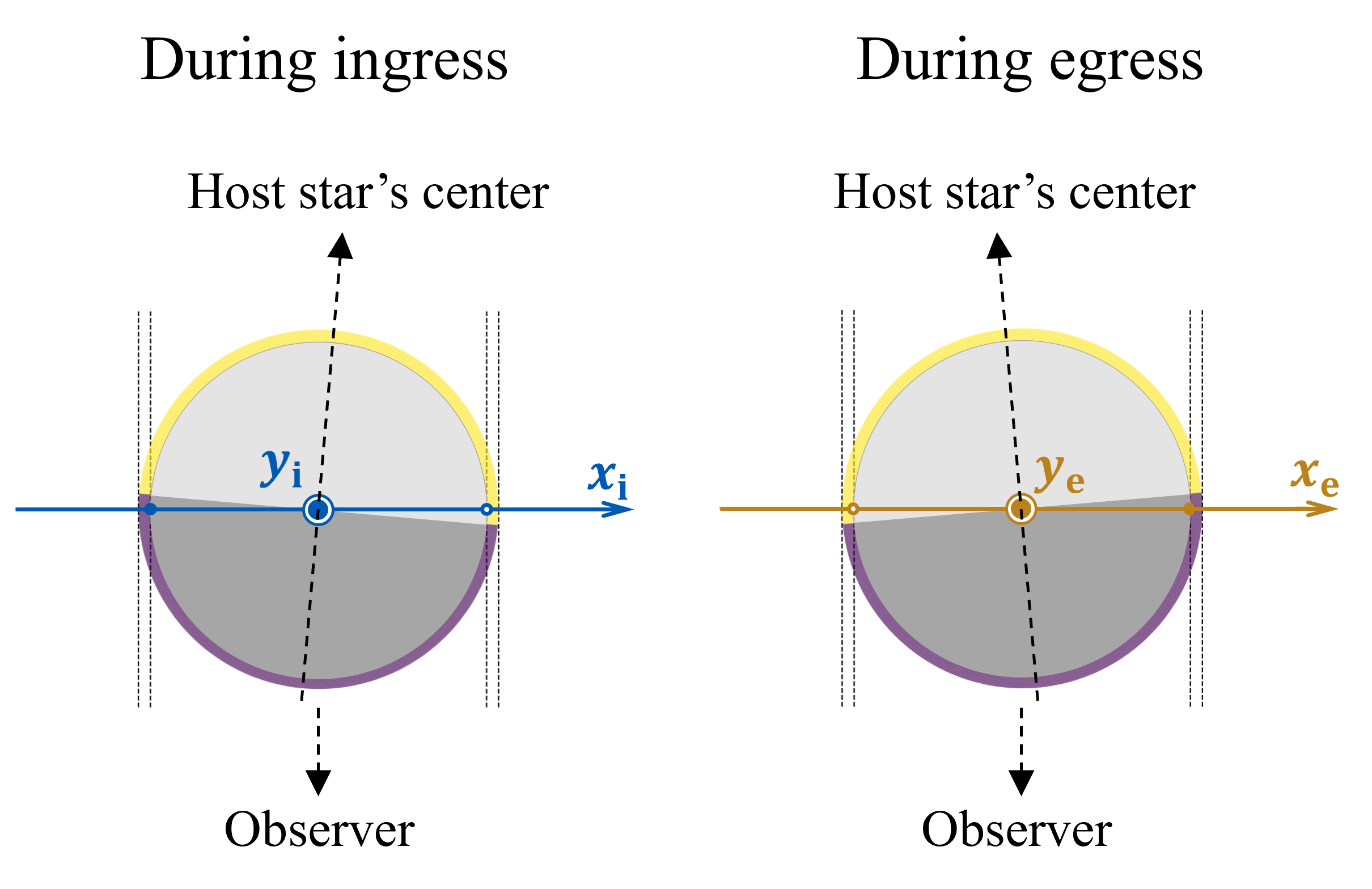}
\caption{Light conditions of the atmosphere transmitted by the host star's light during ingress (left) and egress (right), as expected from the orbital parameters of WASP-39 b. The $\yi$--axis (for ingress) and $\ye$--axis (for egress) are normal to the plane of the paper and directed out of the page.
The white and gray semicircles represent the planet's dayside and nightside, respectively. The yellow and purple regions represent the atmosphere of the dayside and nightside, respectively. 
The blue and orange points indicate where the limbs observed during ingress or egress intersect the $\xii$--axis or $\xe$--axis. White-filled points represent positions in the positive $\xii$ or negative $\xe$ directions, while filled points represent those in the negative $\xii$ or positive $\xe$ directions.
Dashed arrows show the direction toward the center of the host star and the observer, which are not aligned. Therefore, the light conditions of the atmosphere at the positive $\xii$ limb versus the negative $\xii$ limb, or at the positive $\xe$ limb versus the negative $\xe$ limb are different. The atmosphere between the dashed lines on each limb contributes to the transmission spectrum.
\label{fig:ponchi_atm}}
\end{figure}

As shown in Figure \ref{fig:3d_wasp39b}, the direction in which the dayside is visible changes between ingress and egress. This direction corresponds to the positive $\xii$ direction at ingress and the negative $\xe$ direction at egress. 

The difference in the light conditions of the atmosphere at two opposing limbs observable during transit is most pronounced along the $\xii$--axis at ingress and the $\xe$--axis at egress. Figure \ref{fig:ponchi_atm} illustrates the differences in light conditions between these limbs. At the limbs in the positive $\xii$ direction and the negative $\xe$ direction, light primarily passes through the atmosphere slightly offset from the terminator on the dayside. In contrast, at the limbs in the negative $\xii$ direction and the positive $\xe$ direction, light primarily passes through the atmosphere slightly offset from the terminator on the nightside. The angle between the terminator plane and the $\xii$ axis at ingress or the $\xe$ axis at egress is given by $\arcsin(\Rs/a)$, where $\Rs$ is the host star's radius and $a$ is the semi-major axis.

\subsection{Formulation}\label{subsec:formulation}

Next, we formulate this approach and derive the relationship between asymmetries and the following parameters: the contact times $\tI, \tII, \tIII, \tIV$, and the transit depth $k^2$.
In our model, atmospheric asymmetries are represented by the slight displacement of the center of the planet's shadow disk from the center of mass. The shadow disk has a specific radius at each wavelength. This modeling allows the use of conventional transit model light curve fitting for each wavelength. 
Therefore, the asymmetry at wavelength $\lambda$ is expressed by a two-dimensional displacement vector $\Delta \bm{c}(\lambda)$ and its radius $\Rp(\lambda)$. We aim to obtain $\Delta \bm{c}(\lambda)$ as a function of $\tI (\lambda), \tII (\lambda)$, and $k(\lambda)$ at ingress, and $\tIII (\lambda), \tIV (\lambda)$, and $k(\lambda)$ at egress. 
Since we use the planet's center of mass as a reference for displacement, the assumption about the orbit of the planet's center of mass is important. We will discuss this point further in \S \ref{subsec:center_of_mass}. In this section, we assume the orbit of the planet's center of mass is known.

The coordinate system of our formulation is shown in Panel (b) of Figure \ref{fig:definition}. The length is normalized by the stellar radius at a reference wavelength $\lambda_0$. In our formulation, we take the wavelength dependence of the stellar radius $\Rs(\lambda)$ into account, by using the ratio of the stellar radii at wavelength $\lambda$ and $\lambda_0$,
\begin{align}
\alpha(\lambda) &= \frac{\Rs(\lambda)}{\Rs(\lambda_{0})}.
\end{align}
The observed planetary radius $\Rp(\lambda)$ divided by $\Rs(\lambda_0)$ is expressed as
\begin{align}
\frac{\Rp(\lambda)}{\Rs(\lambda_0)} &= \alpha(\lambda)k(\lambda),
\end{align}
where $k(\lambda) = \Rp(\lambda)/\Rs(\lambda)$.

First, let's consider the case of ingress. Since $\tI (\lambda)$ and $\tII (\lambda)$ are defined as the times when the planet is externally tangent to and internally tangent to the stellar disk, respectively, we obtain 
\begin{align}
\label{eq:t1}
    |\bm{c}\left(\lambda, \tI(\lambda)\right)| &= \alpha(\lambda) (1+k(\lambda)) \\
\label{eq:t2}
    |\bm{c}\left(\lambda, \tII(\lambda)\right)| &= \alpha(\lambda) (1-k(\lambda)),
\end{align}
where $\bm{c}\left(\lambda, t\right)$ indicates the trajectory of the center of the planet's shadow disk at wavelength $\lambda$ in the $X$--$Y$ coordinate system.

Assuming that the atmospheric asymmetries at the planetary limb do not change during ingress, $\bm{c}\left(\lambda, t\right)$ is expressed as
\begin{align}
\label{eq:cb_deltac}
\bm{c}(\lambda, t) &= \bm{c_{\mathrm{b}}}(t) + \Delta \bm{c}(\lambda),
\end{align}
where $\bm{c_{\mathrm{b}}}(t)$ is the trajectory of the center of mass of the planet.
The change of planetary phase during ingress is ignored on this assumption, however, this is reasonable because the duration of ingress is much shorter than the orbital period of planets.
We define
\begin{align}
\bm{m}(\lambda; \tI(\lambda), \tII(\lambda)) &= \frac{\bm{c_{\mathrm{b}}}(\tI(\lambda)) + \bm{c_{\mathrm{b}}}(\tII(\lambda))}{2} \label{eq:m}\\
\bm{d}(\lambda; \tI(\lambda), \tII(\lambda)) &= \frac{\bm{c_{\mathrm{b}}}(\tI(\lambda)) - \bm{c_{\mathrm{b}}}(\tII(\lambda))}{2} \label{eq:d} .
\end{align}
Here, $\bm{m}$ represents the midpoint between the coordinates of the planet's center of mass at $\tI$ and $\tII$, and $\bm{d}$ represents half of the displacement between the coordinates of the planet's center of mass at $\tI$ and $\tII$.
If we assume that planets move in uniform linear motion during ingress, $\bm{m}(\lambda; \tI(\lambda), \tII(\lambda))$ is proportional to $\ti = (\tI + \tII)/2$, and $\bm{d}(\lambda; \tI(\lambda), \tII(\lambda))$ is proportional to $\taui = \tII - \tI$. 

Using equations (\ref{eq:t1}), (\ref{eq:t2}) and (\ref{eq:cb_deltac}), we obtain the simultaneous equations,
\begin{equation}
\label{eq:simul_eqs}
\begin{cases}
\begin{aligned}
    |\bm{m} + \bm{d} + \Delta \bm{c}| &= \alpha (1+k) \\
    |\bm{m} - \bm{d} + \Delta \bm{c}| &= \alpha (1-k),
\end{aligned}
\end{cases}
\end{equation}
where the arguments are omitted for readability. We convert these equations into
\begin{equation}
\label{eq:simul_eqs_c}
\begin{cases}
\begin{aligned}
    \bm{d}\cdot(\bm{m} + \Delta \bm{c}) &= \alpha^2 k \\
    |\bm{m} + \Delta \bm{c}|^2 &= \alpha^2(1+k^2) - |\bm{d}|^2 .
\end{aligned}
\end{cases}
\end{equation}
Defining a rotation matrix by
\begin{equation}
\mathcal{R}_{d}
=
\frac{1}{|\bm{d}|}
\begin{pmatrix}
d_{X} & -d_{Y} \\
d_{Y} & d_{X}
\end{pmatrix},
\end{equation}
where $(d_X, d_Y)^\top = \bm{d}$,
we obtain the following expression from equation (\ref{eq:simul_eqs_c})
\begin{equation}
\label{eq:pm}
\mathcal{R}_{d}^{\top}(\bm{m} + \Delta \bm{c})
=
\bm{s}
\end{equation}
where
\begin{align}
\bm{s} &= \left(\alpha^2 \frac{k}{|\bm{d}|},\ \pm \sqrt{\alpha^2(1+k^2) - |\bm{d}|^2 - \left(\alpha^2 \frac{k}{|\bm{d}|}\right)^2} \right)^{\top} \nonumber \\
&= \alpha\left(\frac{\alpha k}{|\bm{d}|},\ \pm \sqrt{\left(1 - \frac{|\bm{d}|^2}{\alpha^2}\right)\left(1 -\frac{\alpha^2 k^2}{|\bm{d}|^2}\right)} \right)^{\top} \label{eq:vec_s}.
\end{align}
When the inclination of the orbit is not $90\tcdegree$, we can choose smaller $\Delta \bm{c}$ by adopting the positive sign in equation (\ref{eq:vec_s}). However, if the inclination is almost $90\tcdegree$, it is difficult to determine which sign is correct. In such cases, we cannot determine the direction of atmospheric asymmetries along the $Y$--axis. 

By solving equation (\ref{eq:pm}) for $\Delta \bm{c}$, we finally obtain 
\begin{equation}
\label{eq:result}
\Delta \bm{c}
=
-\bm{m} + \mathcal{R}_{d}\bm{s}.
\end{equation}
The vector $\bm{m}$ represents the midpoint of the planet's center of mass at $\tI$ and $\tII$, while $\mathcal{R}_{d}\bm{s}$ represents the midpoint of the center of the planet's shadow disk at wavelength $\lambda$ at $\tI$ and $\tII$.
$\bm{m}$ is primarily determined by $\ti$, while $\mathcal{R}_{d}\bm{s}$ is primarily determined by $\taui$.

The displacement for egress can also be determined by defining $\bm{m}(\lambda)$ and $\bm{d}(\lambda)$ as
\begin{align}
\bm{m}(\lambda; \tIII(\lambda), \tIV(\lambda)) &= \frac{\bm{c_{\mathrm{b}}}(\tIV(\lambda)) + \bm{c_{\mathrm{b}}}(\tIII(\lambda))}{2} \\
\bm{d}(\lambda; \tIII(\lambda), \tIV(\lambda)) &= \frac{\bm{c_{\mathrm{b}}}(\tIV(\lambda)) - \bm{c_{\mathrm{b}}}(\tIII(\lambda))}{2},
\end{align}
and solving simultaneous equations (\ref{eq:simul_eqs}).
For egress, we have to adopt the negative sign in $\bm{s}$ (equation (\ref{eq:vec_s})) to choose smaller $\Delta \bm{c}$.

We can convert the displacement vector $\Delta \bm{c}$ into the length from the planet’s center of mass to the edge of the planetary shadow in the direction of the vector $\bm{e}_{\theta} = (\cos \theta, \sin \theta)^\top$, denoted as $\Rp^{\theta}(\lambda)/\Rs(\lambda_0)$. Here, $\theta$ is the angle measured counterclockwise from the $X$-axis.
From the relationship 
\begin{equation}
\left|\frac{\Rp^{\theta}(\lambda)}{\Rs(\lambda_0)}\bm{e}_{\theta} - \Delta \bm{c}(\lambda)\right| = \alpha(\lambda) k(\lambda),
\end{equation}
we obtain 
\begin{equation}\label{eq:rp_theta}
\frac{\Rp^{\theta}(\lambda)}{\Rs(\lambda_0)} = \bm{e}_{\theta}\cdot\Delta \bm{c}+\sqrt{\alpha^2 k^2 - |\Delta \bm{c}|^2 + (\bm{e}_{\theta}\cdot\Delta \bm{c})^2},
\end{equation}
where the arguments are omitted for readability.

However, as mentioned in \S \ref{subsec:concept}, the sensitivity of the light curve shape to the magnitude of displacement along the $Y$ direction is much lower than that along the $\xii$ direction (about $10$ times lower for the WASP-39 b case). The current data quality is insufficient to determine the $Y$ component accurately. This applies to egress as well.
Therefore, in this paper, we focus on the information obtained from the $\xii$ component of $\Delta \bm{c}$ during ingress and the $\xe$ component of $\Delta \bm{c}$ during egress.

For ingress, to transform the $X$--$Y$ coordinate system into the $\xii$--$\yi$ coordinate system, we define the following rotation matrix as 
\begin{equation}
\mathcal{R}_{\mathrm{i}}
=
\begin{pmatrix}
c_{\mathrm{b},X}(t_{\mathrm{ib}}) & -c_{\mathrm{b},Y}(t_{\mathrm{ib}}) \\
c_{\mathrm{b},Y}(t_{\mathrm{ib}}) & c_{\mathrm{b},X}(t_{\mathrm{ib}})
\end{pmatrix},
\end{equation}
where $\bm{c_{\mathrm{b}}}(t) = (c_{\mathrm{b},X}(t), c_{\mathrm{b},Y}(t))^\top$, $t_{\mathrm{ib}}$ is the time when the planet's center of mass intersects the edge of the host star during ingress. The displacement vector in the $\xii$--$\yi$ coordinate system is then given by
\begin{equation}
\Delta \bm{c_{\mathrm{i}}} 
=
-\mathcal{R}_{\mathrm{i}}^{\top}\Delta \bm{c} .
\end{equation}

Similarly, for egress, the displacement vector in the $\xe$--$\ye$ coordinate system is written as
\begin{equation}
\Delta \bm{c_{\mathrm{e}}}
=
\mathcal{R}_{\mathrm{e}}^{\top}\Delta \bm{c},
\end{equation}
where 
\begin{equation}
\mathcal{R}_{\mathrm{e}}
=
\begin{pmatrix}
c_{\mathrm{b},X}(t_{\mathrm{eb}}) & -c_{\mathrm{b},Y}(t_{\mathrm{eb}}) \\
c_{\mathrm{b},Y}(t_{\mathrm{eb}}) & c_{\mathrm{b},X}(t_{\mathrm{eb}})
\end{pmatrix}
\end{equation}
is the rotation matrix for the egress case, and $t_{\mathrm{eb}}$ is the time when the planet's center of mass intersects the edge of the host star during egress. 

For ingress, the $\xii$ component of the displacement can be converted into the lengths $\Rpxip$ and $\Rpxin$ as shown in Figure \ref{fig:def_Rp}.
Similarly, for egress, the $\xe$ component of the displacement can be converted into the lengths $\Rpxep$ and $\Rpxen$.
Note that $\Rpxip$, $\Rpxin$, $\Rpxep$, and $\Rpxen$ are not exact radii, but projected lengths of the planet's shadow disk onto the $\xii$--axis or $\xe$--axis, measured from the planet's center of mass. 
The conversions are given by
\begin{equation}\label{eq:rp_spectra}
\begin{aligned}
\frac{\Rpxip(\lambda)}{R_{s}(\lambda_0)} &=  \alpha(\lambda)k(\lambda) + \Delta c_{\mathrm{i},x\mathrm{i}}(\lambda; \tI(\lambda), \tII(\lambda), k(\lambda)) \\
\frac{\Rpxin(\lambda)}{R_{s}(\lambda_0)} &=  \alpha(\lambda)k(\lambda) - \Delta c_{\mathrm{i},x\mathrm{i}}(\lambda; \tI(\lambda), \tII(\lambda), k(\lambda)) \\
\frac{\Rpxep(\lambda)}{R_{s}(\lambda_0)} &=  \alpha(\lambda)k(\lambda) + \Delta c_{\mathrm{e},x\mathrm{e}}(\lambda; \tIII(\lambda), \tIV(\lambda), k(\lambda)) \\
\frac{\Rpxen(\lambda)}{R_{s}(\lambda_0)} &=  \alpha(\lambda)k(\lambda) - \Delta c_{\mathrm{e},x\mathrm{e}}(\lambda; \tIII(\lambda), \tIV(\lambda), k(\lambda)),
\end{aligned}
\end{equation}
where $\Delta c_{\mathrm{i},x\mathrm{i}}(\lambda)$ is the $\xii$ component of $\Delta \bm{c_{\mathrm{i}}}(\lambda)$, and $\Delta c_{\mathrm{e},x\mathrm{e}}(\lambda)$ is the $\xe$ component of $\Delta \bm{c_{\mathrm{e}}}(\lambda)$. 
This means that we can derive the spectra of $\Rpxip$ and $\Rpxin$ from the chromatic variations in the parameters $\tI$, $\tII$, and $k^2$, and the spectra of $\Rpxep$ and $\Rpxen$ from the chromatic variations in the parameters $\tIII$, $\tIV$, and $k^2$.

\subsection{Constraints Imposed by Using the Conventional Transit Model}\label{subsec:taui=taue}

We use the conventional transit model to infer $\tI$, $\tII$, $\tIII$, $\tIV$, and $k^2$ from observed data. However, it introduces certain constraints. 
Ideally, we would like to treat ingress and egress independently in the above formulation.
This means that the parameters for ingress $\tI$, $\tII$, and $\ki^2$, and the parameters for egress $\tIII$, $\tIV$, and $\ke^2$, should be independent. Here, $\ki^2$ and $\ke^2$ represent the transit depths for ingress and egress, respectively.
However, the conventional transit model imposes the constraint $\ki^2 = \ke^2 = k^2$, which means the apparent planetary radius is assumed to be the same for both ingress and egress at each wavelength.
If the difference between $\ki^2$ and $\ke^2$ becomes significant enough to pose a problem, the light curve depths during ingress and egress would differ. In such cases, verifying that the shape of the light curve remains sufficiently symmetric could serve as a way to assess whether this assumption is valid.

Additionally, the four contact times are not treated independently. 
If we assume a circular orbit, it imposes the constraint $\taui = \taue$, meaning that the durations of ingress and egress are the same.
In this case, the four contact times can be expressed by three parameters, $ \Ttot, \Tfull$, and $t_0$ (Panel (a) of Figure \ref{fig:definition}), as follows:
\begin{equation}\label{eq:params_convert}
\begin{aligned}
\tI &= t_0 - \frac{\Ttot}{2} \\
\tII &= t_0 - \frac{\Tfull}{2} \\
\tIII &= t_0 + \frac{\Tfull}{2} \\
\tIV &= t_0 + \frac{\Ttot}{2}.
\end{aligned}
\end{equation}
Even when considering a non-zero orbital eccentricity, the difference between $\taui$ and $\taue$ remains small \citep{2010exop.book...55W}.
Since $\taui$ and $\taue$ reflect the $Y$ component of $\Delta \bm{c}(\lambda)$, the condition $\taui = \taue$ imposes the constraint that the $Y$ component of $\Delta \bm{c}(\lambda)$ must be the same for both ingress and egress at each wavelength. However, given the sensitivity of $\taui$ and $\taue$ to the magnitude of displacement and the current data quality, the impact of this constraint is likely not significant.

A more flexible model than the conventional transit model is required to remove these constraints. Exploring such models is beyond the scope of this paper.

\subsection{Impact of the Planet's Center of Mass Orbit on the Spectra} \label{subsec:center_of_mass} 

Since we use the planet’s center of mass as a reference for the displacement of the
planet’s shadow disk, accurately determining the orbital parameters of the planet’s center of mass is crucial for investigating atmospheric asymmetries from transit light curves, as highlighted by \citet{2024Natur.632.1017E}. While the orbital parameters determined from transit data at wavelengths that exhibit small radii are considered to be close to those of the center of mass \citep{2012ApJ...751...87D}, these parameters cannot be accurately determined solely from transit light curves because they are affected by atmospheric asymmetries. Although the values measured using the radial velocity method are reliable, achieving the precision required for our analysis is not feasible with current observational capabilities. Therefore, it is essential to discuss how variations in these orbital parameters could impact the inferred spectra.

For simplicity, we assume a circular orbit with zero eccentricity. A detailed analysis considering the effects of orbital eccentricity is presented in \S \ref{sec:validation}.  Under this assumption, the trajectory of the planet's center of mass $\bm{c_{\mathrm{b}}}(t)$ can be expressed as
\begin{equation}
\label{eq:cb_circular}
\bm{c_{\mathrm{b}}}(t)
=
\begin{pmatrix}
a\sin{\left(2\pi\frac{t-t_{0\mathrm{b}}}{P}\right)} \\
-b\cos{\left(2\pi\frac{t-t_{0\mathrm{b}}}{P}\right)}
\end{pmatrix},
\end{equation}
where $P$ is the orbital period, $a$ is the semi-major axis, $b$ is the impact parameter, and $t_{0\mathrm{b}}$ is the central transit time of the center of mass. 

For ingress, $\bm{m}(\lambda)$ can be approximated as follows
\begin{align}
\bm{m}(\lambda)
&= \frac{\bm{c_{\mathrm{b}}}(\tI(\lambda)) + \bm{c_{\mathrm{b}}}(\tII(\lambda))}{2} \nonumber \\
&= 
\begin{pmatrix}
a\sin{\left(2\pi\frac{\ti(\lambda)-t_{0\mathrm{b}}}{P}\right)}\cos{\left(\pi\frac{\taui(\lambda)}{P}\right)} \\
-b\cos{\left(2\pi\frac{\ti(\lambda)-t_{0\mathrm{b}}}{P}\right)}\cos{\left(\pi\frac{\taui(\lambda)}{P}\right)}
\end{pmatrix} \nonumber \\
&\approx
\begin{pmatrix}
2\pi a\frac{\ti(\lambda)-t_{0\mathrm{b}}}{P} \\
-b\left(1-\frac{1}{2}\left(2\pi\frac{\ti(\lambda)-t_{0\mathrm{b}}}{P}\right)^2-\frac{1}{2}\left(\pi\frac{\taui(\lambda)}{P}\right)^2\right)
\end{pmatrix},
\end{align}
where $2\pi(\ti(\lambda)-t_{0\mathrm{b}})/P \ll 1$ and $\pi\taui(\lambda)/P \ll 1$ were used, retaining second-order terms in the approximation.
Interpreting $\bm{m}$ as a function of the orbital parameters $a$, $b$, and $t_{0\mathrm{b}}$, and considering small changes $\Delta a$, $\Delta b$, and $\Delta t_{0\mathrm{b}}$, we have
\begin{align}
\Delta \bm{m} &= \frac{\partial \bm{m}}{\partial a} \Delta a + \frac{\partial \bm{m}}{\partial b} \Delta b + \frac{\partial \bm{m}}{\partial t_{0\mathrm{b}}} \Delta t_{0\mathrm{b}}
 \nonumber \\
&\approx
\begin{pmatrix}
2\pi \left(\frac{\ti(\lambda)-t_{0\mathrm{b}}}{P} \Delta a - \frac{a}{P} \Delta t_{0\mathrm{b}}\right) \\
-\Delta b + 2\pi b \frac{\ti(\lambda)-t_{0\mathrm{b}}}{P} \Delta t_{0\mathrm{b}}
\end{pmatrix},
\end{align}
This indicates that variations in $\bm{m}$ are largely wavelength-independent, as the wavelength dependence of $(\ti(\lambda)-t_{0\mathrm{b}})/P$ is minimal compared to that of $\Delta \bm{c}(\lambda)$.
Given that the term $\mathcal{R}_{d}\bm{s}$ in equation (\ref{eq:result}) is determined by $\bm{d}$ and is approximately positioned on the host star's edge, variations in $\bm{d}$ have a negligible impact on the $\xii$ component of $\Delta \bm{c}(\lambda)$, and consequently on the $\Rpxip$ and $\Rpxin$ spectra. 
Therefore, changes in $a$, $b$, and $t_{0\mathrm{b}}$ primarily result in offsets in the $\xii$ component of $\Delta \bm{c}(\lambda)$, affecting the offsets but not the shape of the $\Rpxip$ and $\Rpxin$ spectra. The same applies to the spectra from egress, $\Rpxep$ and $\Rpxen$.

If the $\xii$ component of $\Delta \bm{m}$ increases, the offset of the $\Rpxip$ spectra decreases, while the offset of the $\Rpxin$ spectra increases. Similarly, if the $\xe$ component of $\Delta \bm{m}$ increases, the offset of the $\Rpxep$ spectra decreases, while that of the $\Rpxen$ spectra increases. As a result, negative values of $\Delta a$, $\Delta b$, or $\Delta t_{0\mathrm{b}}$ decrease the offset of the $\Rpxip$ spectra and increase the offset of the $\Rpxin$ spectra at ingress. Positive values of $\Delta a$ or $\Delta b$, or negative values of $\Delta t_{0\mathrm{b}}$ decrease the offset of the $\Rpxep$ spectra and increase the offset of the $\Rpxen$ spectra at egress.
Therefore, inaccuracies in $a$ or $b$ affect north-south asymmetries, while inaccuracies in $t_{0\mathrm{b}}$ affect morning-evening asymmetries.

Thus, even if the estimated orbital parameters of the planet's center of mass are inaccurate, the shape of the $\Rpxip$, $\Rpxin$, $\Rpxep$, and $\Rpxen$ spectra remains largely unaffected. However, careful consideration is required when interpreting inferred values, such as the VMRs of molecules or temperature, as spectral offsets can influence these results.

\subsection{Wavelength Dependence of Host Star's Radius}\label{subsec:stelar_radius} 
In the formulation, the wavelength dependence of the host star's radius, denoted as $\alpha(\lambda)$, is taken into account. Simulations based on a stellar atmospheric model suggest a wavelength-dependent variation in stellar radii (Appendix \ref{ap:phoenix}).
The wavelength dependence of apparent stellar radii has not been well studied observationally, especially in the near-infrared region. While observations of giant stars, such as Mira variables, suggest a general increase in radius with wavelength in the near-infrared \citep[e.g.,][]{2005ApJ...620..961M, 2008A&A...479L..21W}, similar measurements have not been conducted for main-sequence stars, including the Sun, within this wavelength range \citep{2016IAUS..320..342R}. 

However, transit duration provides valuable information. We can express $R_s(\lambda)$ as
\begin{equation}
\label{eq:Rs_lambda_duration}
R_{s}(\lambda) = a\sqrt{\sin^2{\left(\frac{\pi T_{\mathrm{bc}}(\lambda)}{P}\right)} + \cos^2{\left(\frac{\pi T_{\mathrm{bc}}(\lambda)}{P}\right)}\cos^2{i}},
\end{equation}
where $a$ is the semi-major axis, $i$ is the inclination, 
and $T_{\mathrm{bc}}(\lambda)$ is the transit duration for the center of mass of the planet. 
However, due to atmospheric asymmetries, the exact values of $T_{\mathrm{bc}}(\lambda)$ are not directly measurable, making it challenging to disentangle the wavelength dependence of $T_{\mathrm{bc}}(\lambda)$ from transit data alone. For simplicity, in \S \ref{sec:wasp39b}, we use $(\Ttot + \Tfull)/2$ as a proxy for $T_{\mathrm{bc}}(\lambda)$ (see Appendix \ref{ap:duration}) and consider an illustrative case in which the linear trend of $(T_{\rm tot}+T_{\rm full})/2$ is attributed entirely to $\alpha(\lambda)$.

Given these difficulties,  it is essential to evaluate how inaccuracies in $\alpha(\lambda)$ may affect the inferred spectra of $\Rpxip$, $\Rpxin$, $\Rpxep$, and $\Rpxen$.
Inaccuracies in $\alpha(\lambda)$ primarily influence the inferred spectra through their impact on $\bm{s}$, as defined in equation \eqref{eq:vec_s}. When $\alpha(\lambda) > 1$, the length of $\bm{s}$ increases, leading to a smaller $\xii$ component or a larger $\xe$ component of $\Delta \bm{c}$ compared to the case where $\alpha(\lambda) = 1$. As a result, $\Rpxip$ and $\Rpxen$ decrease, while $\Rpxin$ and $\Rpxep$ increase compared to the case where $\alpha(\lambda) = 1$. 
Similarly, when $\alpha(\lambda) < 1$, $\Rpxip$ and $\Rpxen$ increase, while $\Rpxin$ and $\Rpxep$ decrease compared to the case where $\alpha(\lambda) = 1$. 
Notably, the effect of $\alpha(\lambda)$ is opposite between the spectra of the north and south limbs.
These changes in the spectra of $\Rpxip$, $\Rpxin$, $\Rpxep$, and $\Rpxen$ can affect the inferred volume mixing ratios of molecules and atmospheric temperatures during atmospheric retrievals. Therefore, careful consideration is required when interpreting these inferred values.

Although this does not apply to WASP-39, which we investigate in \S \ref{sec:wasp39b}, a potential method for constraining the wavelength dependence of the host star's radius is to observe the transit durations of planets with negligible atmospheric asymmetries, such as those with thin atmospheres. For example, transits of Mercury and Venus have been used to estimate the solar radius \citep{2012ApJ...750..135E, 2015ApJ...798...48E}.

\section{Validation of the Method}\label{sec:validation}
This section validates the method by recovering atmospheric asymmetry signals from simulated transit light curves. Additionally, we examine potential false signals of asymmetry introduced by the method. For simplicity, $\alpha = \Rs(\lambda)/\Rs(\lambda_{0})$ is set to 1 in the following simulations, assuming no wavelength dependence of the stellar radius. While the effect of $\alpha$ was discussed in \S \ref{subsec:stelar_radius}, we revisit this effect in \S \ref{s:recovery}. The simulations presented here utilized the Python packages {\sf jaxoplanet} \citep{2024zndo..10736936H} and {\sf karate} (\url{https://github.com/sh-tada/karate}). The {\sf karate} package provides functions for converting the planetary radius and contact times into $\Delta \bm{c}$ during ingress and egress, or $\Rpxip$, $\Rpxin$, $\Rpxep$, and $\Rpxen$.

\subsection{Recovering Morning-Evening Asymmetries}\label{ss:catwoman}
\begin{figure*}[htb!]
\centering
\includegraphics[width=0.8\linewidth]{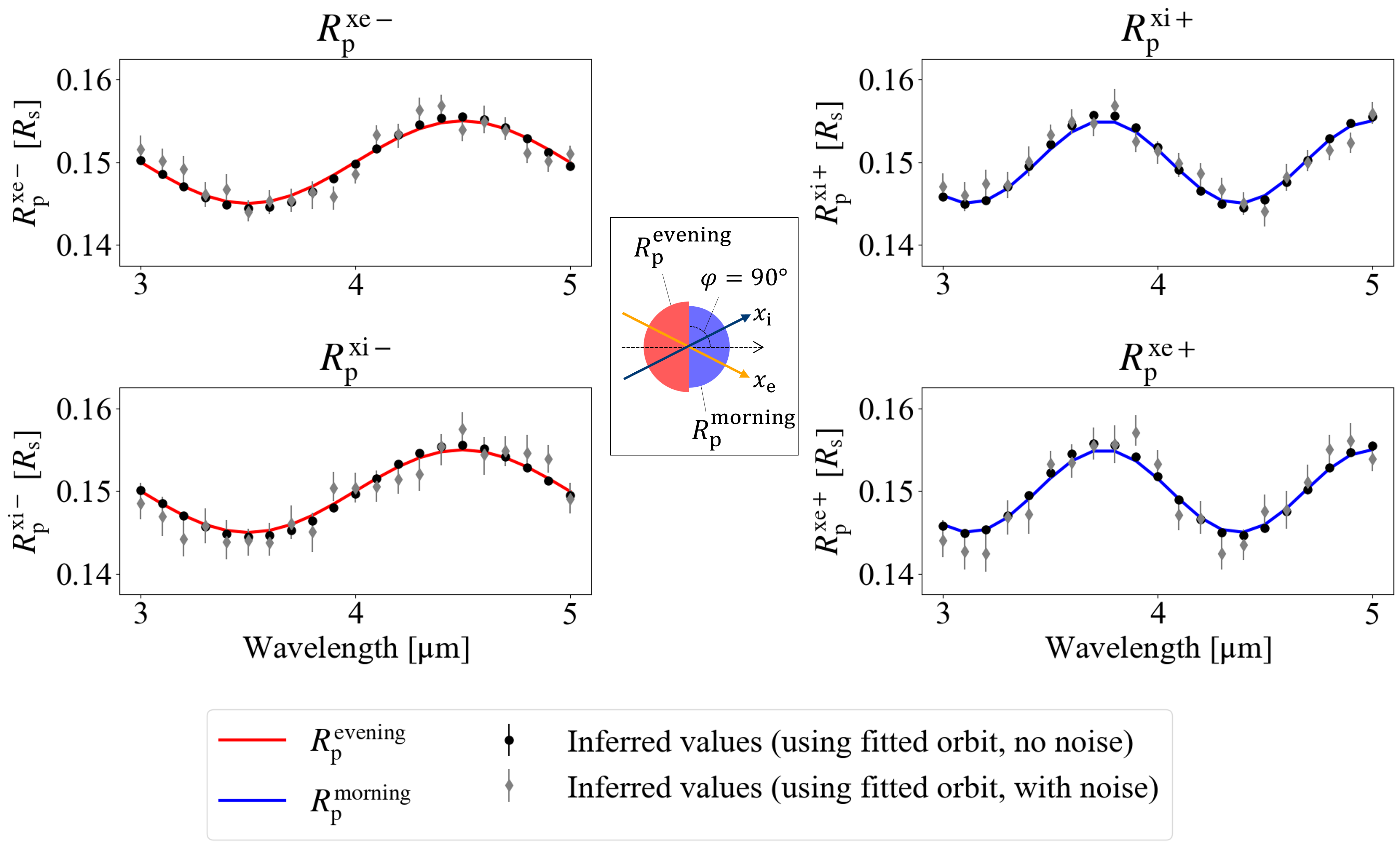}
\caption{Inferred values of $\Rpxip$, $\Rpxin$, $\Rpxep$, and $\Rpxen$ based on the MCMC analysis of light curves simulated using {\sf catwoman} \citep{2020JOSS....5.2382J, 2021AJ....162..165E}. The values were converted using the estimated orbital parameters. The dots represent the median values from the MCMC sampling, while the error bars indicate the 68\% credible intervals. Results are shown for both noiseless data and data with added Gaussian noise (mean = 0, standard deviation = 250 ppm). The true values of $\Rp^{\mathrm{morning}}$ (blue) and $\Rp^{\mathrm{evening}}$ (red) are shown as colored lines. The arrangement of the panels reflects the positional relationship between the regions on the planet contributing to each spectrum, as viewed from the night side.
The central panel shows the shape of the planetary shadow, along with the $\xii$--axis, $\xe$--axis. $\varphi$ is the angle between the direction of the planet's orbital motion and the boundary of the semi-circles, set to $90^{\circ}$ to model a morning-evening asymmetry. The radius of the morning-side semi-circle is denoted as $\Rp^{\mathrm{morning}}$, and the radius of the evening-side semi-circle is denoted as $\Rp^{\mathrm{evening}}$.
\label{fig:sim_catwoman_ecc0_rp4}}
\end{figure*}

First, we examine whether the wavelength-dependent morning-evening asymmetries of the planetary shadow can be accurately recovered from simulated transit light curves. 
Light curves were simulated using {\sf catwoman} \citep{2020JOSS....5.2382J, 2021AJ....162..165E}, which models the planetary shadow as two semi-circles with independently adjustable radii to represent atmospheric asymmetries between the morning and evening terminators. In this simulation, we set $\varphi$, the angle between the direction of the planet's orbital motion and the boundary of the semi-circles (Figure \ref{fig:sim_catwoman_ecc0_rp4}), to $90^{\circ}$ to model a morning-evening asymmetry.

We focused on a planet with zero eccentricity. The remaining orbital parameters were based on WASP-39 b, with an orbital period $P = 4.06\ \mathrm{days}$, semi-major axis $a = 11.4\ \Rs$, and impact parameter $b = 0.45$ \citep[]{2011A&A...531A..40F}. The time of inferior conjunction $t_0$ was set to 0, and the coefficients of a quadratic limb-darkening model $(u_1,\ u_2)$ were set to $(0.1,\ 0.1)$. 

\begin{table}[bt!]
\centering
\caption{Parameters and prior distributions for the transit light curve model under the assumption of a circular orbit, used in the MCMC analysis of simulated light curves.\label{tab:sim_params_lc_e0}}
\begin{tabular}{lll}
\hline\hline
Symbol & Description & Prior\\
\hline
$k^2$ & transit depth & $\mathcal{U}(0, 0.1)$ \\
$t_{0}$ & central transit time [s] & $\mathcal{U}(-5000, 5000)$ \\
$\Ttot$ & total transit duration [s] & $\mathcal{U}(5000, 15000)$ \\
$\Tfull$ & full transit duration [s] & $\mathcal{U}(1, \Tfull^{\mathrm{max}})$ \\
$c_{\mathrm{base}}$ & baseline coefficient & $\mathcal{U}(0.99, 1.01)$ \\
$jitter$ & jitter error & $\mathcal{U}(0, 0.01)$ \\
$u_{1}$ & limb darkening coefficient & $\mathcal{U}(-3,3)$ \\
$u_{2}$ & limb darkening coefficient & $\mathcal{U}(-3,3)$ \\
\hline
\end{tabular}
\tablecomments{$\mathcal{U}(a,b)$: the uniform distribution from $a$ to $b$. $\Tfull^{\mathrm{max}}$ is defined in equation \eqref{eq:Tfullmax}.}
\end{table}

The radius of the morning-side semi-circle, $\Rp^{\mathrm{morning}}(\lambda)$, and the radius of the evening-side semi-circle, $\Rp^{\mathrm{evening}}(\lambda)$ (Figure \ref{fig:sim_catwoman_ecc0_rp4}), varied from $0.145$ to $0.155\ \Rs$ independently, generating 21 normalized transit light curves. For ease of visualization, these light curves were assigned wavelengths ranging from $3$ to $5\ \mathrm{\mu m}$.
The integration time was set to 65 seconds, closely matching the 63.14 seconds used in observational data for WASP-39 b obtained with JWST NIRSpec/G395H \citep[][]{2023Natur.614..664A, 2016PASP..128i4401S, 2018PASP..130k4402B}, analyzed in \S \ref{sec:wasp39b}.
Both noiseless data and data with added Gaussian noise (mean = 0, standard deviation = 250 ppm) were generated to simulate observational uncertainties. The noise level, with a standard deviation of 250 ppm, is comparable to the most precise light curves analyzed in \S \ref{sec:wasp39b}.

First, we measured the planetary radius $k$ and contact times $\tI$, $\tII$, $\tIII$, and $\tIV$ from the simulated light curves using the conventional transit light curve model assuming a circular orbit. 
The analysis employed the Hamiltonian Monte Carlo - No U-Turn Sampler (HMC-NUTS), a specific Markov Chain Monte Carlo (MCMC) algorithm, implemented in {\sf NumPyro} \citep{2019arXiv191211554P}. The parameters of the transit light curve model and their prior distributions are summarized in Table \ref{tab:sim_params_lc_e0}. The light curve baseline was assumed to be constant ($c_{base}$). The likelihood of the light curve data was modeled as a normal distribution, with the mean defined as the product of the transit light curve model and the baseline, and the variance modeled as $jitter^2$. In this analysis, the orbital period $P$ was fixed to its true value.

The semi-major axis $a$ and inclination $i$ were derived from the transit depth $k^2$, the central transit time $t_0$, the total transit duration $\Ttot$, and the full transit duration $\Tfull$ as follows
\begin{align}
\label{eq:a_from_durations}
a &= \sqrt{\frac{(1+k)^2\cos^2\left(\frac{\pi \Tfull}{P}\right) - (1-k)^2\cos^2\left(\frac{\pi \Ttot}{P}\right)}{\sin\left(\frac{\pi \Ttot + \pi \Tfull}{P}\right) \sin\left(\frac{\pi \Ttot - \pi \Tfull}{P}\right)}} \\
\label{eq:cosi_from_durations}
\cos{i} &= \sqrt{\frac{-(1+k)^2\sin^2\left(\frac{\pi \Tfull}{P}\right) + (1-k)^2\sin^2\left(\frac{\pi \Ttot}{P}\right)}{(1+k)^2\cos^2\left(\frac{\pi \Tfull}{P}\right) - (1-k)^2\cos^2\left(\frac{\pi \Ttot}{P}\right)}}.
\end{align}
Valid values for $a$ and $i$ can only be obtained if $k^2$, $\Ttot$, and $\Tfull$ satisfy the following condition
\begin{equation}
\begin{cases}
\begin{aligned}
    (1-k)^2\sin^2\left(\frac{\pi \Ttot}{P}\right) - (1+k)^2\sin^2\left(\frac{\pi \Tfull}{P}\right) &\geq 0 \\
    \Ttot &> \Tfull.
\end{aligned}
\end{cases}
\end{equation}
To avoid indeterminate solutions, we impose an upper bound on $\Tfull$, which is defined by
\begin{equation}
\label{eq:Tfullmax}
\Tfull^{\mathrm{max}} = \frac{P}{\pi}\arcsin{\left(\frac{1-k}{1+k}\sin{\left(\frac{\pi \Ttot}{P}\right)}\right)}.
\end{equation}

The inferred values of $k$, $\tI$, $\tII$, $\tIII$, and $\tIV$ can be converted to $\Delta \bm{c}$ during ingress and egress by assuming the orbital parameters of the planet's center of mass. Here, we derive these orbital parameters from the inferred values obtained through the MCMC analysis. We calculate the median of the MCMC samples converted into orbital parameters of $a$, $b$, and $t_0$ for each light curve, and then take the mean of these medians as the orbital parameters for the conversion.
For noiseless data and data with added Gaussian noise, the resulting orbital parameters were $(a, b, t_0) = (11.404\ \mathrm{\Rs},\ 0.449,\ 0.783\ \mathrm{s})$ and $(11.409\ \mathrm{\Rs},\ 0.449,\ 2.86\ \mathrm{s})$, respectively. These values deviate from the true values due to the asymmetries in the planetary shadow, even in the case of noiseless data. The orbital period $P$ used for the conversion was the true value.

We can convert $k$, $\Delta c_{\mathrm{e},x\mathrm{e}}$, and $\Delta c_{\mathrm{i},x\mathrm{i}}$ to $\Rpxip$, $\Rpxin$, $\Rpxep$, and $\Rpxen$ using equation \eqref{eq:rp_spectra}.
The resulting spectra are shown in Figure \ref{fig:sim_catwoman_ecc0_rp4}. 
Based on the shape of the planetary shadow, $\Rpxip$ and $\Rpxep$ are expected to reflect the spectrum of $\Rp^{\mathrm{morning}}$, while $\Rpxin$ and $\Rpxen$ are expected to reflect the spectrum of $\Rp^{\mathrm{evening}}$.
The inferred values align closely with these expectations. These results demonstrate that this method can effectively capture morning-evening asymmetries, even when the shape of the planetary shadow deviates from being perfectly circular.

\subsection{Symmetric Planets with Non-Zero Eccentricity}\label{s:nosignal}

Next, we confirm that no asymmetries are detected from the transit light curves of symmetric planets. To assess the impact of planetary eccentricity on asymmetry detection, we analyzed simulated transit light curves for three planets with different eccentricities $e$ and arguments of periastron $\omega$. The three cases considered are $(e\cos\omega,\ e\sin\omega) = (0,\ 0)$, $(0.1,\ 0)$, and $(0,\ 0.1)$.
We simulated normalized transit light curves by varying $k$, the ratio of the planetary radius to the stellar radius, from 0.145 to 0.155, generating 21 light curves for each planet. All other parameters for the light curve simulation were identical to those described in \S \ref{ss:catwoman}. 
Both noiseless data and data with added Gaussian noise (mean = 0, standard deviation = 250 ppm) were generated to simulate observational uncertainties.
The analysis procedures were identical to those described in \S \ref{ss:catwoman}.
It should be noted that the transit light curves of planets with elliptical orbits were also fitted using the circular orbit model.

\begin{figure}[tb!]
\centering
\includegraphics[width=0.8\linewidth]{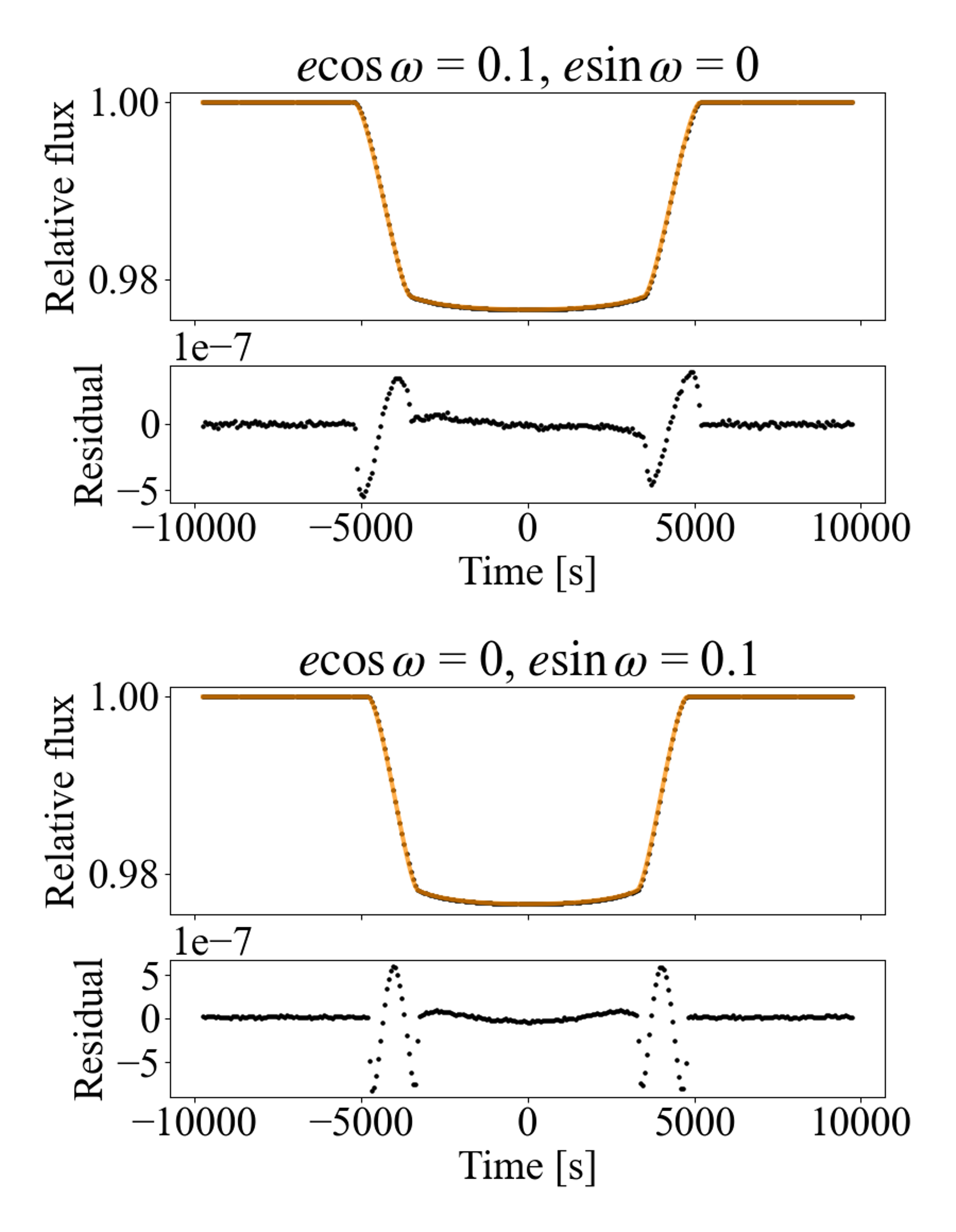}
\caption{Simulated noiseless light curves of symmetric planets with elliptical orbits (black dots) and fitted light curves using the circular orbit model (orange lines). The bottom panels show the residuals. (Top) Case of $(e\cos\omega,\ e\sin\omega) = (0.1,\ 0)$. (Bottom) Case of $(e\cos\omega,\ e\sin\omega) = (0,\ 0.1)$.\label{fig:sim_light_curve_ecosw_esinw}}
\end{figure}

The simulated noiseless light curves of symmetric planets with elliptical orbits, $(e\cos\omega,\ e\sin\omega) = (0.1,\ 0)$ and $(0,\ 0.1)$, along with fitted light curves using the circular orbit model, are shown in Figure \ref{fig:sim_light_curve_ecosw_esinw}. The residuals are small, indicating that light curves with non-zero eccentricity can be well-fitted using the circular orbit model.

\begin{figure*}[tb!]
\centering
\includegraphics[width=0.8\linewidth]{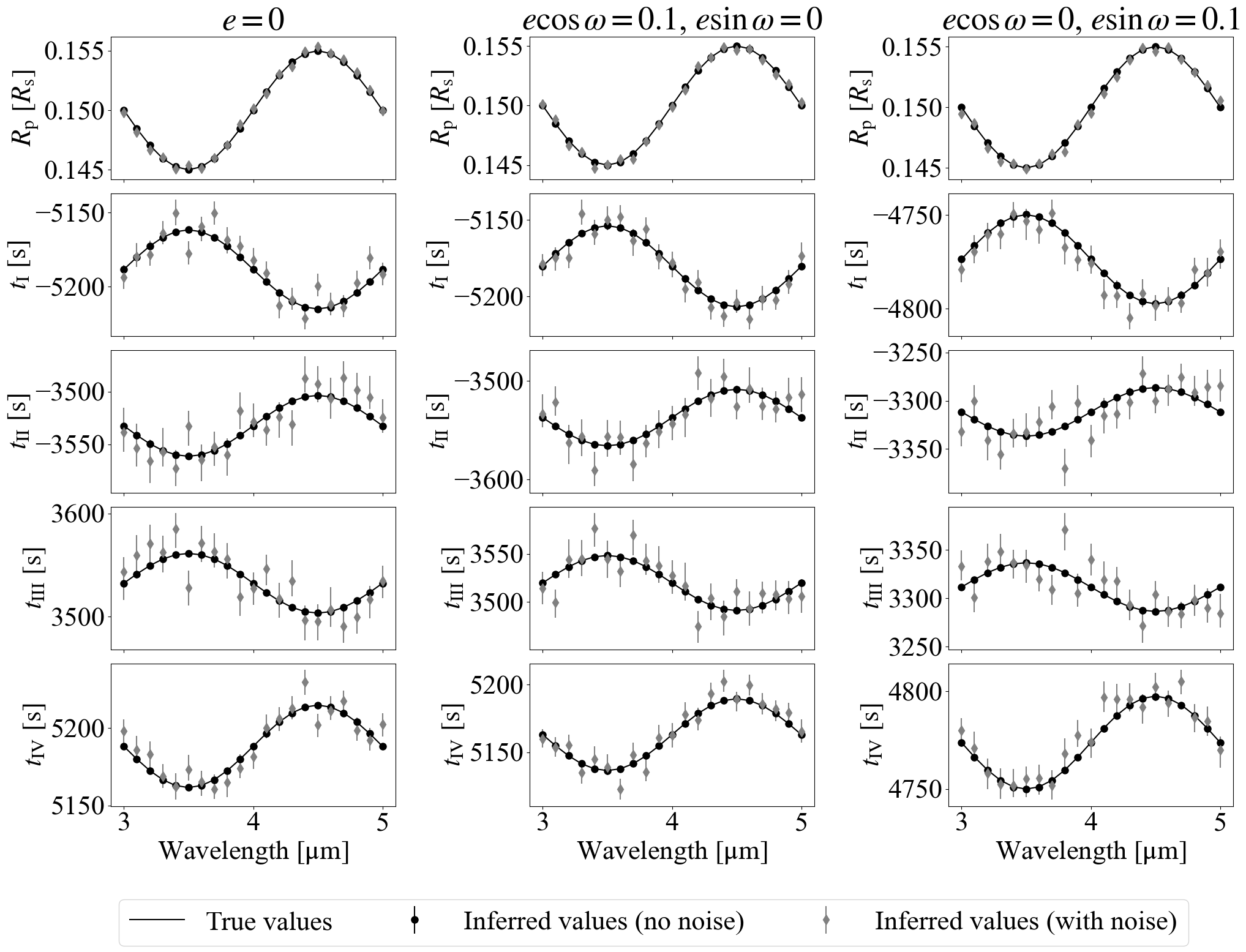}
\caption{Inferred values of $k = \Rp / \Rs$ and the contact times $\tI$, $\tII$, $\tIII$, and $\tIV$ for three symmetric planets with $(e\cos\omega,\ e\sin\omega) = (0,\ 0)$, $(0.1,\ 0)$, and $(0,\ 0.1)$, based on the MCMC analysis of simulated data. The circular orbit model was used to derive these values in the MCMC analysis. The dots represent the median values from the MCMC sampling, while the error bars indicate the 68\% credible intervals. Results are shown for both noiseless data and data with added Gaussian noise (mean = 0, standard deviation = 250 ppm), along with the true values.\label{fig:sim_dc0_ecc0_rpct}}
\end{figure*}

Figure \ref{fig:sim_dc0_ecc0_rpct} shows the inferred values of $k = \Rp / \Rs$ and the contact times $\tI$, $\tII$, $\tIII$, and $\tIV$ for the three planets, along with their true values. Results are presented for both noiseless data and data with added Gaussian noise. The inferred values closely match the true values. This result reflects the fact that the difference in the durations of ingress and egress is approximately $e\cos\omega(\Rs/a)^3$, which remains small even for planets with non-zero eccentricity \citep{2010exop.book...55W}.
Focusing on the noiseless data, the errors in $k$ were nearly zero. For the planet with $e=0$, the errors in the contact times were also nearly zero. For planets with $(e\cos\omega,\ e\sin\omega) = (0.1,\ 0)$ and $(0,\ 0.1)$, the errors in the contact times were $\sim 0.1\ \mathrm{s}$. These results indicate that, for these planets, as long as the uncertainty of the inferred values exceeds approximately $0.1\ \mathrm{s}$, using a circular orbit model to measure these values does not introduce significant errors.

\begin{figure*}[tb!]
\centering
\includegraphics[width=0.8\linewidth]{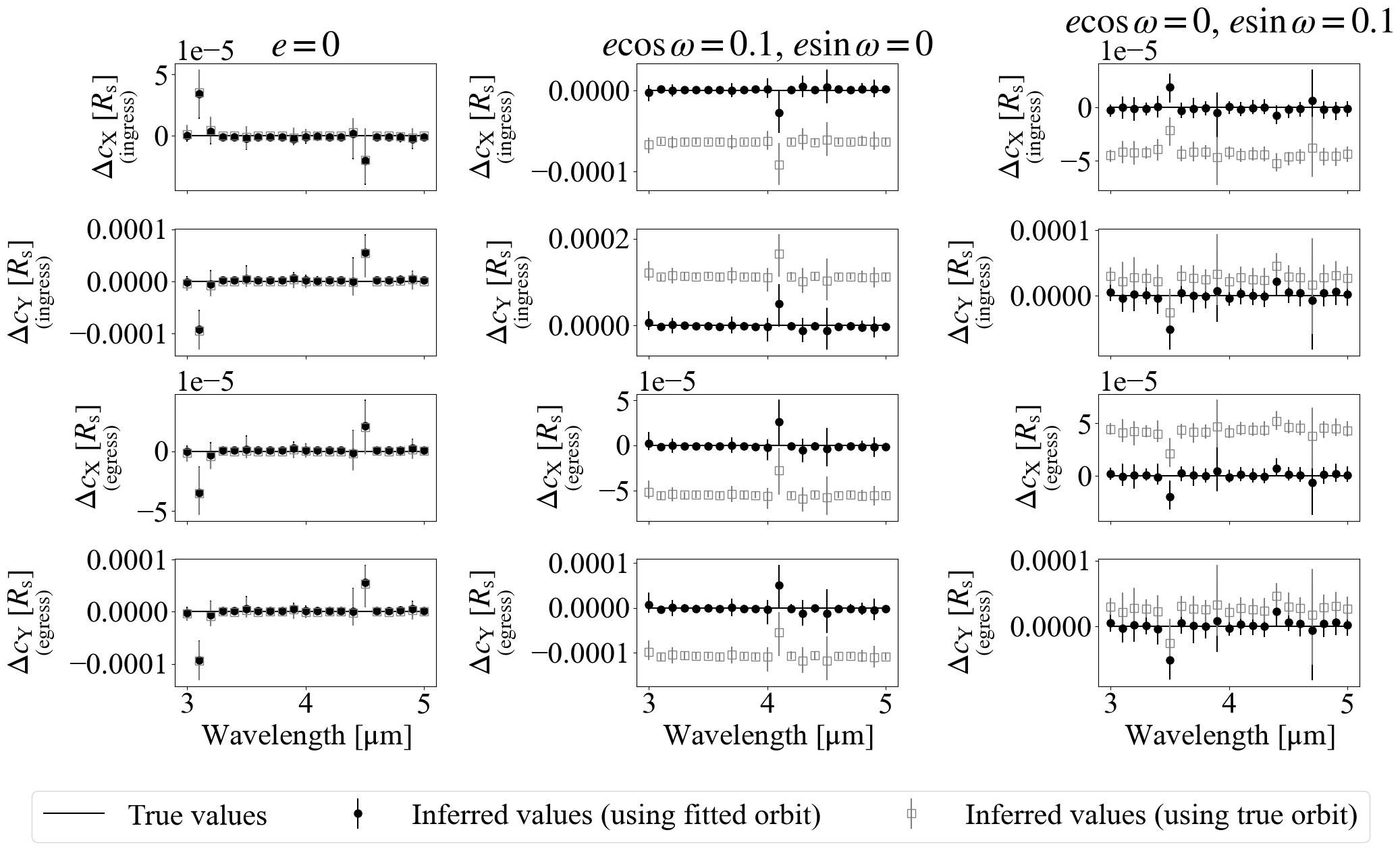}
\caption{Inferred values of $\Delta c_{\mathrm{X}}$ and $\Delta c_{\mathrm{Y}}$ during ingress and egress for three symmetric planets with $(e\cos\omega,\ e\sin\omega) = (0,\ 0)$, $(0.1,\ 0)$, and $(0,\ 0.1)$, based on the MCMC analysis of simulated data.
The values converted using both estimated and true orbital parameters are shown. The circular orbit model was used to derive these values in the MCMC analysis. The dots represent the median values from the MCMC sampling, and the error bars indicate the 68\% credible intervals. Results are shown only for noiseless data, along with the true values.\label{fig:sim_dc0_ecc0_dc}}
\end{figure*}

The inferred values of $k$, $\tI$, $\tII$, $\tIII$, and $\tIV$ can be converted to $\Delta \bm{c}$ during ingress and egress by assuming the orbital parameters of the planet's center of mass. 
The orbital parameters used for these conversions were $(a, b, t_0) = (11.400\ \mathrm{\Rs},\ 0.4500,\ -5.7\times10^{-5}\ \mathrm{s})$, $(11.457\ \mathrm{\Rs},\ 0.4455,\ -8.653\ \mathrm{s})$, and $(12.593\ \mathrm{\Rs},\ 0.4056,\ -3.6\times10^{-5}\ \mathrm{s})$ for planets with $(e\cos\omega,\ e\sin\omega) = (0,\ 0)$, $(0.1,\ 0)$, and $(0,\ 0.1)$, respectively.

\begin{figure*}[bt!]
\centering
\includegraphics[width=0.8\linewidth]{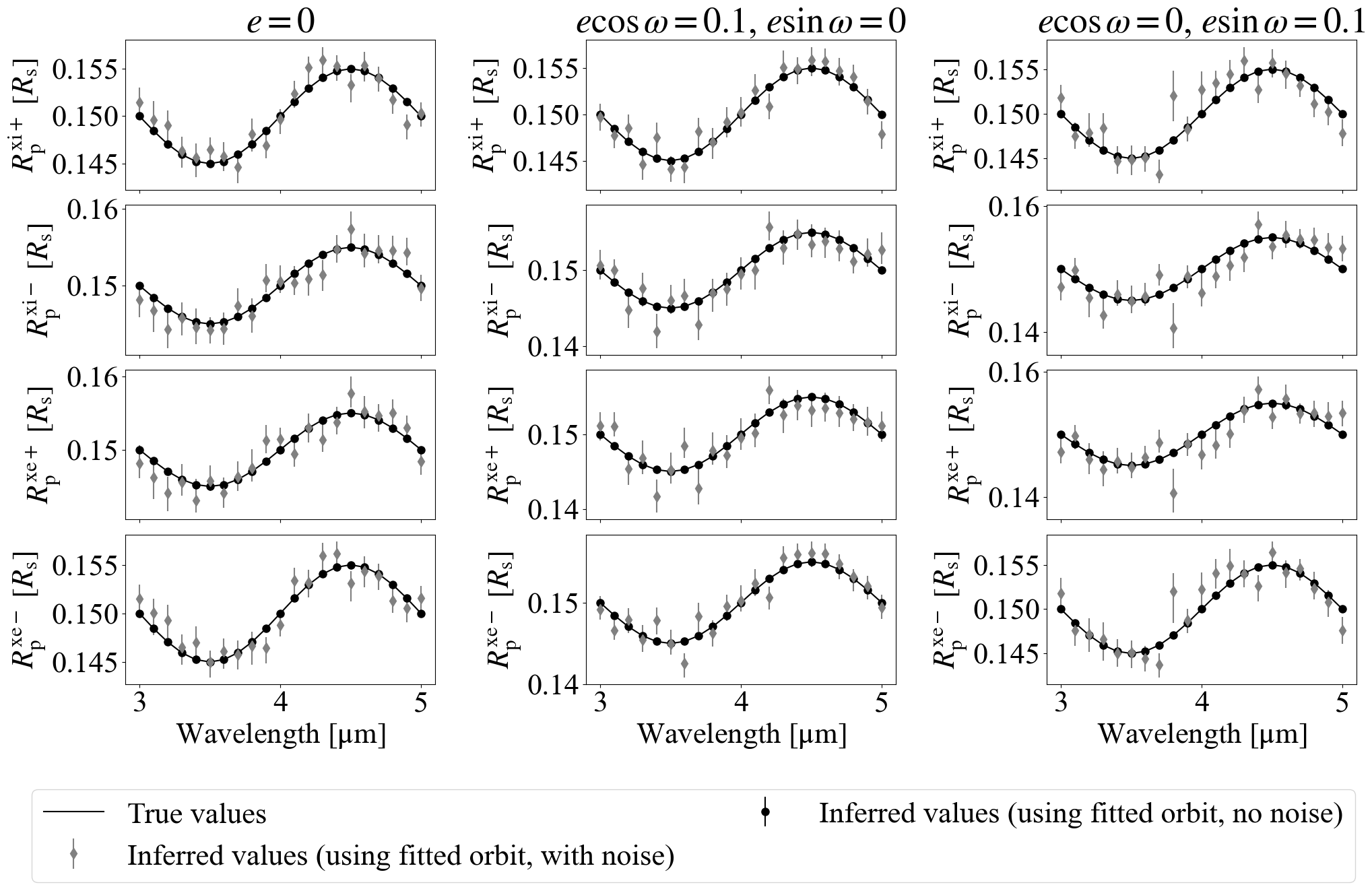}
\caption{Inferred values of $\Rpxip$, $\Rpxin$, $\Rpxep$, and $\Rpxen$ for three symmetric planets with $(e\cos\omega,\ e\sin\omega) = (0,\ 0)$, $(0.1,\ 0)$, and $(0,\ 0.1)$, based on the MCMC analysis of simulated data.
The values were converted using the estimated orbital parameters. The circular orbit model was used in the MCMC analysis to derive these values. The dots represent the median values from the MCMC sampling, while the error bars indicate the 68\% credible intervals. Results are shown for both noiseless data and data with added Gaussian noise (mean = 0, standard deviation = 250 ppm), along with the true values.\label{fig:sim_dc0_ecc0_rp4}}
\end{figure*}

Figure \ref{fig:sim_dc0_ecc0_dc} presents the converted $\Delta c_{\mathrm{X}}$ and $\Delta c_{\mathrm{Y}}$ during ingress and egress, focusing on the noiseless data. For comparison, $\Delta c_{\mathrm{X}}$ and $\Delta c_{\mathrm{Y}}$ converted using the true orbital parameters are also shown. The results indicate that $\Delta \bm{c}$ values converted using the estimated orbital parameters closely match the true value of zero, with errors less than $1\times10^{-4} \Rs$.
Despite the presence of errors $\sim 0.1\ \mathrm{s}$ in the measured contact times for planets with $(e\cos\omega,\ e\sin\omega) = (0.1,\ 0)$ and $(0,\ 0.1)$, the orbital parameters estimated from the MCMC analysis effectively represent an averaged position of the planetary shadow during ingress and egress. Consequently, $\Delta \bm{c}$, the relative position of the planetary shadow, can be measured accurately. In contrast, $\Delta \bm{c}$ values converted using the true orbital parameters show small shifts from zero, reflecting the errors in the measured contact times.

Figure \ref{fig:sim_dc0_ecc0_rp4} shows the inferred values of $\Rpxip$, $\Rpxin$, $\Rpxep$, and $\Rpxen$. For both noiseless data and data with added Gaussian noise, the estimated orbital parameters were used for these conversions. The results demonstrate that the inferred values closely match the true values and do not detect any false asymmetries.

\subsection{Asymmetric Planet with Day-Night Asymmetries}\label{s:recovery}

Here, we examine whether the wavelength-dependent displacement of the planetary shadow can be accurately recovered from simulated transit light curves. In the light curve simulation, the $X$ and $Y$ components of the displacement $\Delta \bm{c}$ varied with wavelength, ranging from $-0.005$ to $0.005\ \Rs$, respectively. These variations were defined independently for ingress and egress. Since $\Delta c_{X}$ and $\Delta c_{Y}$ take different values for ingress and egress, they exhibit time dependence. In this simulation, $\Delta c_{X}$ and $\Delta c_{Y}$ were constant before $\tII$ and after $\tIII$, and varied linearly with time between $\tII$ and $\tIII$.
All other parameters for the light curve simulation were identical to those described in \S \ref{s:nosignal}. 
For this analysis, we focused on a planet with zero eccentricity because, as confirmed in \S \ref{s:nosignal}, the effect of planetary eccentricity is negligible, at least for $e < 0.1$. 
Both noiseless data and data with added Gaussian noise (mean = 0, standard deviation = 250 ppm) were generated to simulate observational uncertainties.
The analysis procedures were identical to those described in \S \ref{ss:catwoman}.

\begin{figure}[tb!]
\centering
\includegraphics[width=\linewidth]{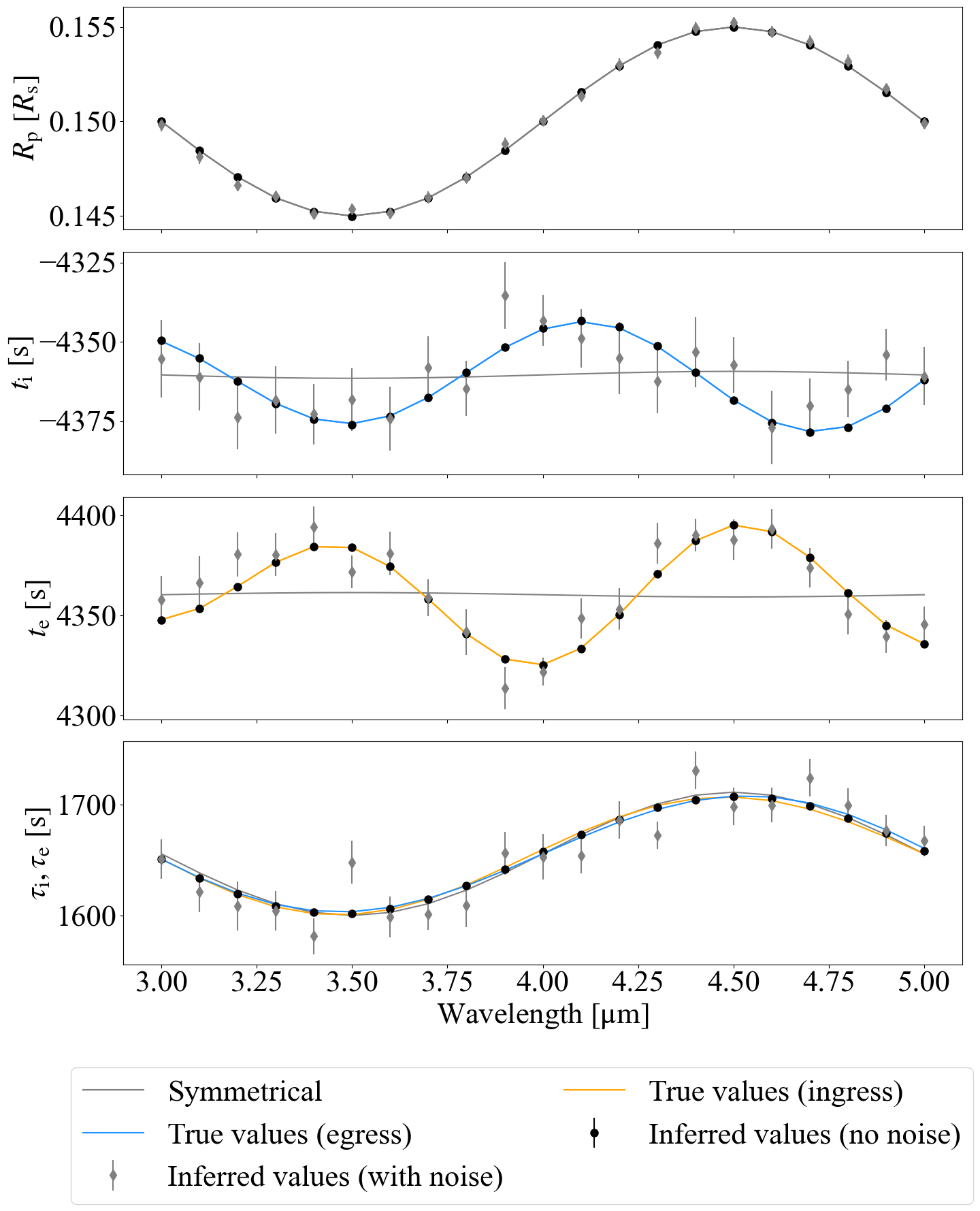}
\caption{Inferred values of $k = \Rp / \Rs$, the timing of ingress $\ti$, the timing of egress $\te$, the duration of ingress $\taui$, and the duration of egress $\taue$ for an asymmetric planet, based on the MCMC analysis of simulated data. The dots represent the median values from the MCMC sampling, while the error bars indicate the 68\% credible intervals. Results are shown for both noiseless data and data with added Gaussian noise (mean = 0, standard deviation = 250 ppm), along with the true values (colored lines). The values for a symmetric planet are also plotted (gray lines). The inferred $\taui$ and $\taue$ are identical because a circular orbit model was used in the analysis. In the panel for $\taui$ and $\taue$, the true values of both are plotted but are difficult to distinguish as the lines are very close and nearly overlap.\label{fig:sim_dc_ecc0_rptie}}
\end{figure}

Figure \ref{fig:sim_dc_ecc0_rptie} shows the inferred values of selected parameters for the planet with the displacement $\Delta \bm{c}$, along with their true values and the corresponding values for a symmetric planet. The results include $k = \Rp / \Rs$, the timing of ingress $\ti$, the timing of egress $\te$, the duration of ingress $\taui$, and the duration of egress $\taue$, as these parameters are closely related to atmospheric asymmetries, as described in \S \ref{sec:formulation}. Results are presented for both noiseless data and data with added Gaussian noise. The inferred values closely match the true values.
In the panel for $\taui$ and $\taue$, the true values of both $\taui$ and $\taue$ are plotted but are difficult to distinguish, as the lines are very close and nearly overlap. This highlights the low sensitivity of $\taui$ and $\taue$ to atmospheric asymmetries, as described in \S \ref{sec:formulation}, implying the challenges of measuring atmospheric asymmetry along the $Y$-axis.

\begin{figure*}[tb!]
\centering
\includegraphics[width=0.8\linewidth]{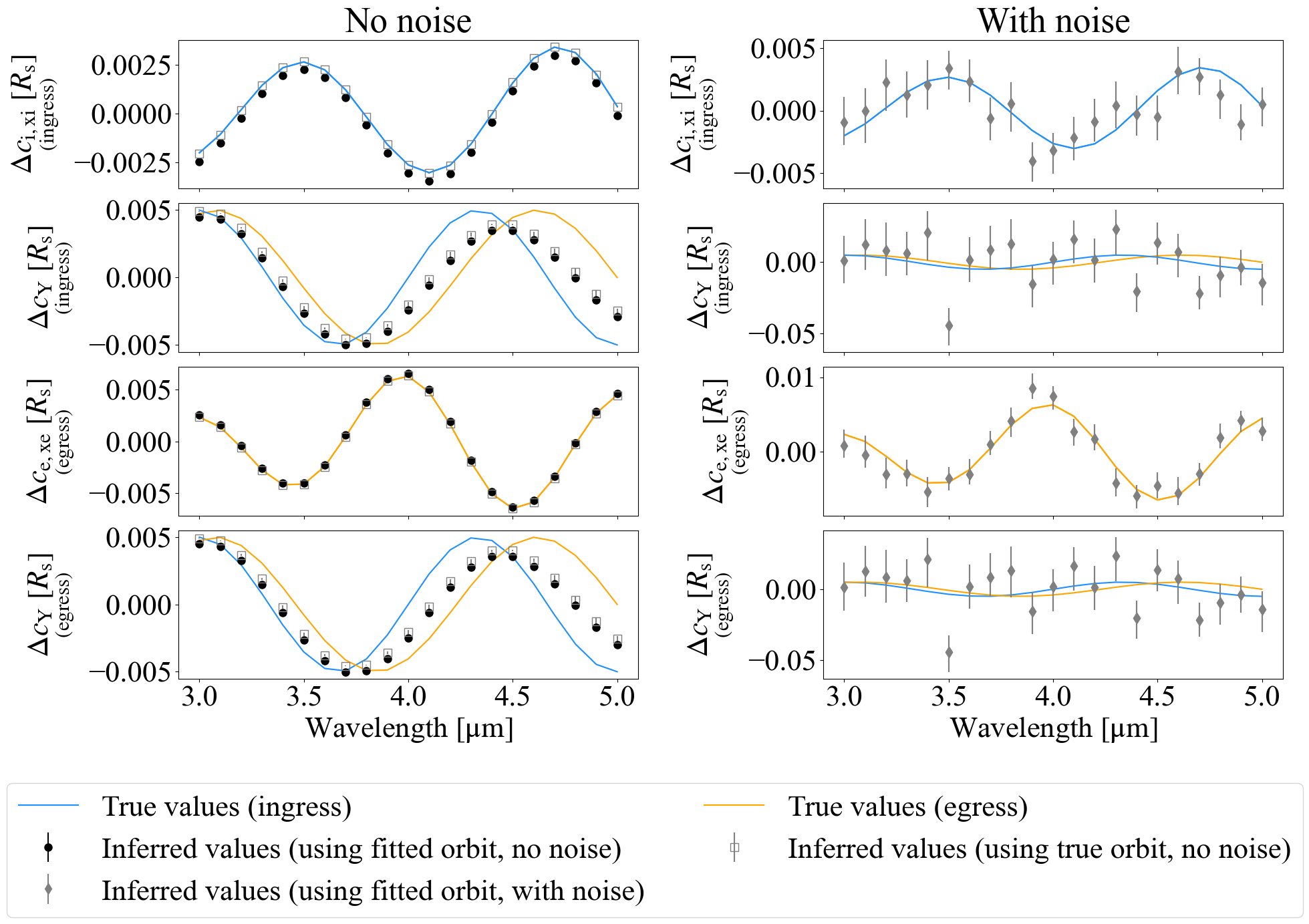}
\caption{Inferred values of $\Delta c_{\mathrm{i},x\mathrm{i}}$, $\Delta c_{\mathrm{e},x\mathrm{e}}$, and $\Delta c_{\mathrm{Y}}$ during ingress and egress for an asymmetric planet, based on the MCMC analysis of simulated data.
The values were converted using the estimated orbital parameters.
The dots represent the median values from the MCMC sampling, and the error bars indicate the 68\% credible intervals. Results are shown for noiseless data (left) and data with added Gaussian noise (mean = 0, standard deviation = 250 ppm) (right), along with the true values.\label{fig:sim_dc_ecc0_dc}}
\end{figure*}

We derived the orbital parameters from the inferred values obtained through the MCMC analysis and converted $k$, $\tI$, $\tII$, $\tIII$, and $\tIV$ to $\Delta \bm{c}$. The orbital parameters used for these conversions were $(a, b, t_0) = (11.399\ \mathrm{\Rs},\ 0.4494,\ -0.663\ \mathrm{s})$ for noiseless data, and $(a, b, t_0)$ $= (11.403\ \mathrm{\Rs}$, $0.4488$, $1.426\ \mathrm{s})$ for data with added Gaussian noise. 

Figure \ref{fig:sim_dc_ecc0_dc} shows the converted values of $\Delta c_{\mathrm{i},x\mathrm{i}}$, $\Delta c_{\mathrm{e},x\mathrm{e}}$, and $\Delta c_{\mathrm{Y}}$ during ingress and egress. For comparison, the values converted using the true orbital parameters are also shown for noiseless data. The converted $\Delta c_{\mathrm{i},x\mathrm{i}}$ and $\Delta c_{\mathrm{e},x\mathrm{e}}$ closely align with the true values. However, for noiseless data, the values converted using the estimated orbital parameters exhibit slight shifts compared to the true values or those derived using the true orbital parameters. This discrepancy arises because the orbital parameters estimated from the MCMC analysis are influenced by atmospheric asymmetries, leading to offset-like errors in the converted values.

The inferred values of $\Delta c_{\mathrm{Y}}$ do not accurately reproduce the true values. This is because the inferred $\Delta c_{\mathrm{Y}}$ values for ingress and egress are nearly identical, as discussed in \S \ref{subsec:taui=taue}. Instead, these values are comparable to the mean of the true $\Delta c_{\mathrm{Y}}$ at ingress and egress. Nevertheless, for data with added noise, the large uncertainties associated with $\Delta c_{\mathrm{Y}}$ effectively mask this limitation, making it negligible in practice.

\begin{figure}[tb!]
\centering
\includegraphics[width=\linewidth]{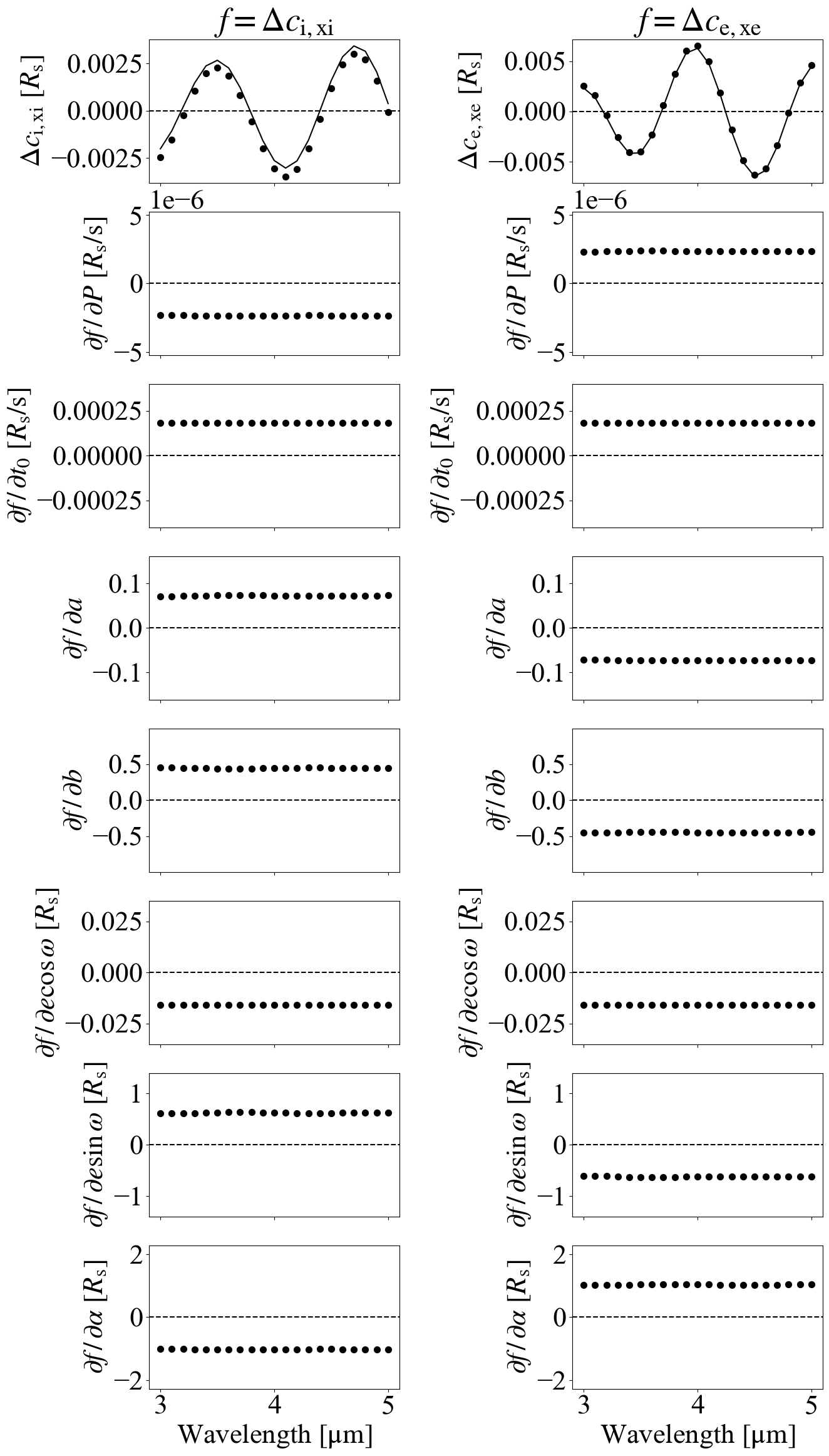}
\caption{Partial derivatives of $\Delta c_{\mathrm{i},x\mathrm{i}}$ (left) and $\Delta c_{\mathrm{e},x\mathrm{e}}$ (right) with respect to each orbital parameter. The functions used to estimate the partial derivatives were $\Delta c_{\mathrm{i},x\mathrm{i}}$ and $\Delta c_{\mathrm{e},x\mathrm{e}}$, which were derived by converting the median values of $k$, $\tI$, $\tII$, $\tIII$, and $\tIV$ obtained from the MCMC analysis using the orbital parameters estimated from the same analysis. $\Delta c_{\mathrm{i},x\mathrm{i}}$ and $\Delta c_{\mathrm{e},x\mathrm{e}}$ are also shown at the top panels, along with the true values.\label{fig:sim_dc_grad}}
\end{figure}

In \S \ref{subsec:center_of_mass}, we investigated the effect of errors in the orbital parameters of the planet's center of mass. We found that these errors have an almost wavelength-independent effect on $\Delta c_{\mathrm{i},x\mathrm{i}}$ and $\Delta c_{\mathrm{e},x\mathrm{e}}$, or $\Rpxip$, $\Rpxin$, $\Rpxep$, and $\Rpxen$. This can be confirmed by examining the partial derivatives of these values with respect to each orbital parameter.
Since the {\sf karate} package is built on {\sf JAX} \citep{jax2018github}, a library for array-oriented numerical computation with automatic differentiation, we can efficiently compute the partial derivatives of these values. 

For simplicity, the functions used to estimate the partial derivatives were $\Delta c_{\mathrm{i},x\mathrm{i}}$ and $\Delta c_{\mathrm{e},x\mathrm{e}}$, which were derived by converting the median values of $k$, $\tI$, $\tII$, $\tIII$, and $\tIV$ obtained from the MCMC analysis using the orbital parameters estimated from the same analysis, $(a, b, t_0) = (11.399\ \mathrm{\Rs},\ 0.4494,\ -0.663\ \mathrm{s})$. These functions were treated as functions of the orbital parameters including the orbital period $P$, $e\cos\omega$, and $e\sin\omega$, and the wavelength dependence of the host star's radius $\alpha$, for the calculation of partial derivatives. Here, $e$ denotes the eccentricity, and $\omega$ is the argument of periastron.

Figure \ref{fig:sim_dc_grad} shows the partial derivatives of $\Delta c_{\mathrm{i},x\mathrm{i}}$ and $\Delta c_{\mathrm{e},x\mathrm{e}}$ with respect to each parameter. These derivatives were evaluated at the values estimated from the MCMC analysis along with other parameters: $P = 4.06\ \mathrm{days}$, $e=10^{-10}$, $\omega=0$, and $\alpha=1$. For the evaluation of partial derivatives with respect to $e\cos\omega$ and $e\sin\omega$, a small non-zero value was assigned to $e$. 
The results indicate that all partial derivatives are nearly wavelength-independent. As described in \S \ref{subsec:center_of_mass} and \ref{subsec:stelar_radius}, inaccuracies in $a$, $b$, and $\alpha = \Rs(\lambda)/\Rs(\lambda_{0})$ primarily affect north-south asymmetries, while inaccuracies in $t_0$ primarily affect morning-evening asymmetries. This interpretation aligns with the opposite signs of the partial derivatives of $\Delta c_{\mathrm{i},x\mathrm{i}}$ and $\Delta c_{\mathrm{e},x\mathrm{e}}$ with respect to $a$ and $b$, and the identical signs of those with respect to $t_0$. Similarly, inaccuracies in $P$ or $e\sin\omega$ affect north-south asymmetries, whereas inaccuracies in $e\cos\omega$ affect morning-evening asymmetries.

\begin{table}[bt!]
\centering
\caption{Parameter errors corresponding to a +0.001 $\Rs$ offset in $\Delta c_{\mathrm{i},x\mathrm{i}}$ and $\Delta c_{\mathrm{e},x\mathrm{e}}$.}
\begin{tabular}{c|cc}
\hline\hline
Parameter & $\delta(\Delta c_{\mathrm{i},x\mathrm{i}}) = 0.001$ & $\delta(\Delta c_{\mathrm{e},x\mathrm{e}}) = 0.001$ \\
\hline
$\delta P$ & $-400\ \mathrm{s}$ & $+400\ \mathrm{s}$ \\
$\delta t_0$& $+5\ \mathrm{s}$ & $+5\ \mathrm{s}$ \\
$\delta a$ & $+0.01\ \Rs$ & $-0.01\ \Rs$\\
$\delta b$ & $+0.002$ & $-0.002$ \\
$\delta (e\cos\omega)$ & $-0.06$ & $-0.06$ \\
$\delta (e\sin\omega)$ & $+0.002$ & $-0.002$ \\
$\delta \alpha$ & $-0.001$ & $+0.001$ \\
\hline
\end{tabular}
\label{tab:delta_Delta_c}
\end{table}

Using the partial derivatives, we can estimate the parameter errors required to produce a +0.001 $\Rs$ offset in $\Delta c_{\mathrm{i},x\mathrm{i}}$ and $\Delta c_{\mathrm{e},x\mathrm{e}}$ (Table \ref{tab:delta_Delta_c}). Note that these values vary for planets with different orbital parameters.

\begin{figure*}[tb!]
\centering
\includegraphics[width=0.8\linewidth]{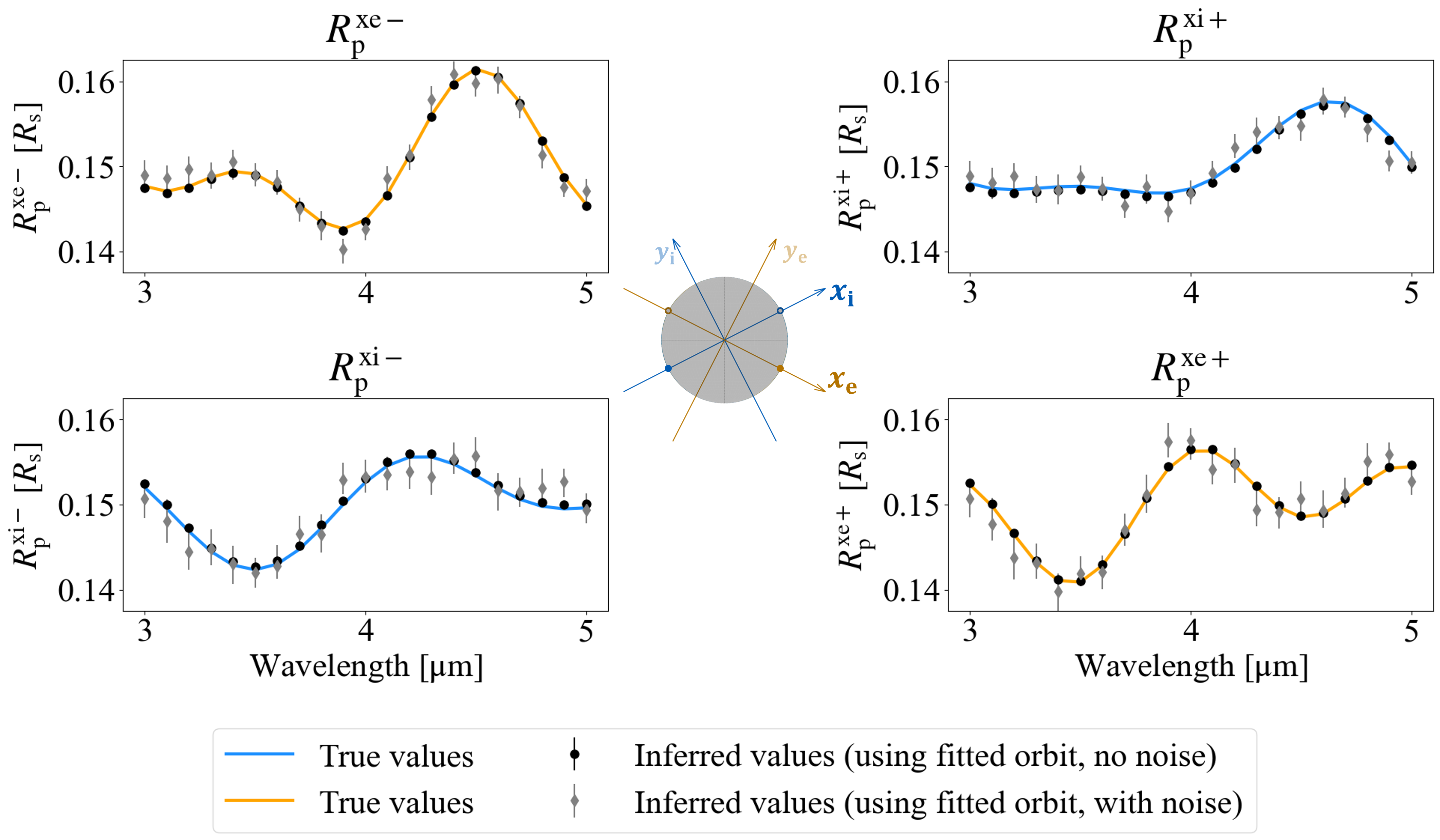}
\caption{Inferred values of $\Rpxip$, $\Rpxin$, $\Rpxep$, and $\Rpxen$ for an asymmetric planet, based on the MCMC analysis of simulated data.
The values were converted using the estimated orbital parameters.
The dots represent the median values from the MCMC sampling, while the error bars indicate the 68\% credible intervals. Results are shown for both noiseless data and data with added Gaussian noise (mean = 0, standard deviation = 250 ppm), along with the true values. The arrangement of the panels reflects the positional relationship between the regions on the planet contributing to each spectrum, as viewed from the night side.
The central panel shows the planet viewed from the night side, along with the $\xii$--axis, $\yi$--axis, $\xe$--axis, and $\ye$--axis.\label{fig:sim_dc_ecc0_rp4}}
\end{figure*}

We can convert $k$, $\Delta c_{\mathrm{i},x\mathrm{i}}$, and $\Delta c_{\mathrm{e},x\mathrm{e}}$ to $\Rpxip$, $\Rpxin$, $\Rpxep$, and $\Rpxen$ using equation \eqref{eq:rp_spectra}. The resulting spectra are shown in Figure \ref{fig:sim_dc_ecc0_rp4}. Results are presented for both noiseless data and data with added Gaussian noise. The inferred values closely align with the true values. These results demonstrate that, despite the limitations imposed by the conventional transit model, this method can effectively capture atmospheric asymmetries in these directions.

\section{Application to JWST NIRS\MakeLowercase{pec} G395H data of WASP-39 \MakeLowercase{b}} 
\label{sec:wasp39b}
In this section, we apply the method described in the previous section to the hot Saturn, WASP-39 b, around a G8 host star, which has a mass of $\sim 0.28\ \mathrm{M_{jup}}$, a radius of $\sim 1.27\ \mathrm{R_{jup}}$, an equilibrium temperature of $\sim 1100\ \mathrm{K}$, and an orbital period of $\sim 4.05\ \mathrm{days}$ \citep[]{2011A&A...531A..40F}. 
As the Introduction mentions, WASP-39 b  was the first to have its $t_0$ chromatic variation reported by \cite{2023Natur.614..659R} using NIRSpec/PRISM observations. Furthermore, \cite{2024Natur.632.1017E} analyzed the morning-evening asymmetries using these data.

\subsection{Data} \label{subsec:data}
We used the data taken with JWST NIRSpec/G395H \citep{2022A&A...661A..80J, 2023PASP..135c8001B} on 2022 July 30–31, as part of the JWST Transiting Exoplanet Community Director’s Discretionary Early Release Science (JTEC ERS) Program \citep[][ERS-1366]{2023Natur.614..664A, 2016PASP..128i4401S, 2018PASP..130k4402B}.
These data have the highest spectral resolution (R $\sim$ 3000) among the data taken in the JTEC ERS Program, and have higher precision than the other data with NIRSpec/PRISM at the same spectral resolution \citep{2024MNRAS.531.2731S}. The wavelength range used in this analysis was $2.8$--$5.0\ \mathrm{\mu m}$, which provides high-precision spectra.

To obtain the spectral time series, we performed data reduction using stages 1 and 2 of the ExoTiC-JEDI pipeline \citep{2023Natur.614..664A}. Stage 1 was almost the same as that of the JWST pipeline \citep{2024zndo..10870758B}, with custom bias subtraction replacing the pipeline's bias subtraction step, and additional group-level destriping steps included as described in \citet{2023Natur.614..664A}. The ramp-jump detection threshold was set to $10\sigma$, while all other settings were kept at their default values, either as determined by the JWST pipeline \citep{2024zndo..10870758B} or as specified in \citet{2023Natur.614..664A} for the custom steps.

We conducted the stage 2 following \citet{2023ApJ...949L..15G}.
For each integration, the data for each row was divided into segments of 100 pixels. Within each segment, outliers were identified and replaced using the following procedure. A fourth-degree polynomial was fitted to the pixel values, and the residuals and their standard deviation $\sigma$ were calculated. A pixel was classified as an outlier if the absolute value of its residual exceeded 4$\sigma$. This process was iteratively repeated, excluding the identified outliers at each step, until no additional outliers were detected. The identified outliers were then replaced with the values predicted by the fitted polynomial.
Following outlier replacement, background subtraction was performed. Each column was fitted with a Gaussian profile, and the centers of the Gaussian fits across all columns were fitted with a second-degree polynomial. Based on this fit, a region centered on the Gaussian peak and extending to 15 times the median of the Gaussian standard deviations was masked. The background value for each column was then estimated as the median of the unmasked region and subtracted. After background subtraction, the outlier cleaning process was repeated.
To extract the one-dimensional spectrum, each column was again fitted with a Gaussian profile. The centers of the Gaussian fits across all columns were fitted with a second-degree polynomial. A region extending to six times the median of the Gaussian standard deviations around the Gaussian center was defined, and the pixel values within this region were summed for each column.
Finally, the extracted time-series spectra were corrected for wavelength shifts by calculating the median of the spectra over time and using cross-correlation to determine the wavelength shift for each time step. These shifts were then applied to align the spectra in the wavelength direction.
Light curves with significant noise were identified by visual inspection and excluded from the set of light curves. 

To verify the impact of data reduction on the resulting spectra, an alternative data reduction process was also performed, yielding nearly identical results (Appendix \ref{ap:reduction}).
Since two independent procedures produced nearly identical spectra, we consider the reduction process to be reliable.

\subsection{Light Curve Fitting} \label{subsec:lcfit}

We applied binning along the wavelength direction to create light curves with a spectral resolution of R $\sim 100$. This resolution was determined by balancing resolution and precision. Each light curve was normalized by the median value before the transit event, as a mirror-tilt event occurred during the observation \citep{2023Natur.614..664A}, altering the baseline before and after the tilt event. The resulting light curves were analyzed by the Hamilton Monte Carlo - No U-Turn Sampler (HMC-NUTS), a specific Markov Chain Monte Carlo (MCMC) algorithm, implemented in NumPyro \citep{2019arXiv191211554P}. For the transit light curve modeling, a modified version of the {\sf jkepler} code was used, adapted for multi-wavelength analysis. {\sf jkepler} uses {\sf exoplanet\_core} library \citep{2021JOSS....6.3285F}, which calculates limb darkened transit light curves based on \citet{2020AJ....159..123A}.

This analysis fixed the orbital period at 4.05528 days \citep{2023ApJS..265....4K}. We assumed the orbital eccentricity of WASP-39 b as zero based on the results by \citet{2015ApJ...810..118K}.
Considering the semi-major axis of WASP-39 b, even if the orbital eccentricity is non-zero, the difference between the durations of ingress and egress is less than 1 second.
Given the precision of the data, this difference is negligible.

\begin{table}[bt!]
\centering
\caption{Model parameters and their prior distributions in the light curve analysis using MCMC. }
\begin{tabular}{lll}
\hline\hline
Symbol & Description & Prior\\
\hline
$k^2$ & Transit depth$\,^\dagger$ & $\mathcal{U}(0, 0.1)$ \\
$t_{0}$ & Central transit time [s]$\,^\ast$$\,^\dagger$ & $\mathcal{U}(9000, 10000)$ \\
$\Ttot$ & Total transit duration [s]$\,^\dagger$ & $\mathcal{U}(9000, 11000)$ \\
$\Tfull$ & Full transit duration [s]$\,^\dagger$ & $\mathcal{U}(1, \Tfull^{\mathrm{max}})$ \\
$c_{\mathrm{base}}$ & Baseline coefficient$\,^\dagger$ & $\mathcal{U}(0.995, 1.005)$ \\
$c_{\mathrm{tilt}}$ & Baseline coefficient$\,^\dagger$ & $\mathcal{U}(0.995, 1.0)$ \\
$jitter$ & Additional jitter error$\,^\dagger$ & $\mathcal{U}(0, 0.03)$ \\
$q_{1\lambda \mathrm{min}}$ & Limb darkening coefficient & $\mathcal{U}(0,1)$ \\
$q_{2\lambda \mathrm{min}}$ & Limb darkening coefficient & $\mathcal{U}(0,1)$ \\
$q_{1\lambda \mathrm{max}}$ & Limb darkening coefficient & $\mathcal{U}(0,1)$ \\
$q_{2\lambda \mathrm{max}}$ & Limb darkening coefficient & $\mathcal{U}(0,1)$ \\
\hline
\end{tabular}
\tablecomments{$\ast$: time from BJD - 59791. $\dagger$: independently sampled for each wavelength. $\mathcal{U}(a,b)$: the uniform distribution from $a$ to $b$. $\Tfull^{\mathrm{max}}$ is defined in equation (\ref{eq:Tfullmax}).}
\label{tab:params_lc}
\end{table}

\begin{figure}[tb!]
\centering
\includegraphics[width=\linewidth]{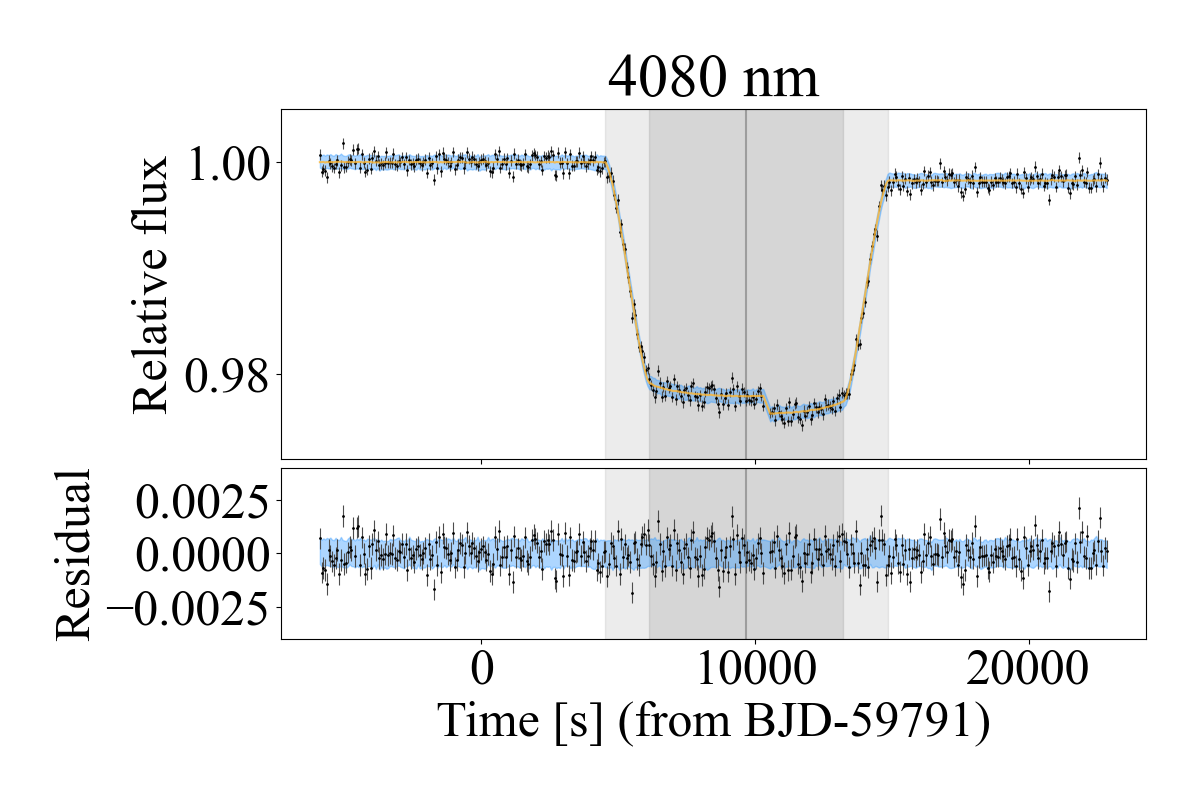}
\caption{Result of light curve fitting at 4080 nm. The black dots represent the data. The orange line shows the median of the MCMC sampling, and the thin blue region represents the 68\% credible interval. The residuals are the differences between the data and the median of the MCMC sampling. The gray vertical line indicates the median of the inferred central transit time, $t_0$. The dark gray region represents the median of the inferred total duration, $\Ttot$. The light gray region represents the median of the inferred full duration, $\Tfull$.
\label{fig:lightcurve}}
\end{figure}

\begin{figure*}[ht!]
\centering
\includegraphics[width=\linewidth]{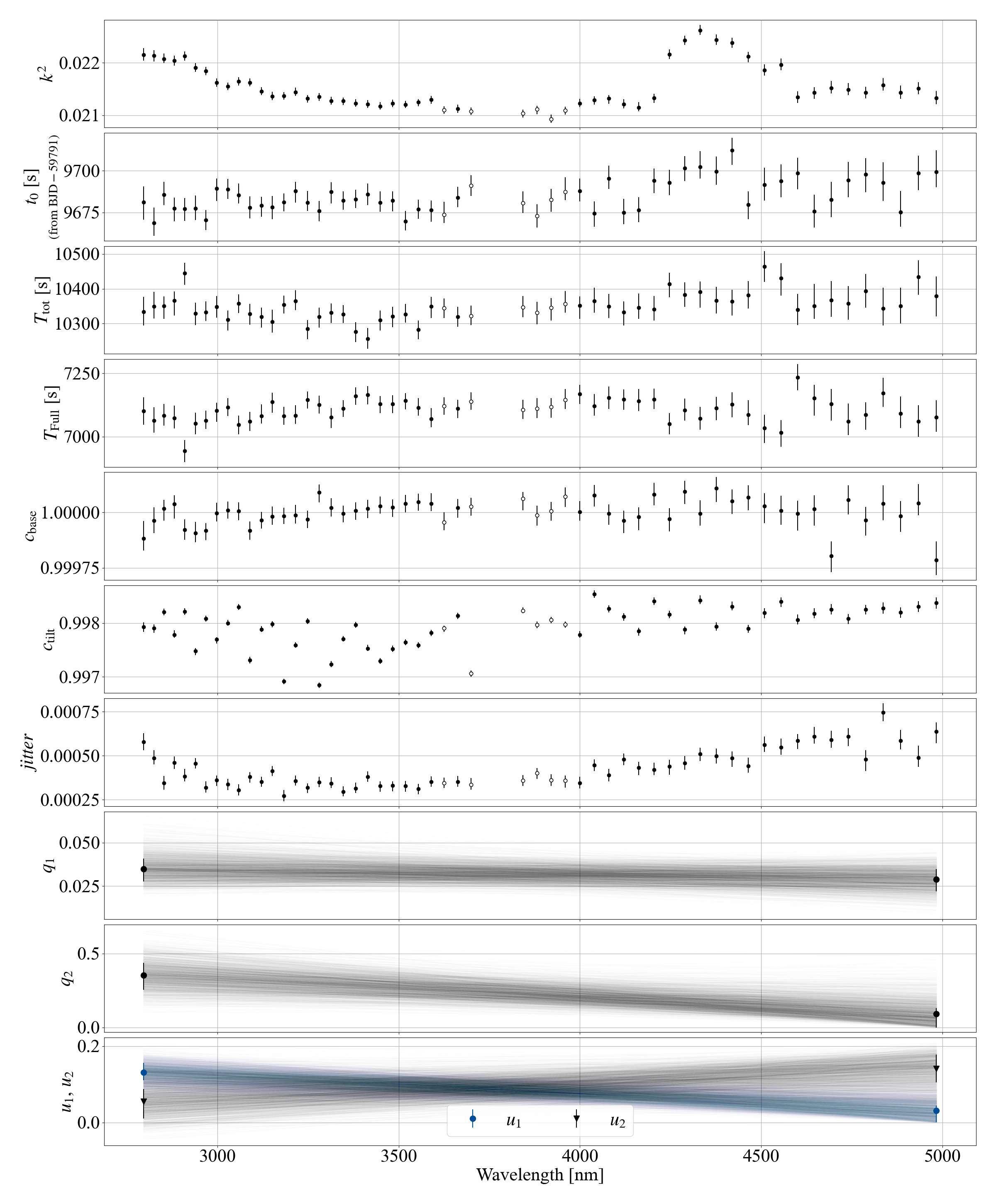}
\caption{Inferred values of all the parameters in the light curve model obtained from the MCMC sampling. For the limb darkening coefficients, the inferred values at the minimum and maximum wavelengths, along with the wavelength dependence for each sampling, are shown. $u_1$ and $u_2$, converted from $q_1$ and $q_2$, are also displayed. The dots represent the median values of the MCMC sampling, and the error bars indicate the 68\% credible intervals.
The white-filled dots indicate the wavelengths that exhibit radii smaller than the 10th percentile and are used to determine the orbit of the planet's center of mass.
\label{fig:params_lightcurve}}
\end{figure*}

The model parameters are summarized in Table \ref{tab:params_lc}. 
The transit depth $k^2$, the central transit time $t_0$, the total duration $\Ttot$, and the full duration $\Tfull$ were treated as independent parameters for each wavelength. To model the transit light curve, we then calculate the semi-major axis $a_{\lambda}$ and inclination $i_{\lambda}$ at each wavelength using equation \eqref{eq:a_from_durations} and \eqref{eq:cosi_from_durations}.
To avoid indeterminate solutions, we impose an upper bound on $\Tfull$ (equation \eqref{eq:Tfullmax}).

We used a quadratic limb darkening model. 
$q_1$ and $q_2$ from \citet{2013MNRAS.435.2152K} were sampled and then converted to $u_1$ and $u_2$. Since $q_1$ and $q_2$ are expected to exhibit similar values for nearby wavelengths, we incorporated this by assuming a simple linear wavelength dependence for $q_1$ and $q_2$. To implement this, $q_1$ and $q_2$ at minimum wavelength, $q_{1\lambda \mathrm{min}}$ and $q_{2\lambda \mathrm{min}}$, and $q_1$ and $q_2$ at maximum wavelength, $q_{1\lambda \mathrm{max}}$ and $q_{2\lambda \mathrm{max}}$, were sampled using uniform distributions between 0 and 1. 

Regarding the parameterization of limb darkening coefficients, \citet{2024AJ....168..227C} demonstrated that using $q_1$ and $q_2$ could introduce wavelength-dependent biases in the inferred transit depths. We address this issue in Appendix \ref{ap:ldc} by comparing the results with those obtained from additional light curve fits, where $u_1$ and $u_2$ were treated as free parameters for each wavelength. We did not detect the wavelength-dependent biases in the inferred transit depths. In addition, while the uncertainty in the resulting spectra was larger when no wavelength dependence was assumed for $u_1$ and $u_2$, the overall shape of the spectra remained consistent with the results obtained using the linear wavelength dependence for $q_1$ and $q_2$.

The baseline of light curves was assumed to be constant.
However, the mirror-tilt event during the observation \citep{2023Natur.614..664A} changed the baseline before and after the tilt event.
We set the baseline before the tilt event as $c_{base}$, and after the tilt event as $c_{base} \times c_{tilt}$. 
These baseline constants were also treated as parameters independently for each wavelength.

The likelihood of the light-curve data was modeled as a product of normal distributions for each wavelength. The means of these distributions were given by (transit light curve model)$\times$(baseline) and the variances are given by $(\mathrm{data\ error})^2 + (jitter)^2$. The $jitter$ was treated as an independent parameter for each wavelength.

An example of the fitted light curve and its residuals obtained by the above procedure is shown in Figure \ref{fig:lightcurve}. The variance of the residuals remains nearly constant before and after the transit, as well as before and after the tilt event, indicating that the light curve is well-fitted. 

Figure \ref{fig:params_lightcurve} shows the inferred values of all the parameters in the light curve model obtained from the MCMC sampling. The dots represent the median values of the MCMC sampling, and the error bars indicate the 68\% credible intervals. For parameters independently sampled for each wavelength, the values for all wavelength bins are shown, allowing us to observe the chromatic variation of $k^2$, $t_{0}$, $\Ttot$, and $\Tfull$. For the limb darkening coefficients, the inferred values at the minimum and maximum wavelengths, along with the wavelength dependence for each sampling, are shown. Additionally, $u_1$ and $u_2$, converted from $q_1$ and $q_2$, are also displayed.

\subsection{Planetary Radius Spectra in Four Directions} \label{subsec:4limbs}

By assuming the orbit of the planet's center of mass and the wavelength dependence of the host star's radius, we can obtain the spectra of $\Rpxip$, $\Rpxin$, $\Rpxep$, and $\Rpxen$ from the results of the light curve fitting. 

The orbit of the planet's center of mass cannot be determined solely by transit light curves because they are affected by atmospheric asymmetries. However, as discussed in \citet{2012ApJ...751...87D}, the orbital parameters determined from the data at wavelengths that exhibit small radii are considered to be close to the orbital parameters of the center of mass. Here, we use wavelengths that exhibit radii smaller than the 10th percentile to determine the orbit of the planet's center of mass. For each of those wavelength bins, we calculated the median of the orbital parameters obtained from the MCMC sampling and then used the average of these values as the orbital parameters of the center of mass. As a result, we determined the mean wavelength of these bins as $\lambda_0 = 3821.5\ \mathrm{nm}$, the semi-major axis as $ 11.367\ \mathrm{R_{s}(\lambda_0)}$, the impact parameter as $0.45284$, and the central transit time as $9681.4\ \mathrm{s}$ from BJD - 59791 as the orbital parameters of the planet's center of mass.

\begin{figure}[tb!]
\centering
\includegraphics[width=\linewidth]{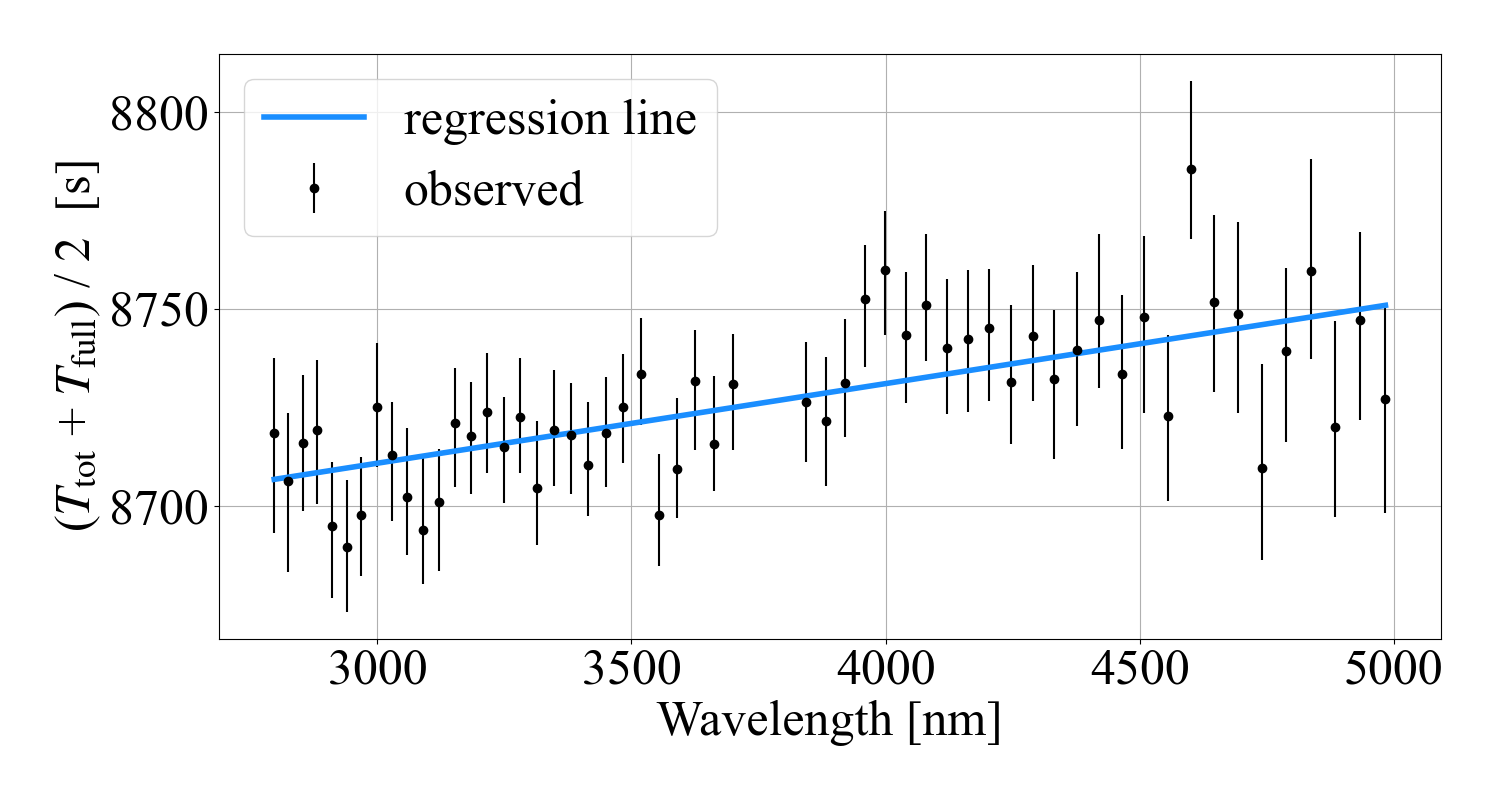}
\caption{Inferred value of $(\Ttot + \Tfull)/2$ and its linear regression line (blue solid line). The dots with error bars represent the median values of the MCMC sampling and the 68\% credible intervals. 
\label{fig:TtotTfullmean}}
\end{figure}

Although the wavelength-dependent stellar radius, $R_s(\lambda)$, is challenging to estimate with the precision required for this method, the transit duration provides important information about it (\S \ref{subsec:stelar_radius}). We observed that the transit duration gradually increases with wavelength, as shown in Figure \ref{fig:TtotTfullmean}. 

The natural logarithms of the Bayesian evidences for the constant model (M1) and the linear model (M2: constant + slope $\times$ wavelength) were $-246.2$ and $-234.0$, respectively, indicating that M2 is preferred over M1 \citep{2005blda.book.....G}. 
The prior distributions used in this analysis were as follows. For the constant term, which represents the value at $\lambda_0$, a uniform prior was set as $\mathcal{U}(\mathrm{min}(\bm{y}), \mathrm{max}(\bm{y})) = \mathcal{U}(8690.1\ \mathrm{s}, 8786.2\ \mathrm{s})$. For the slope, the prior distribution was set as  
\begin{align*} 
&\mathcal{U}\left(-2 \cdot \frac{\mathrm{max}(\bm{y}) - \mathrm{min}(\bm{y})}{\mathrm{max}(\bm{x}) - \mathrm{min}(\bm{x})},\ 2 \cdot \frac{\mathrm{max}(\bm{y}) - \mathrm{min}(\bm{y})}{\mathrm{max}(\bm{x}) - \mathrm{min}(\bm{x})}\right) \\
=\ &\mathcal{U}(-87.86\ \mathrm{s/\mu m},\ 87.86\ \mathrm{s/\mu m}), 
\end{align*} 
where $\bm{y}$ is the mean of the samples of $(\Ttot + \Tfull) / 2$ for each wavelength, and $\bm{x}$ is the wavelength. 
This linear trend can arise from a combination of the wavelength dependence of the host star's radius, atmospheric asymmetries of the planet, and the treatment of limb darkening. To examine the sensitivity of the inferred spectra to the wavelength dependence of the stellar radius, we consider an illustrative case in which the entire linear trend of $(\Ttot+\Tfull)/2$ is attributed to $\alpha(\lambda)$.

We can use equation \eqref{eq:Rs_lambda_duration} to derive $\alpha(\lambda)$. For $T_{\mathrm{bc}}(\lambda)$, we used the linear regression of $(\Ttot + \Tfull)/2$ (see Appendix \ref{ap:duration}). 
To perform a linear regression robust to outliers, we used the Student's t-distribution as the likelihood function for the MCMC sampling. The free parameters in this regression were the slope of the line, the value at $\lambda=\lambda_{0}$, and the scale and degrees of freedom of the Student's t-distribution. 
Under this assumption, we then obtained $\alpha (\lambda) = R_{s}(\lambda)/R_{s}(\lambda_0) \sim 1 + 1.83\times 10^{-6}(\lambda - \lambda_{0})$, where the wavelengths $\lambda$ and $\lambda_0$ are in units of nanometers. 
If interpreted entirely as a change in the stellar radius, this corresponds to a rate of $0.183\%\ \mathrm{\mu m^{-1}}$ in the wavelength range of approximately 3 to 5 $\mathrm{\mu m}$.

However, as discussed in Appendix \ref{ap:phoenix}, the {\sf PHOENIX} specific intensities \citep{2013A&A...553A...6H} for a star with an effective temperature of $5500\,\mathrm{K}$, $\log g\ \mathrm{[cgs]}=4.5$, and $\mathrm{[Fe/H]}=0.0$ imply a much smaller wavelength dependence of the projected stellar radius, approximately $0.009\%\ \mathrm{\mu m^{-1}}$ over this wavelength range. WASP-39, a G8-type star, has an effective temperature of $5485\pm 50\ \mathrm{K}$, surface gravity $\log g\ \mathrm{[cgs]} = 4.41\pm 0.15$, and metallicity $\mathrm{[Fe/H]} = 0.01\pm 0.09$ \citep{2018A&A...613A..41M}. 
Therefore, the observed linear trend in $(\Ttot+\Tfull)/2$ is unlikely to be explained by chromatic changes in the stellar radius alone, although such a contribution cannot be ruled out entirely. Atmospheric asymmetries of WASP-39 b and the treatment of the limb-darkening parameters may also contribute to the observed trend.
Accordingly, we use the wavelength-dependent $\alpha(\lambda)$ constructed above only as an illustrative prescription for assessing its impact on the inferred spectra, rather than as an independent measurement of the wavelength dependence of the stellar radius.

In the following discussion, we use this illustrative $\alpha(\lambda)$ as the fiducial case, while also showing the results obtained assuming $\alpha(\lambda)=1$. The differences in the inferred spectra between the two cases are within the current uncertainties.

\begin{figure*}[hbt!]
\centering
\includegraphics[width=\linewidth]{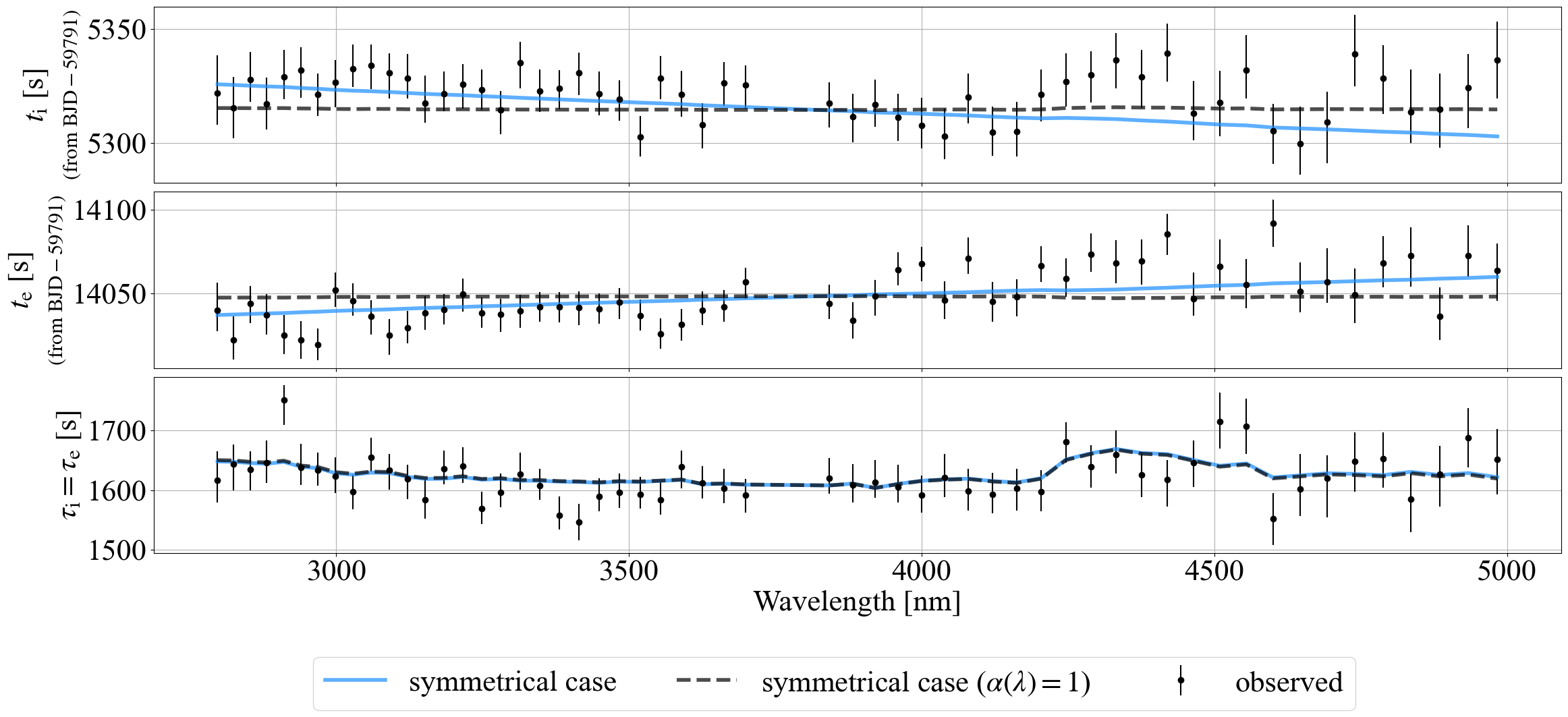}
\caption{Inferred values of the timing of ingress $\ti$, the timing of egress $\te$, and the duration of ingress and egress $\tau$. The dots with error bars represent the median values from the MCMC sampling with 68\% credible intervals. The solid blue lines (using the illustrative $\alpha(\lambda)$ constructed from $(\Ttot + \Tfull)/2$) and the dashed gray lines (using $\alpha(\lambda)=1$) represent these values in the case of a completely symmetrical atmosphere, derived from the orbital parameters of the planet's center of mass and the inferred planetary radii.
\label{fig:titetau}}
\end{figure*}

Figure \ref{fig:titetau} shows the time-related parameters of the conventional transit model, the timings of ingress ($\ti$), the timing of egress ($\te$), and the duration of ingress or egress ($\taui = \taue$). We plot the predicted values for the case of no asymmetries (a completely symmetrical atmosphere), which were calculated using eqs. (14) and (15) in \citet{2010exop.book...55W} with the median values of $k^2$ from MCMC sampling, along with the semi-major axis, impact parameter, and central transit time of the planet's center of mass. As shown in this figure, the deviations from the symmetrical atmosphere are observed to be larger in the timings of ingress and egress than in the duration.

\begin{figure*}[ht!]
\centering
    \begin{minipage}{0.494\linewidth}
        \centering
        \includegraphics[width=\linewidth]{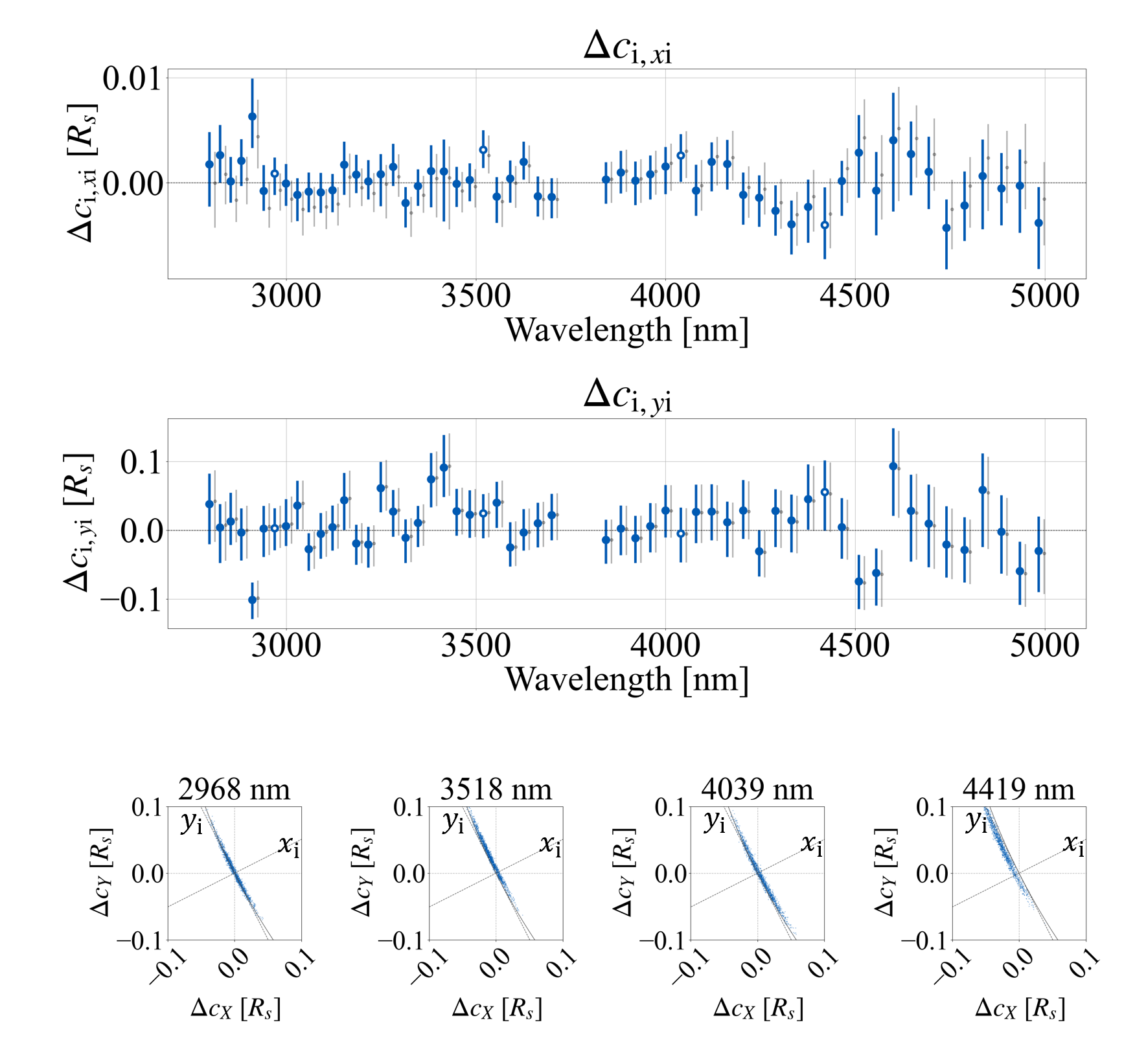}
    \end{minipage}
    \hfill
    \begin{minipage}{0.494\linewidth}
        \centering
        \includegraphics[width=\linewidth]{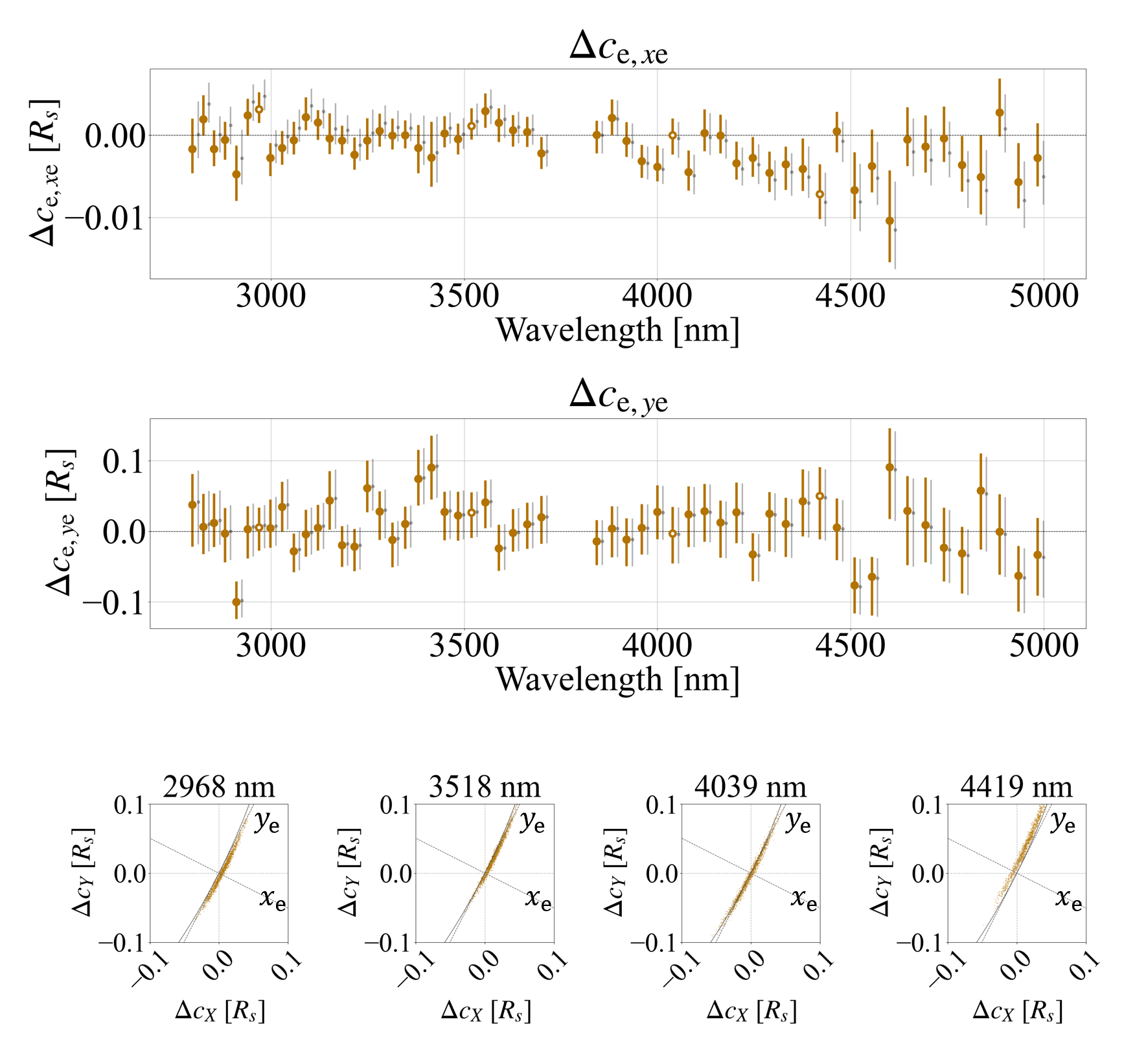}
    \end{minipage}
\caption{Inferred center displacements of the planet's shadow disk: $\Delta c_{\mathrm{i},x\mathrm{i}}$, $\Delta c_{\mathrm{i},y\mathrm{i}}$, $\Delta c_{\mathrm{e},x\mathrm{e}}$ and $\Delta c_{\mathrm{e},y\mathrm{e}}$. The dots represent the median values from MCMC sampling, and the error bars represent the 68\% credible intervals. The colored dots indicate the case using the illustrative $\alpha(\lambda)$ constructed from $(\Ttot + \Tfull)/2$, while the small gray dots indicate the case using $\alpha(\lambda)=1$.
The bottom panels show the distribution of displacement values converted from MCMC sampling at selected wavelengths, which are indicated by white-filled dots in the top and second panels. The dotted lines represent the $X$--axis and $Y$--axis, and the dashed lines represent the $\xii$--axis and $\yi$--axis for the left-side four panels, and the $\xe$--axis and $\ye$--axis for the right-side four panels. The edge of the host star is plotted as black solid lines, which are nearly aligned with the $\yi$--axis or $\ye$--axis. The distribution along the host star's edge is clearly visible.
\label{fig:cxcy}}
\end{figure*}

Figure \ref{fig:cxcy} shows the center displacements of the planet's shadow disk, $\Delta c_{\mathrm{i},x\mathrm{i}}$, $\Delta c_{\mathrm{i},y\mathrm{i}}$, $\Delta c_{\mathrm{e},x\mathrm{e}}$ and $\Delta c_{\mathrm{e},y\mathrm{e}}$. The colored dots indicate the case using the illustrative $\alpha(\lambda)$ constructed from $(\Ttot + \Tfull)/2$, while the small gray dots indicate the case using $\alpha(\lambda)=1$. 
The precision of $\Delta c_{\mathrm{i},x\mathrm{i}}$ and $\Delta c_{\mathrm{e},x\mathrm{e}}$ is much higher than that of $\Delta c_{\mathrm{i},y\mathrm{i}}$ and $\Delta c_{\mathrm{e},y\mathrm{e}}$.
By comparing the wavelength dependence of the plots in Figures \ref{fig:titetau} and \ref{fig:cxcy}, it becomes clear that the shapes of $\ti$ and $\Delta c_{\mathrm{i},x\mathrm{i}}$, as well as $\te$ and $\Delta c_{\mathrm{e},x\mathrm{e}}$, are almost identical, although their signs are reversed.
The bottom panels of Figure \ref{fig:cxcy} show the distribution of displacement values converted from MCMC sampling at selected wavelengths. The clear distribution along the host star's edge indicates that $\Delta c_{\mathrm{i},x\mathrm{i}}$ and $\Delta c_{\mathrm{e},x\mathrm{e}}$ are well determined by the data.

\begin{figure*}[ht!]
\centering
\includegraphics[width=\linewidth]{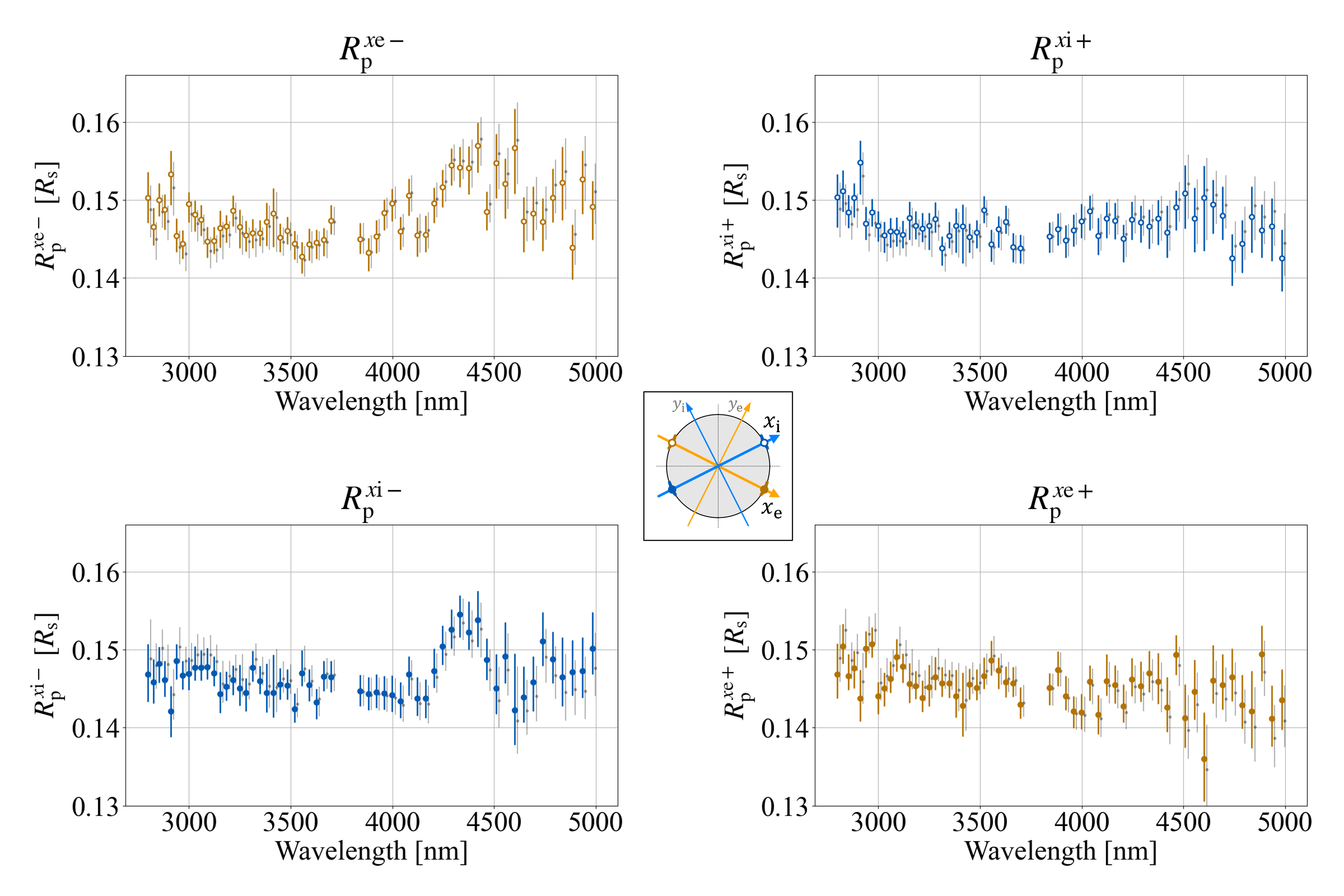}
\caption{
Spectra of $\Rpxen$ (top left), $\Rpxip$ (top right), $\Rpxin$ (bottom left), and $\Rpxep$ (bottom right).
The dots with error bars represent the median values from the MCMC sampling with 68\% credible intervals. The colored dots indicate the case using the illustrative $\alpha(\lambda)$ constructed from $(\Ttot + \Tfull)/2$, while the small gray dots indicate the case using $\alpha(\lambda)=1$.
The central panel shows the positions on the planet that correspond to these spectra, along with the $\xii$--axis, $\yi$--axis, $\xe$--axis, and $\ye$--axis. The color of points indicating those positions corresponds to the color of dots in each spectrum. Since these spectra are not exact radii but rather projected lengths of the planet’s shadow disk onto the $\xii$--axis or $\xe$--axis, measured from the planet’s center of mass (Figure \ref{fig:def_Rp}), the uncertainties of $\Delta \bm{c}$ shown in Figure \ref{fig:cxcy} are also illustrated to indicate the uncertainties of positions from which these values are projected.
\label{fig:4limbs}}
\end{figure*}

Finally, we convert $\Delta c_{\mathrm{i},x\mathrm{i}}$ to the spectra of $\Rpxip$ and $\Rpxin$, and $\Delta c_{\mathrm{e},x\mathrm{e}}$ to the spectra of $\Rpxep$ and $\Rpxen$, as shown in Figure \ref{fig:4limbs}. The central panel of Figure \ref{fig:4limbs} illustrates the corresponding positions on the planet for these spectra. Since these values represent projected lengths of the planet’s shadow disk onto the $\xii$--axis or $\xe$--axis, measured from the planet’s center of mass, rather than exact radii (Figure \ref{fig:def_Rp}), the uncertainties in $\Delta \bm{c}$ shown in Figure \ref{fig:cxcy} are also illustrated to indicate the uncertainties in the positions from which these values are projected.

Clear differences in the shape of these spectra are observed. In particular, $\Rpxip$ and $\Rpxep$ (the morning side) have flatter spectra than $\Rpxen$ and $\Rpxin$ (the evening side). 
We can also see differences between the spectra of $\Rpxen$ and $\Rpxip$ (the north side) and those of $\Rpxin$ and $\Rpxep$ (the south side). We will discuss these differences using atmospheric models in the next section.

\section{Implication to the planetary atmosphere} \label{sec:atmospheric}
In the previous section, we decomposed the light curve of WASP-39 b into spectra in four directions (Figure \ref{fig:4limbs}). In this section, we analyze each spectrum using atmospheric retrieval and interpret the results. We use the spectra calculated with the illustrative $\alpha(\lambda)$ constructed by attributing the linear trend of $(\Ttot+\Tfull)/2$ entirely to the wavelength dependence of the stellar radius.
Given the current uncertainties in the spectra, the impact of this choice of \(\alpha(\lambda)\) on the retrieved atmospheric parameters is expected to be small, although the treatment of \(\alpha(\lambda)\) should be considered carefully in future analyses.

\subsection{Atmospheric Retrieval}

\begin{table}[htb!]
\centering
\caption{Model parameters and their prior distributions in the atmospheric retrieval analysis using MCMC. }
\begin{tabular}{lll}
\hline\hline
Symbol & Description & Prior\\
\hline
$R_\mathrm{s}$ $\,^\ast$  & Host star's radius ($R_{\odot}$) & $\mathcal{N}(0.939, 0.022)$ \\
$M_\mathrm{p}$ $\,^\ast$  & Planetary mass ($M_{\mathrm{J}}$) & $\mathcal{N}(0.281, 0.032)$ \\
$R_\mathrm{p}$ at 10 bar $\,^\ast$  & height at 10 bar ($R_{\mathrm{J}}$) & $\mathcal{U}(1.0, 1.5)$ \\
$T$ & temperature ($\mathrm{K}$) & $\mathcal{U}(500, 2000)$ \\
$\log P_\mathrm{cloud}$ & cloud top pressure (bar) & $\mathcal{U}(-11.0, 1.0)$\\
$\log\mathrm{H_{2}O}$ & VMR of $\mathrm{H_{2}O}$ & $\mathcal{U}(-15.0, 0.0)$\\
$\log\mathrm{CO}$ & VMR of $\mathrm{CO}$ & $\mathcal{U}(-15.0, 0.0)$\\
$\log\mathrm{CO_{2}}$ & VMR of $\mathrm{CO_{2}}$ & $\mathcal{U}(-15.0, 0.0)$ \\
$\log\mathrm{SO_{2}}$ & VMR of $\mathrm{SO_{2}}$ & $\mathcal{U}(-15.0, 0.0)$\\
$\log\mathrm{H_{2}S}$ & VMR of $\mathrm{H_{2}S}$ & $\mathcal{U}(-15.0, 0.0)$\\
$\log\mathrm{CH_{4}}$ & VMR of $\mathrm{CH_{4}}$ & $\mathcal{U}(-15.0, 0.0)$\\
$\log\mathrm{NH_{3}}$ & VMR of $\mathrm{NH_{3}}$ & $\mathcal{U}(-15.0, 0.0)$ \\
$\log\mathrm{HCN}$ & VMR of $\mathrm{HCN}$ & $\mathcal{U}(-15.0, 0.0)$\\
$\log\mathrm{C_{2}H_{2}}$ & VMR of $\mathrm{C_{2}H_{2}}$ & $\mathcal{U}(-15.0, 0.0)$.\\
\hline
\end{tabular}
\tablecomments{$\ast$: common for all the four limbs. $\mathcal{N}(a, b)$: the normal distribution with a mean of $a$ and a standard deviation of $b$, truncated to ensure that the values are non-negative. $\mathcal{U}(a,b)$: the uniform distribution from $a$ to $b$. $\log X$: $\log_{10} X$}
\label{tab:prior_exojax}
\end{table}

\begin{figure*}[htb!]
\centering
\includegraphics[width=\linewidth]{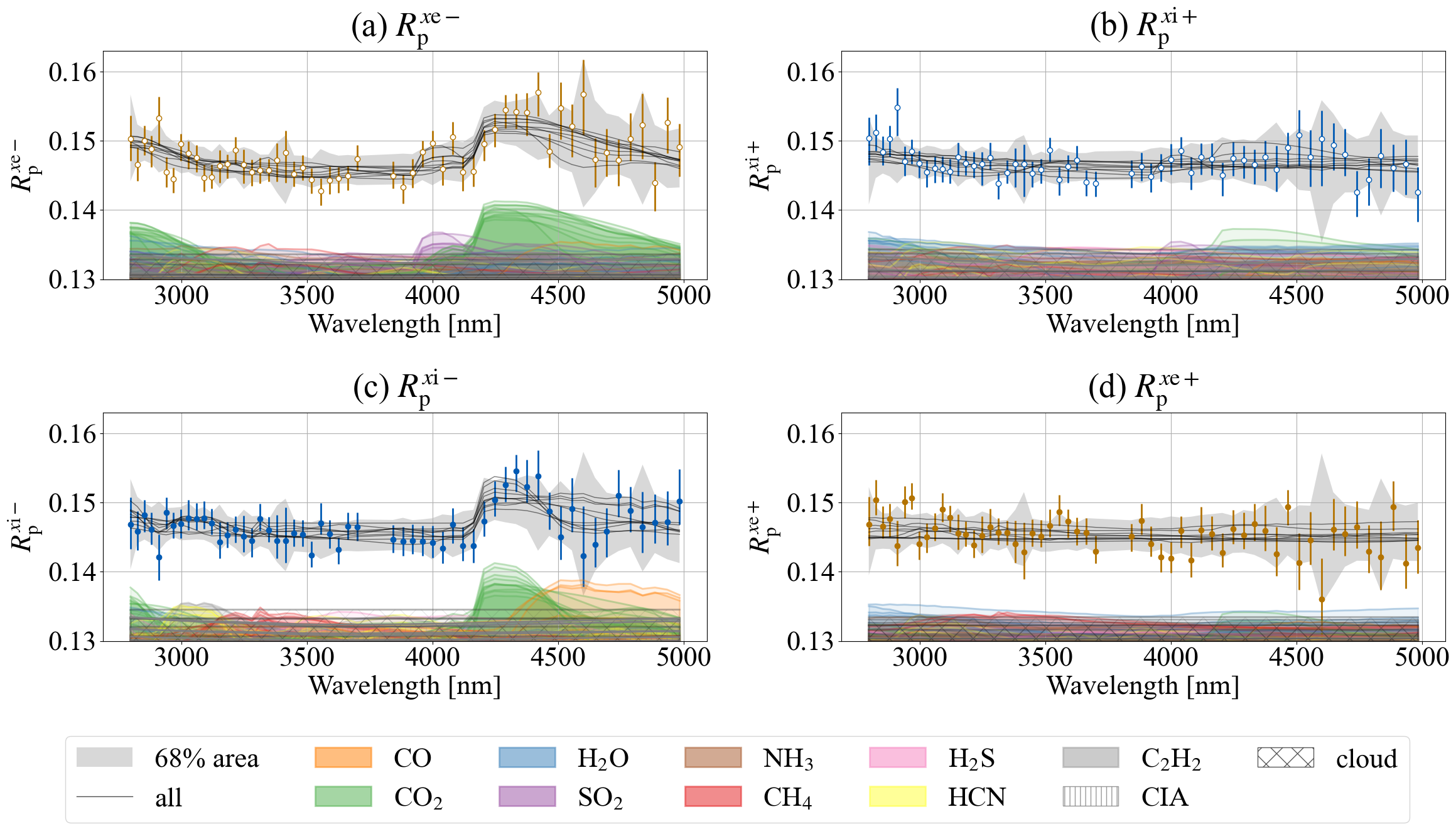}
\caption{Results of MCMC sampling for the spectra: (a) $\Rpxen$, (b) $\Rpxip$, (c) $\Rpxin$, and (d) $\Rpxep$. The dots with error bars represent the spectra shown in Figure \ref{fig:4limbs}, calculated using the illustrative $\alpha(\lambda)$ constructed from $(\Ttot + \Tfull)/2$. The grey regions in each panel indicate the 68\% credible intervals for each spectrum from the MCMC sampling. The black lines in each panel represent ten randomly selected sampled models. The contributions of each molecule, grey clouds, and CIA to these models are also shown as colored or hatched shading, with an offset of -0.01 $R_s$.
\label{fig:4limbs_hmc}}
\end{figure*}

\begin{figure*}[ht!]
\centering
\includegraphics[width=\linewidth]{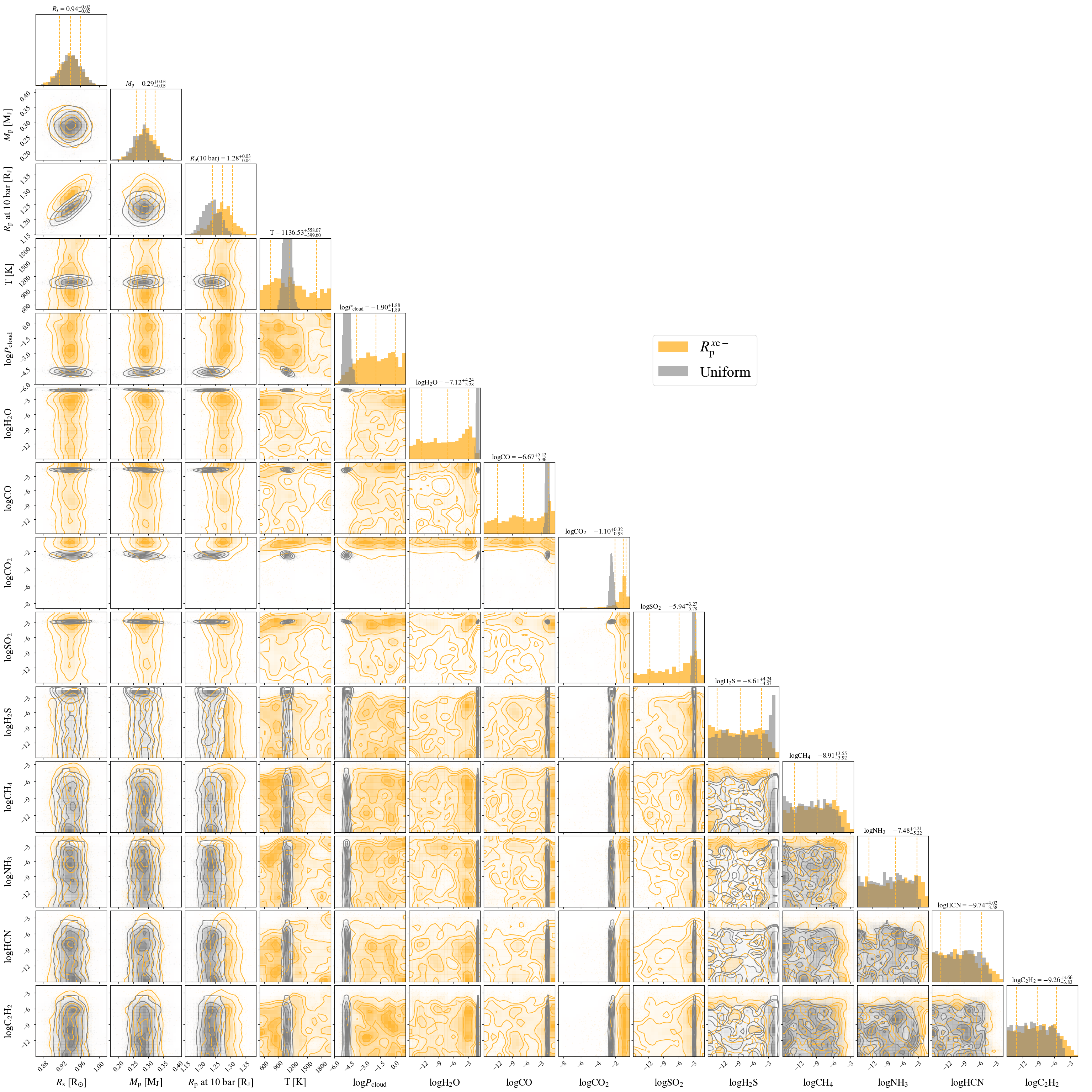}
\caption{Posterior distributions of each parameter for the $\Rpxen$ spectrum (orange). The $\Rp$ at $10$ bar is common for all four limbs (Figure \ref{fig:corner_xen} -- \ref{fig:corner_xep}). The posterior distributions from the analysis with the intrinsic spectral resolution ($\sim$ 2700), assuming a uniform atmosphere (Kawahara et al. submitted), are overlaid for reference (gray). \label{fig:corner_xen}}
\end{figure*}

\begin{figure*}[ht!]
\centering
\includegraphics[width=\linewidth]{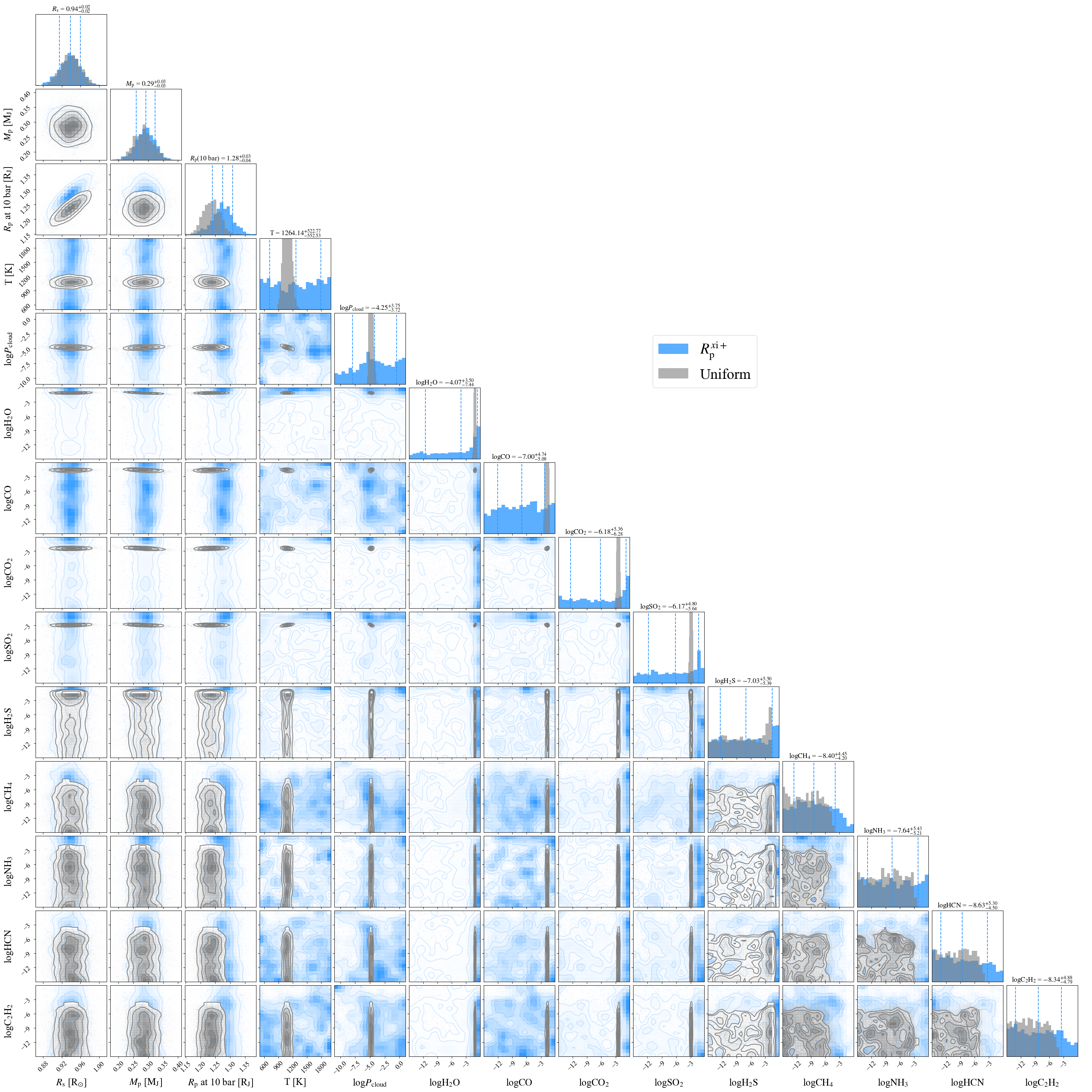}
\caption{Posterior distributions of each parameter for the $\Rpxip$ spectrum (blue). The $\Rp$ at $10$ bar is common for all four limbs (Figure \ref{fig:corner_xen}--\ref{fig:corner_xep}). The posterior distributions from the analysis with the intrinsic spectral resolution ($\sim$ 2700), assuming a uniform atmosphere (Kawahara et al. submitted), are overlaid for reference (gray). \label{fig:corner_xip}}
\end{figure*}

\begin{figure*}[ht!]
\centering
\includegraphics[width=\linewidth]{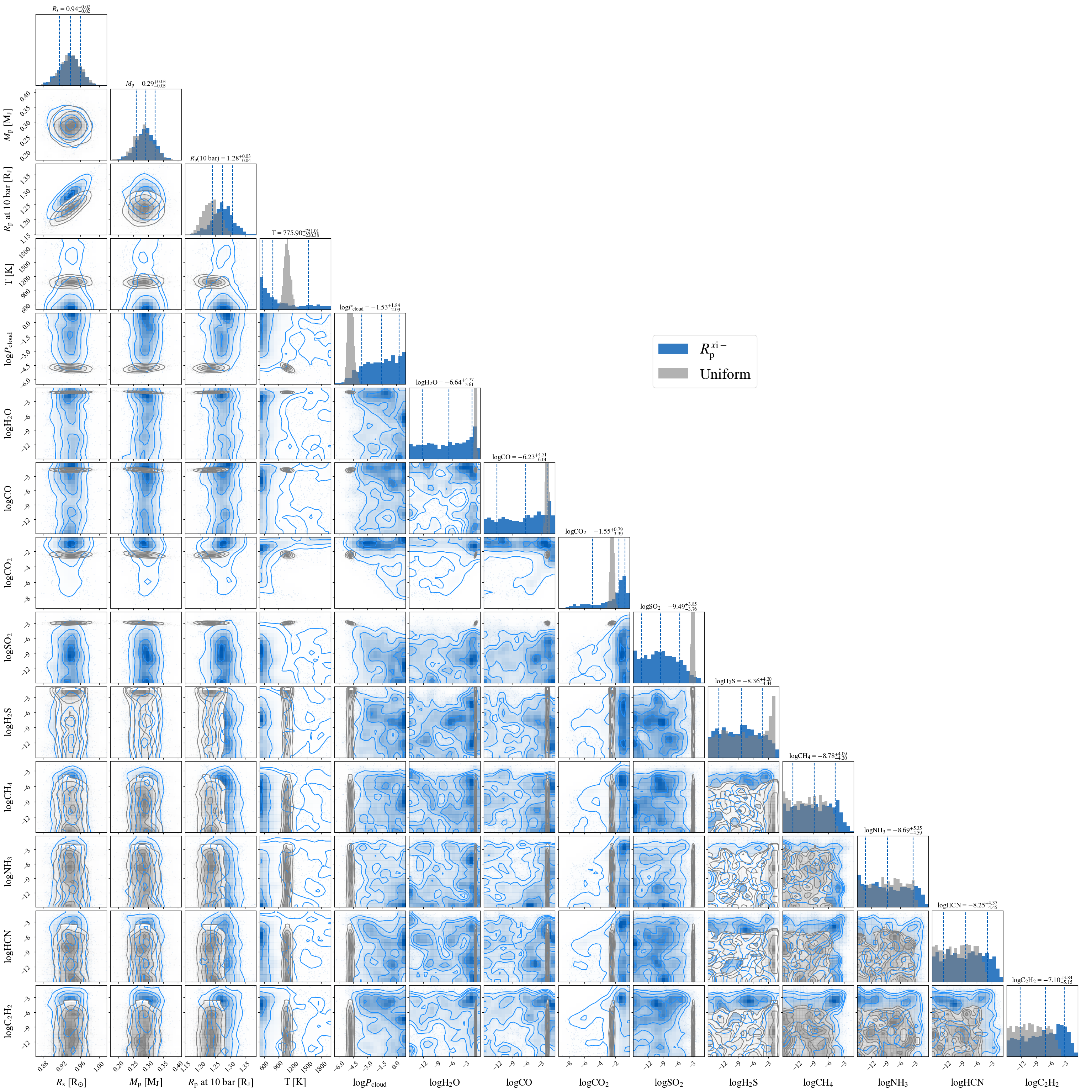}
\caption{Posterior distributions of each parameter for the $\Rpxin$ spectrum (dark blue). The $\Rp$ at $10$ bar is common for all four limbs (Figure \ref{fig:corner_xen}--\ref{fig:corner_xep}). The posterior distributions from the analysis with the intrinsic spectral resolution ($\sim$ 2700), assuming a uniform atmosphere (Kawahara et al. submitted), are overlaid for reference (gray). \label{fig:corner_xin}}
\end{figure*}

\begin{figure*}[ht!]
\centering
\includegraphics[width=\linewidth]{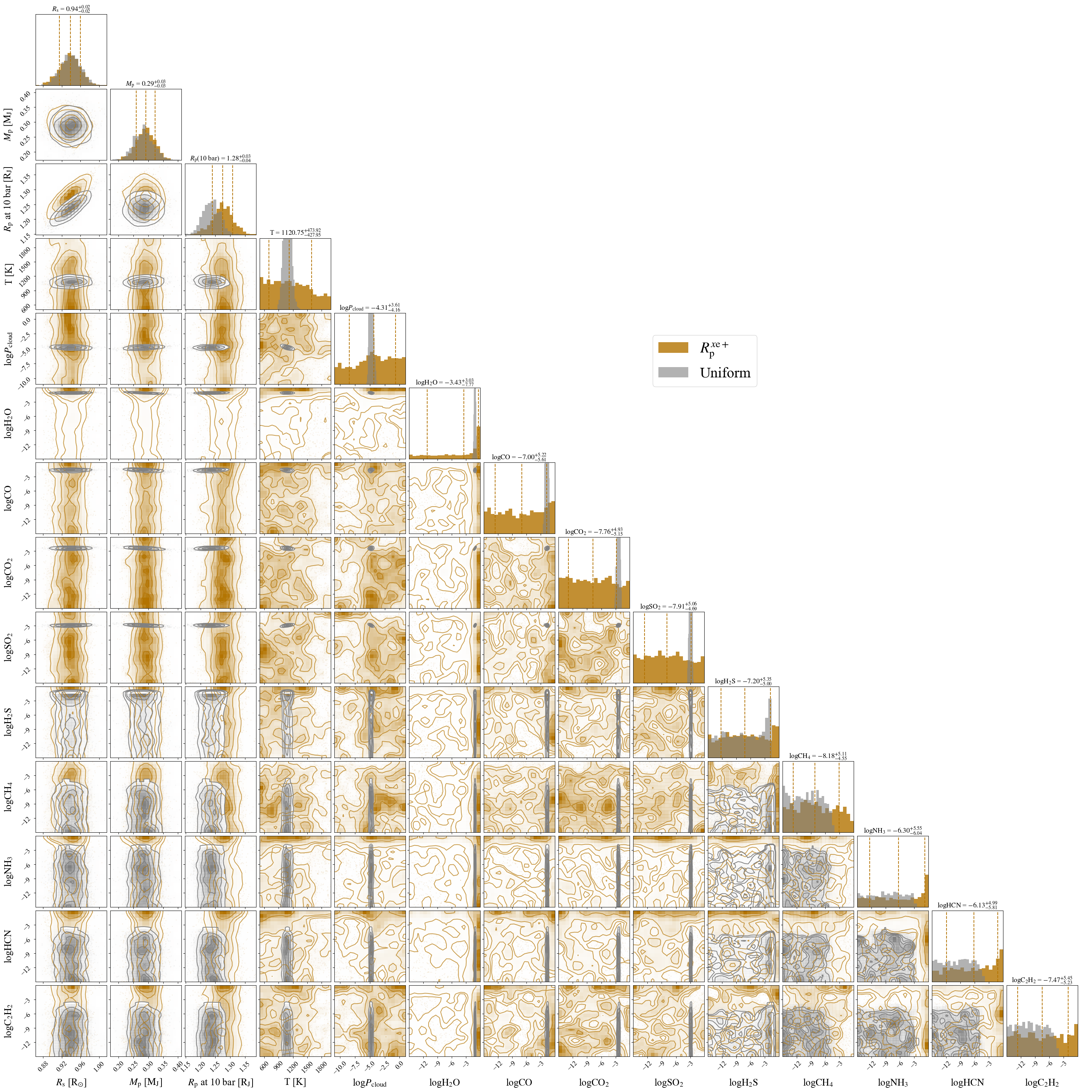}
\caption{Posterior distributions of each parameter for the $\Rpxep$ spectrum (dark orange). The $\Rp$ at $10$ bar is common for all four limbs (Figure \ref{fig:corner_xen}--\ref{fig:corner_xep}). The posterior distributions from the analysis with the intrinsic spectral resolution ($\sim$ 2700), assuming a uniform atmosphere (Kawahara et al. submitted), are overlaid for reference (gray). \label{fig:corner_xep}}
\end{figure*}

The spectra in four directions were analyzed by HMC-NUTS, implemented in NumPyro \citep{2019arXiv191211554P}.
For the transmission spectrum modeling, we used {\sf ExoJAX}, a differentiable planet spectral model \citep[][Kawahara et al. submitted]{2022ApJS..258...31K}.

We assumed a constant temperature profile characterized by $T$ and a gray cloud that is opaque below a certain pressure, independent of wavelength.
The collision-induced absorption (CIA) of the hydrogen and helium atmosphere was also incorporated \citep[$\mathrm{H_2}$--$\mathrm{H_2}$ and $\mathrm{H_2}$--$\mathrm{He}$,][]{2012JQSRT.113.1276R, 2019Icar..328..160K}. 
Molecular species included were $\mathrm{CO}$, $\mathrm{CO_2}$, $\mathrm{H_2O}$, and $\mathrm{CH_4}$ \citep[HITEMP,][]{2010JQSRT.111.2139R, 2020ApJS..247...55H}, $\mathrm{H_2S}$ \citep[ExoMol/AYT2,][]{2016MNRAS.460.4063A, 2018JQSRT.218..178C}, $\mathrm{SO_2}$ \citep[ExoMol/ExoAmes,][]{2016MNRAS.459.3890U, 2018JQSRT.208..152T, 2013JQSRT.130....4R}, $\mathrm{NH_3}$ \citep[ExoMol/CoYuTe,][]{2015JQSRT.161..117A, 2019MNRAS.490.4638C}, $\mathrm{HCN}$ \citep[ExoMol/Harris,][]{2006MNRAS.367..400H, 2014MNRAS.437.1828B, 1985CP.....93..115M, 1973JChPh..58..442C, 1980JChPh..73.1494C}, and $\mathrm{C_2H_2}$ \citep[ExoMol/aCeTY,][]{2020MNRAS.493.1531C, 2016JQSRT.168..193W, 2017JQSRT.203....3G}.

The free parameters in the MCMC sampling were the temperature $T$, the logarithm of the pressure at the cloud top $\log P_\mathrm{cloud}$, the logarithm of the volume mixing ratio (VMR) of nine molecular species, the radius of the host star $R_\mathrm{s}$, the planetary mass $M_\mathrm{p}$, and the planetary radius $R_\mathrm{p}$ at an altitude of $10$ bar. $R_\mathrm{s}$, $M_\mathrm{p}$, and $R_\mathrm{p}$ at 10 bar were assumed to be common across all four limbs, while the other parameters were treated independently for each limb. The VMRs of $\mathrm{H_2}$ and $\mathrm{He}$ were assumed to be the remaining $6/7$ and $1/7$, respectively, after subtracting the VMRs of the nine aforementioned molecules.

The prior distributions of each parameter are summarized in Table \ref{tab:prior_exojax}. The prior distributions for $R_\mathrm{s}$ and $M_\mathrm{p}$ were based on \citet{2018A&A...613A..41M}.
We fixed the radial velocity of the system at $-87.3\ \mathrm{km/s}$ \citep{2023ApJ...955L..19E}. 

The model spectra were derived from the opacity of the atmosphere, which ranged from $10\ \mathrm{bar}$ to $10^{-11}\ \mathrm{bar}$ and was divided logarithmically into 120 layers. The spectra were initially calculated at a spectral resolution of 6000 and then broadened to a resolution of 2700, approximately matching the instrumental spectral resolution, using Gaussian functions. Finally, by applying binning, we reduced the model spectra to a spectral resolution of approximately 100.

The likelihood function for this MCMC sampling was modeled as the product of multivariate normal distributions for each spectrum. The means of these distributions were the model spectra for each direction. The covariance matrices of these distributions represent the variances of the values in each wavelength bin along the diagonal and the covariances between pairs of wavelength bins in the off-diagonal elements. These covariance matrices for each spectrum were obtained from the MCMC sampling in the previous section.

Figure \ref{fig:4limbs_hmc} shows the 68\% credible intervals of each spectrum from the MCMC sampling. The arrangement of $\Rpxen$ (Panel (a)), $\Rpxip$ (Panel (b)), $\Rpxin$ (Panel (c)), and $\Rpxep$ (Panel (d)) is the same as in Figure \ref{fig:4limbs}. The black lines in each panel represent ten randomly selected sampled models. The contributions of each molecule, gray clouds, and CIA to these models are also shown as colored or hatched shading, with an offset of -0.01 $R_s$.

Figure \ref{fig:corner_xen} ($\Rpxen$), \ref{fig:corner_xip} ($\Rpxip$), \ref{fig:corner_xin} ($\Rpxin$), and \ref{fig:corner_xep} ($\Rpxep$) show the posterior distributions of each parameter for each spectrum. In these figures, the posterior distributions from the analysis with the intrinsic spectral resolution ($\sim$ 2700), assuming a uniform atmosphere (Kawahara et al. submitted), are overlaid for reference.

\subsection{Insights from the Atmospheric Retrieval} \label{subsec:atm}
We observe that the spectra on the morning side (Panel (b) $\Rpxip$ and (d) $\Rpxep$ of Figure \ref{fig:4limbs_hmc}) are flatter than those on the evening side (Panel (a) $\Rpxen$ and (c) $\Rpxin$ of Figure \ref{fig:4limbs_hmc}). In both evening-side spectra, absorption features from $\mathrm{CO_2}$ are evident.

Recently, \citet{2024Natur.632.1017E} constrained the temperature of the morning limb to $889^{+54}_{-65}$ and that of the evening limb to $1068^{+43}_{-55}$. In contrast, our analysis did not yield constraints on the limb temperatures in any direction. This discrepancy may arise from differences in the model used for retrieval; for instance, we employed a free chemistry model, whereas \citet{2024Natur.632.1017E} assumed chemical equilibrium.

The atmospheric retrieval constrained the $\log(\mathrm{CO_2})$ in the negative $\xe$ direction to $-1.1^{+0.3}_{-0.9}$ (Figure \ref{fig:corner_xen}) and in the negative $\xii$ direction to $-1.55^{+0.8}_{-3.4}$ (Figure \ref{fig:corner_xin}). Although some models in Figure \ref{fig:4limbs_hmc} suggest the presence of other molecules, such as $\mathrm{SO_2}$ (around $4000\ \mathrm{nm}$ in Panel (a) $\Rpxen$) and $\mathrm{CO}$ (around $4500 \sim 5000\ \mathrm{nm}$ in Panel (c) $\Rpxin$), the VMRs of these molecules were not constrained.
These VMRs might be constrained with increased data precision or by using spectra from other wavelength ranges.

Although the inferred VMRs of $\mathrm{CO_2}$ are relatively high, they are consistent with the high metallicity reported in previous studies. For instance, \citet{2023Natur.614..664A} suggests 3 to 10 times solar metallicity, and \citet{2024MNRAS.530.3252C} reports $20.1^{+10.5}_{-8.1}\times$ and $28.2^{16.3}_{-12.1}\times$ solar abundances for $\mathrm{O/H}$ and $\mathrm{C/H}$, respectively. In our other analysis of these data, using the native spectral resolution of NIRSpec/G395H ($\sim 2700$) and assuming a uniform atmosphere (Kawahara et al. submitted), we also found a relatively high VMR of $\log(\mathrm{CO_2}) = -2.4\pm0.2$.

The VMR of $\mathrm{CO_2}$ was not well constrained in the positive $\xii$ and $\xe$ directions, which correspond to the morning side. This suggests that $\mathrm{CO_2}$ is more abundant on the evening limb than on the morning limb.
In this context, the higher VMRs of $\mathrm{CO_2}$ observed on the evening limb, compared to those expected under the assumption of a uniform atmosphere, can be naturally understood as a result of $\mathrm{CO_2}$ being more concentrated on the evening side.
Since $\mathrm{CO_2}$ is thought to be produced both thermochemically and photochemically \citep{2020ApJ...899..147F}, the higher abundance of $\mathrm{CO_2}$ on the evening limb could be attributed to photochemical production on the dayside, followed by transport via eastward zonal wind.

However, the posterior distribution of the VMR of $\mathrm{CO_2}$ in the positive $\xii$ direction (Figure \ref{fig:corner_xip}) indicates a higher probability of elevated VMRs. Given that differences in temperature can affect the strength of absorption features in the transmission spectrum \citep{2014A&A...564A..73A}, the observed differences in the $\mathrm{CO_2}$ absorption features might be attributed to temperature differences between the morning and evening limbs.

Next, we consider $\mathrm{SO_2}$, which has been detected in analyses assuming a uniform atmosphere \citep[e.g.][Kawahara et al. submitted]{2023Natur.614..664A, 2024Natur.626..979P}. Simulations have shown that the formation of $\mathrm{SO_2}$ can be explained by photochemistry \citep{2023Natur.617..483T}.

The posterior distributions suggest that $\mathrm{SO_2}$ is more likely to be present on the north limb (Figure \ref{fig:corner_xen} $\Rpxen$ and \ref{fig:corner_xip} $\Rpxip$) than on the south limb (Figure \ref{fig:corner_xin} $\Rpxin$ and \ref{fig:corner_xep} $\Rpxep$), although the difference is slight. As discussed in \S \ref{subsec:lightcondition}, the north limb receives more intense illumination from the host star than the south limb, which could result in a higher abundance of $\mathrm{SO_2}$ on the brighter north limb due to enhanced photochemical production.
However, this hypothesis contrasts with the prediction by \citet{2023ApJ...959L..30T}, who suggested that $\mathrm{SO_2}$ would accumulate on the nightside due to the effects of horizontal transport by zonal wind. 

If $\mathrm{CO_2}$ and $\mathrm{SO_2}$ are both abundant on the evening limb slightly offset from the terminator on the dayside (negative $\xe$ direction), while $\mathrm{CO_2}$ remains abundant but $\mathrm{SO_2}$ is depleted on the evening limb slightly offset from the terminator on the nightside (negative $\xii$ direction), one possible explanation, though speculative, is a difference in the dissociation timescales of these molecules. In this scenario, the dissociation timescale of $\mathrm{CO_2}$ would be longer than that of $\mathrm{SO_2}$, allowing $\mathrm{CO_2}$ to be transported from the dayside to the nightside, while $\mathrm{SO_2}$ would dissociate before reaching the nightside.
Further investigation will be required to explore this possibility and gain a deeper understanding of the atmospheric dynamics.

\section{Discussion} \label{sec:Discussion}
We explored the connection between chromatic transit variation (CTV) and atmospheric asymmetries on planetary limbs. Understanding this connection helps leverage high-precision transmission spectroscopic data, such as those obtained with JWST. While more flexible models may become necessary as data quality improves, this understanding provides a valuable foundation for developing optimized models to analyze atmospheric inhomogeneities using transmission spectroscopy.

In \S \ref{sec:formulation}, we discussed two potential sources of uncertainty in this method: the orbit of the planet’s center of mass and the wavelength dependence of the host star’s radius. The analysis in \S \ref{sec:wasp39b} did not account for errors in these values. As noted in \S \ref{sec:formulation}, estimating these uncertainties is currently nearly impossible. However, incorporating these uncertainties into inferred quantities, such as temperatures and volume mixing ratios (VMRs), is essential for making the results more robust. Addressing this challenge is an important task for future work.

The method introduced in this paper enables the investigation of atmospheric properties, particularly the distribution of molecules, around the terminator of close-in exoplanets. Thanks to the wide wavelength coverage and exceptional precision of JWST's transmission spectra, we can explore the distribution of key molecules, such as $\mathrm{SO_2}$, across the planet's atmosphere. By combining this method with complementary observational techniques, such as phase curves and secondary eclipse observations, we can significantly enhance our understanding of exoplanetary atmospheres.

\section{Summary} \label{sec:Summary}

This paper presented a new method for exploring atmospheric inhomogeneity in exoplanets by analyzing chromatic transit variation (CTV). We found that the timings of ingress and egress reflect asymmetries perpendicular to the host star’s edge, while the durations of ingress and egress reflect asymmetries perpendicular to the planet's orbital motion. Based on these insights, we derived formulations that convert the transit depth $k^2$ and contact times $\tI$, $\tII$, $\tIII$, and $\tIV$ into the spectra of projected lengths of the planet's shadow disk, measured from the planet's center of mass in four different directions. 
This approach enables us to divide both the morning and evening limbs into two regions each: one slightly offset from the day-night terminator on the dayside, and one on the nightside.

We applied the method to JWST's NIRSpec/G395H observation of WASP-39b and found a higher abundance of $\mathrm{CO_2}$ on the evening limb compared to the morning limb. Our results also indicate a greater probability of $\mathrm{SO_2}$ on the limb slightly offset from the terminator on the dayside relative to the nightside. These results should be interpreted in the context of the photochemical processes and atmospheric circulation.


\vspace{\baselineskip}
This work is based on observations made with the NASA/ESA/CSA James Webb Space Telescope. The data were obtained from the Mikulski Archive for Space Telescopes at the Space Telescope Science Institute, which is operated by the Association of Universities for Research in Astronomy, Inc., under NASA contract NAS 5-03127 for JWST. The specific observations analyzed can be accessed via \dataset[10.17909/7559-ne05]{https://doi.org/10.17909/7559-ne05}. These observations are associated with program ERS-1366.
The authors acknowledge the Transiting Exoplanet Community Early Release Science Program team for developing their observing program with a zero-exclusive-access period.
We thank Yui Kasagi, Shota Miyazaki, and Hibiki Yama for the fruitful discussions. This study was supported by JSPS KAKENHI grant nos. 21H04998, 23H00133, 23H01224 (H.K.), 21K13984, 22H05150, 23H01224 (Y.K.), 21H04998 (K.M.)


%

\vspace{5mm}
\facility{JWST}


\software{
    ExoTIC-JEDI \citep{2022zndo...7185855A}, jwst \citep{2024zndo..10870758B}, jkepler, ExoJAX \citep{2022ApJS..258...31K}, jaxoplanet \citep{2024zndo..10736936H}, JAX \citep{jax2018github}, Numpyro \citep{bingham2019pyro, 2019arXiv191211554P}, Numpy \citep{harris2020array}, xarray \citep{hoyer_2024_10895413}, matplotlib \citep{Hunter:2007}}, catwoman \citep{2020JOSS....5.2382J, 2021AJ....162..165E}



\appendix

\section{Timing of Ingress and Egress}\label{ap:duration}
We defined the timing of ingress as 
\begin{align}
\ti &= (\tI + \tII)/2 \nonumber \\
&= t_{0} - \frac{\Ttot + \Tfull}{4}, 
\end{align}
and the timing of egress as
\begin{align}
\te &= (\tIII + \tIV)/2 \nonumber \\
&= t_{0} + \frac{\Ttot + \Tfull}{4}.
\end{align}
These values differ slightly from $\ti'$ and $\te'$, which are the times when the center of the planet’s shadow disk intersects the host star's edge. While this difference does not affect the formulation in \S \ref{sec:formulation}, we discuss it here for a better understanding of the method.

For simplicity, we assume the planet's shadow at wavelength $\lambda$ has a transit depth of $k(\lambda)^2$ and follows a circular orbit with a semi-major axis $a_{\mathrm{\lambda}}$, impact parameter $b_{\mathrm{\lambda}}$, and central transit time $t_0(\lambda)$. The values of $a_{\mathrm{\lambda}}$ and $b_{\mathrm{\lambda}}$ can be estimated from $\Ttot(\lambda)$, $\Tfull(\lambda)$, and $k(\lambda)^2$.
From eqs. (14) and (15) in \citet{2010exop.book...55W}, $\Ttot(\lambda)$, $\Tfull(\lambda)$ are given by
\begin{align}
\Ttot &= \frac{P}{\pi}\sin^{-1}\left[\sqrt{\frac{(1+k)^2 - b_{\mathrm{\lambda}}^2}{a_{\mathrm{\lambda}}^2 - b_{\mathrm{\lambda}}^2}}\right], \\
\Tfull &= \frac{P}{\pi}\sin^{-1}\left[\sqrt{\frac{(1-k)^2 - b_{\mathrm{\lambda}}^2}{a_{\mathrm{\lambda}}^2 - b_{\mathrm{\lambda}}^2}}\right],
\end{align}
where arguments for $\lambda$ are omitted for readability.
In contrast, the duration $T_{\mathrm{c}}$ during which the center of the planetary shadow transits the host star is
\begin{align}
T_{\mathrm{c}} &= \frac{P}{\pi}\sin^{-1}\left[\sqrt{\frac{1 - b_{\mathrm{\lambda}}^2}{a_{\mathrm{\lambda}}^2 - b_{\mathrm{\lambda}}^2}}\right].
\end{align}
Thus, $T_{\mathrm{c}} \ne (\Ttot + \Tfull)/2$, meaning $\ti' = t_0 - T_{\mathrm{c}}/2 \ne \ti$ and $\te' = t_0 + T_{\mathrm{c}}/2 \ne \te$.  
For WASP-39 b, assuming $P = 4\ \mathrm{days}$, $k = 0.145$, $a = 11.4\ \Rs$, and $b = 0.45$, we find $T_{\mathrm{c}} - (\Ttot + \Tfull)/2 \sim 30\ \mathrm{s}$ and $\ti - \ti' \sim -15\ \mathrm{s}$.

Additionally, if the planetary atmosphere is perfectly symmetric, $T_\mathrm{c}$ corresponds to the transit duration for the center of mass of the planet, $T_\mathrm{bc}$. In \S \ref{sec:wasp39b}, we approximated $T_\mathrm{bc}$ using $(\Ttot + \Tfull)/2$, even though $T_{\mathrm{c}} \ne (\Ttot + \Tfull)/2$. While this approximation has a negligible impact on current data when estimating wavelength dependence alone, adjustments may be needed as data precision improves.

\section{Stellar Radii Based on a Stellar Atmospheric Model}\label{ap:phoenix}

\begin{figure}[htb!]
\centering
\includegraphics[width=0.8\linewidth]{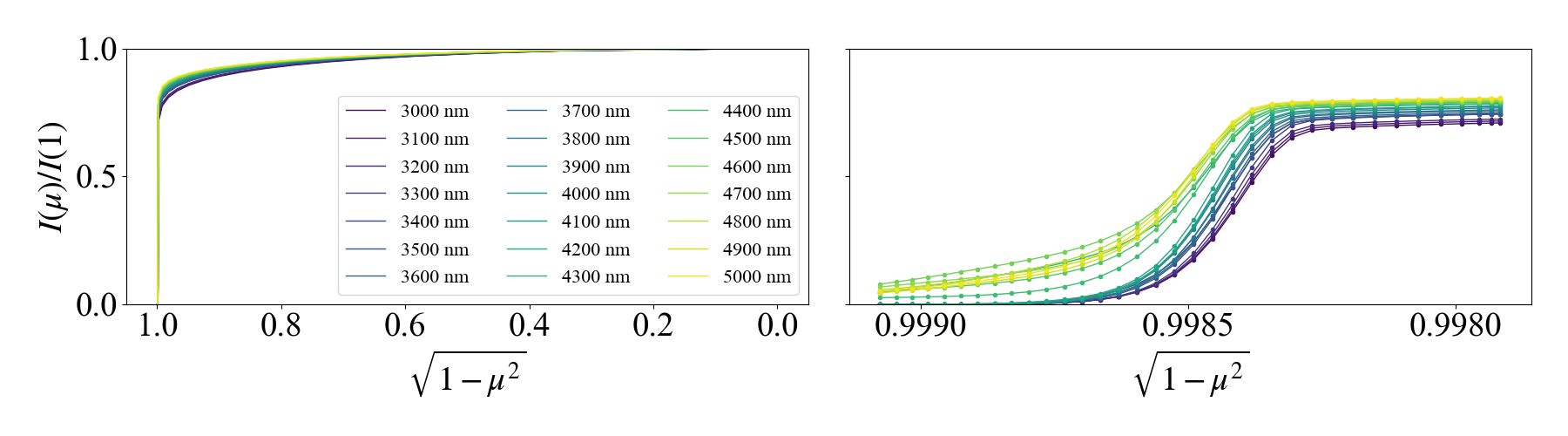}
\caption{
Normalized specific intensities for a star with an effective temperature of $5500\,\mathrm{K}$, $\log g=4.5$ (cgs), and metallicity $\mathrm{[Fe/H]}=0.0$, computed by \citet{2013A&A...553A...6H} using the spherically symmetric model of {\sf PHOENIX}. The horizontal axes show $\sqrt{1-\mu^2}$. The right panel is a magnified view near the limb, where the intensity begins to drop steeply.
\label{fig:specint}}
\end{figure}

\begin{figure}[htb!]
\centering
\includegraphics[width=0.5\linewidth]{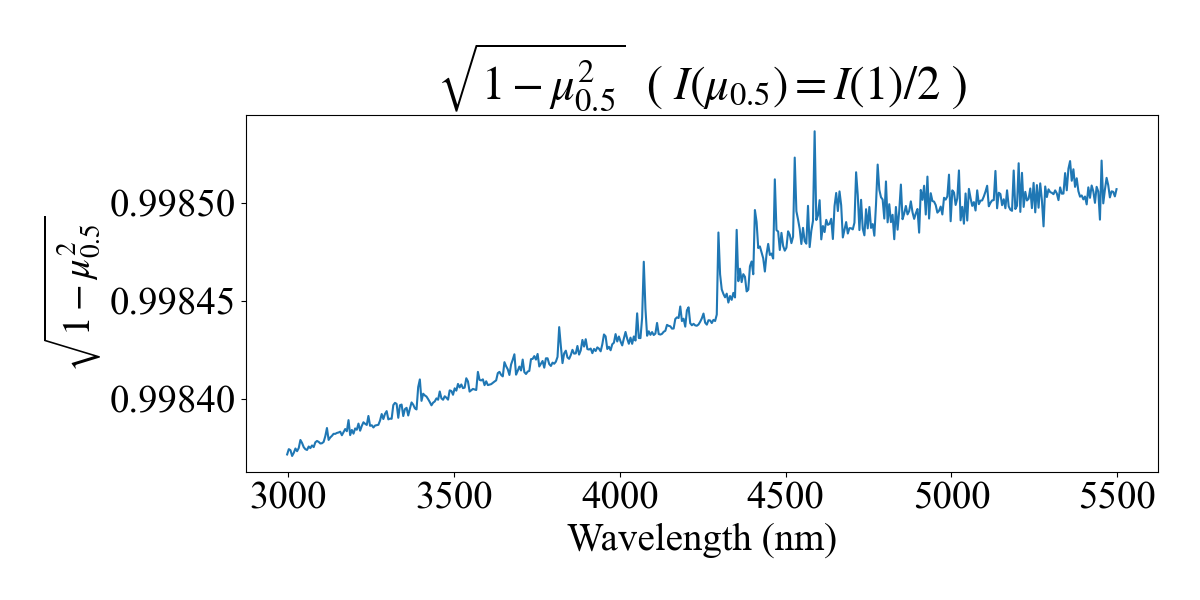}
\caption{
Variation of $\sqrt{1-\mu_{0.5}^2}$ with wavelength. Here $\mu_{0.5}$ is defined by $I(\mu_{0.5})=I(1)/2$. The values are computed from the same specific intensities as in Figure~\ref{fig:specint} and binned into 5\,nm intervals.
\label{fig:specint_half_mu}}
\end{figure}

In the formulation, the wavelength dependence of the host star's radius, denoted as $\alpha(\lambda)$, is taken into account. Figure \ref{fig:specint} shows the normalized specific intensities of a star with an effective temperature of $5500\ \mathrm{K}$, surface gravity $\log g\ \mathrm{[cgs]}=4.5$, and metallicity $\mathrm{[Fe/H]}=0.0$, computed by \citet{2013A&A...553A...6H} using the spherically symmetric model of {\sf PHOENIX}. Here, $\mu=\cos\theta$, where $\theta$ is the angle between the line of sight and the normal to the stellar surface, and the corresponding projected stellar radius is given by $r/R_\star=\sqrt{1-\mu^2}$.
We define $\mu_{0.5}$ by $I(\mu_{0.5})=I(1)/2$ and use $\sqrt{1-\mu_{0.5}^2}$ as a proxy for the projected stellar radius. Figure \ref{fig:specint_half_mu} shows its wavelength dependence. The values were calculated from the specific intensities binned into $5\ \mathrm{nm}$ intervals. Over the wavelength range of $3$--$5\ \mathrm{\mu m}$, the inferred wavelength dependence of the stellar radius is approximately $0.009\%\ \mathrm{\mu m^{-1}}$.

\section{Alternative Data Reduction and Comparison of Spectra}\label{ap:reduction}

To verify the impact of data reduction on the resulting spectra, an alternative data reduction process, different from that described in \S \ref{subsec:data}, was performed. Hereafter, we refer to the data reduction described in \S \ref{subsec:data} as ``Reduction A" and the alternative process as ``Reduction B." This process primarily follows the approach applied to the data labeled "ExoTiC-JEDI [V1]" in \citet{2023Natur.614..664A}. The ExoTiC-JEDI pipeline \citep{2023Natur.614..664A} was used for Reduction B, with some necessary customizations.

To obtain the spectral time series, we first used stages 1 of the ExoTiC-JEDI pipeline \citep{2023Natur.614..664A}. Stage 1 closely followed the JWST pipeline \citep{2024zndo..10870758B}, with the addition of a group-level destriping step as described in \citet{2023Natur.614..664A}. In Reduction B, the bias subtraction step of the JWST pipeline was used. The ramp-jump detection threshold was set to the default value of $4\sigma$, and all other settings were kept at their defaults, either as determined by the JWST pipeline \citep{2024zndo..10870758B} or specified in \citet{2023Natur.614..664A} for the group-level destriping step.

The resulting data was processed following \citet{2023Natur.614..664A}.
We utilized the data-quality flags generated by the JWST Calibration Pipeline to identify problematic pixels, including those flagged as bad, saturated, dead, hot, or exhibiting low quantum efficiency or missing gain values. These pixels were replaced with the median value of their surrounding pixels. Additionally, we performed a search within each integration to identify spatial outliers, defined as pixels deviating from the median of the surrounding 20 pixels in the same row by more than 6$\sigma$, and temporal outliers, defined as pixels differing from the median of that pixel across the surrounding ten integrations by more than 20$\sigma$. Any detected outliers were replaced with the corresponding median values.
To determine the trace position and width, we fitted a Gaussian to each column within an integration. A fourth-order polynomial was then applied to the trace centers and standard deviations values, smoothed using a median filter, to define the aperture region. This aperture extended vertically by five standard deviations from the center of the trace. For background subtraction, the median value of the unilluminated region in each column was subtracted, excluding pixels located within five pixels of the aperture.
For each integration, the pixel counts within the aperture region were summed across columns using an intrapixel extraction method. This process yielded 1D stellar spectra for each integration, with uncertainties calculated based on photon noise and readout noise.

\begin{figure*}[ht!]
\centering
\includegraphics[width=0.8\linewidth]{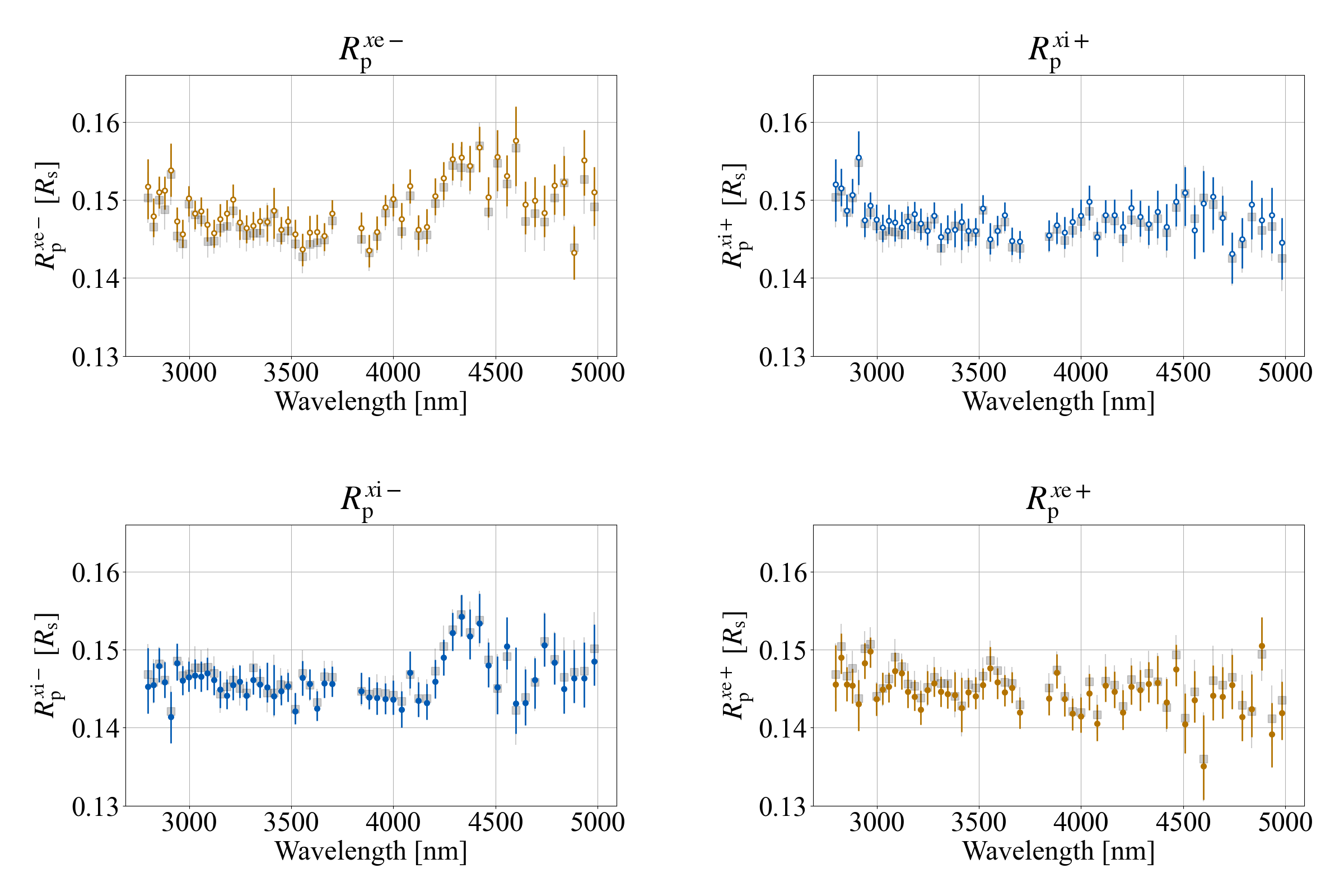}
\caption{
Spectra of $\Rpxen$ (top left), $\Rpxip$ (top right), $\Rpxin$ (bottom left), and $\Rpxep$ (bottom right).
The dots with error bars represent the median values from the MCMC sampling with 68\% credible intervals. The colored dots indicate the Reduction B, while the gray dots indicate the Reduction A.
\label{fig:4limbs_reduction}}
\end{figure*}

We obtained the spectra of $\Rpxen$, $\Rpxip$, $\Rpxin$, and $\Rpxep$ using the procedure described in \S \ref{sec:wasp39b}. Figure \ref{fig:4limbs_reduction} shows these spectra (colored dots) alongside those obtained with Reduction A (gray dots). The spectral shapes derived from both reductions are nearly identical, with only minor differences. Small offsets are observed between them, which result from differences in the orbital parameters estimated through MCMC sampling for wavelengths corresponding to radii smaller than the 10th percentile.
Since two independent procedures produced nearly identical spectra, we consider the reduction process to be reliable.

\section{Discussion of Limb Darkening Coefficients Parametrization and Wavelength Dependence}\label{ap:ldc}

\begin{figure*}[ht!]
\centering
\includegraphics[width=0.8\linewidth]{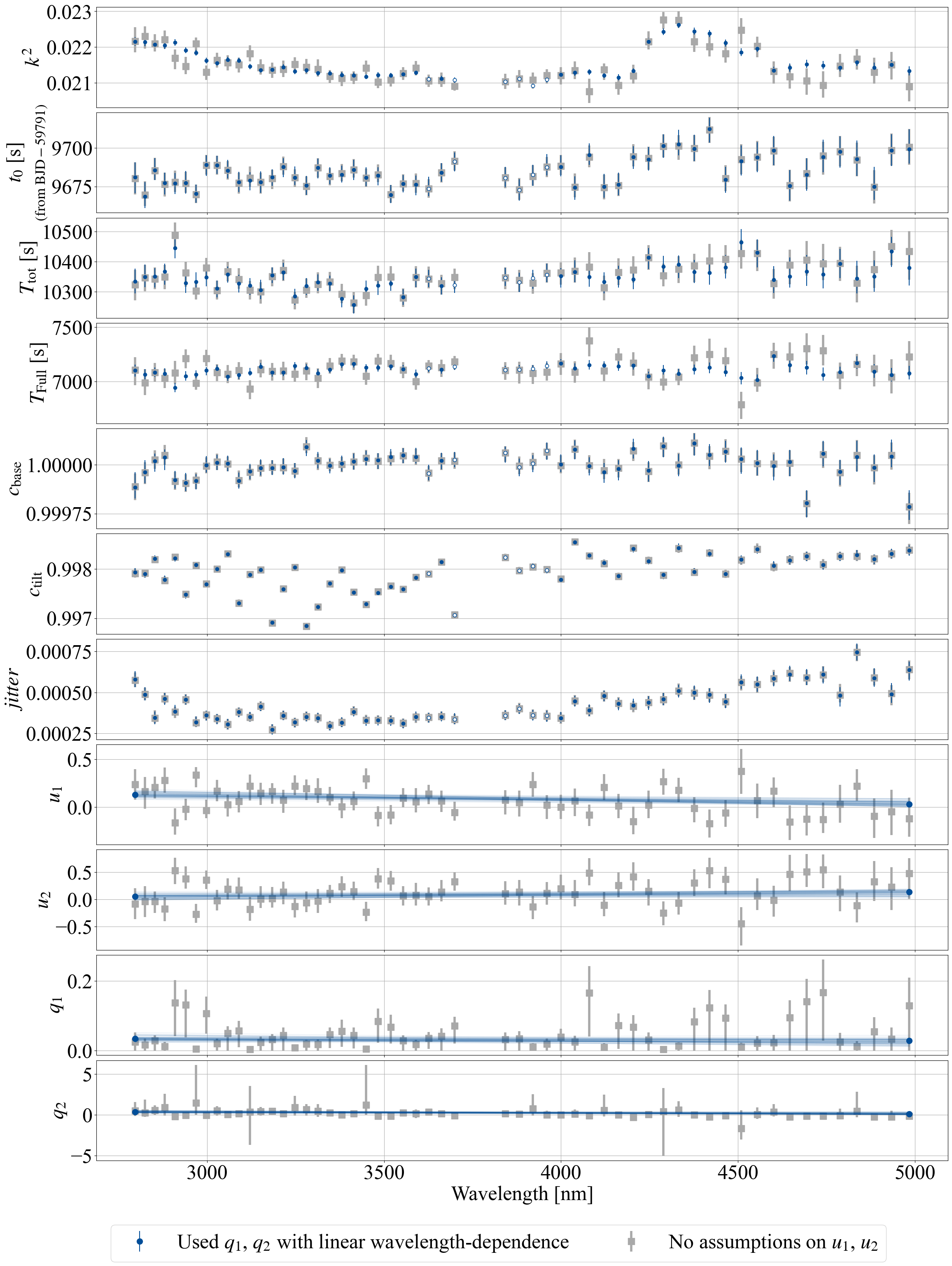}
\caption{Inferred values of all the parameters in the light curve model obtained from the MCMC sampling. The gray dots represent the results using free $u_1$ and $u_2$ for each wavelength, while the blue dots represent the results assuming a linear wavelength dependence for $q_1$ and $q_2$ in the MCMC analysis. For the limb darkening coefficients inferred using the assumption of a linear wavelength dependence for $q_1$ and $q_2$, the inferred values at the minimum and maximum wavelengths, as well as the wavelength dependence for each sampling, are shown. The dots indicate the median values of the MCMC sampling, and the error bars represent the 68\% credible intervals.
\label{fig:params_lightcurve_ldc}}
\end{figure*}

\begin{figure*}[ht!]
\centering
\includegraphics[width=0.85\linewidth]{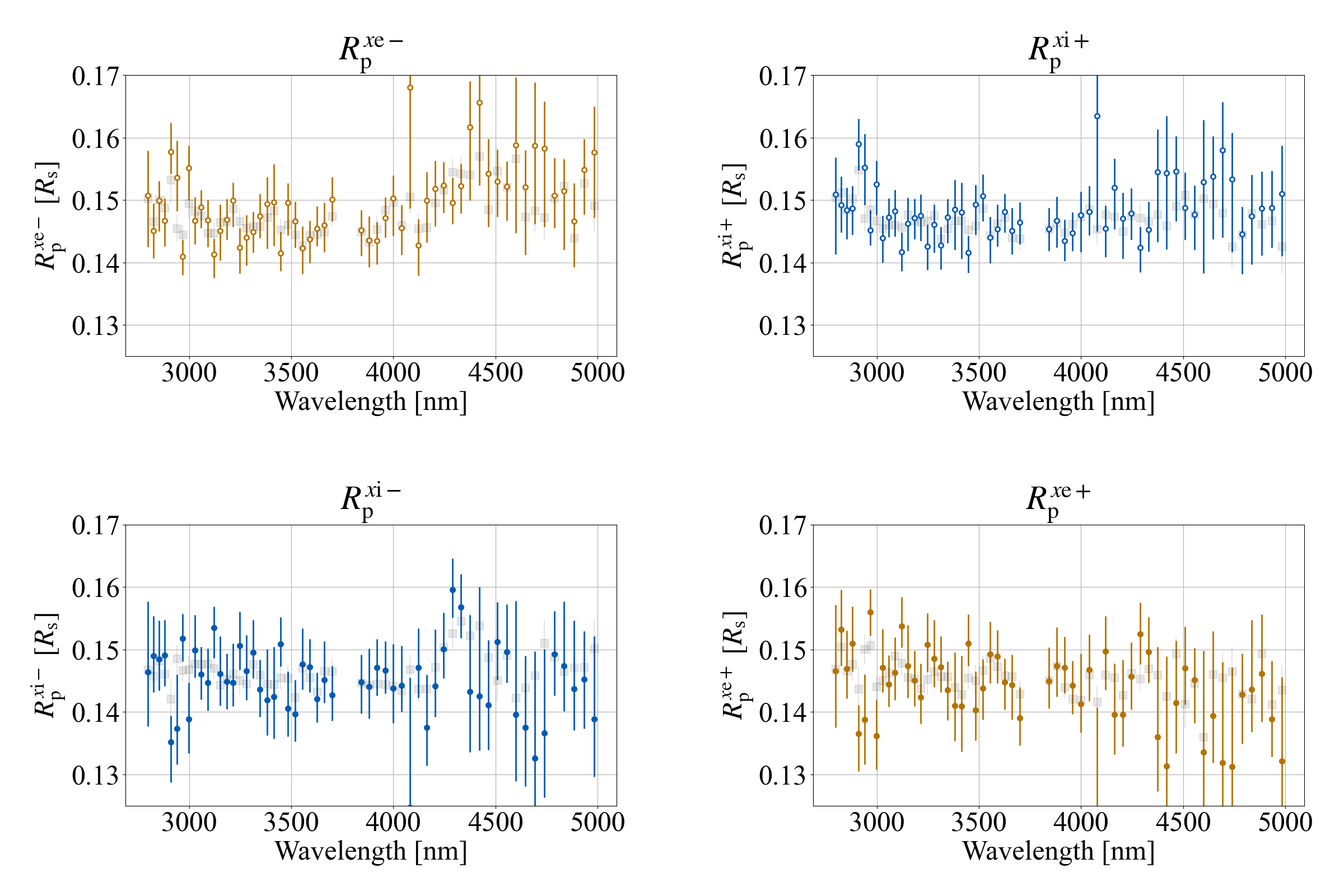}
\caption{
Spectra of $\Rpxen$ (top left), $\Rpxip$ (top right), $\Rpxin$ (bottom left), and $\Rpxep$ (bottom right). The dots with error bars represent the median values from the MCMC sampling, with 68\% credible intervals. The colored dots indicate the results using free $u_1$ and $u_2$ for each wavelength, while the gray dots indicate the results assuming a linear wavelength dependence for $q_1$ and $q_2$ in the MCMC analysis.
\label{fig:4limbs_ldc}}
\end{figure*}

In \S \ref{sec:wasp39b}, we used a quadratic limb darkening model with $q_1$ and $q_2$ values from \citet{2013MNRAS.435.2152K}. Regarding the parameterization of limb darkening coefficients, \citet{2024AJ....168..227C} demonstrated that using $q_1$ and $q_2$ could introduce wavelength-dependent biases in the inferred transit depths. To investigate this issue, we performed additional light curve fits with free $u_1$ and $u_2$ for each wavelength, and compared the results with those in \S \ref{sec:wasp39b}. In this analysis, we employed a wide uniform prior distribution $\mathcal{U}(-3, 3)$ for both $u_1$ and $u_2$, while keeping all other settings and procedures the same as those in \S \ref{sec:wasp39b}.

Figure \ref{fig:params_lightcurve_ldc} shows the inferred values of all the parameters in the light curve model obtained from the MCMC sampling (gray dots), along with those inferred in \S \ref{sec:wasp39b} assuming a linear wavelength dependence for $q_1$ and $q_2$ (blue dots). While the uncertainties increased, we found that almost all parameters remained consistent with the results from \S \ref{sec:wasp39b}. Furthermore, we did not detect the wavelength-dependent biases in the inferred transit depth $k^2$, which were reported by \citet{2024AJ....168..227C} when using the $q_1$ and $q_2$ parameterization. These biases tend to reduce the inferred transit depth. This absence of bias could be due to the low resolution of the spectra analyzed here, which are typically of higher precision, as noted by \citet{2024AJ....168..227C}. Therefore, we conclude that using the $q_1$ and $q_2$ parameterization does not introduce problems in the analysis presented in \S \ref{sec:wasp39b}.

To investigate how different treatments of the limb darkening coefficients affect the results, we obtained the spectra of $\Rpxen$, $\Rpxip$, $\Rpxin$, and $\Rpxep$ (Figure \ref{fig:4limbs_ldc}). For clarity, we used the orbital parameters of the planet's center of mass estimated in \S \ref{sec:wasp39b}, to isolate the effect of the limb darkening treatment on the shape of the spectra.
While the uncertainties in the resulting spectra were larger, the overall shape remained consistent with the results obtained using a linear wavelength dependence for $q_1$ and $q_2$, including absorption features around 4 to 4.5 $\mathrm{\mu m}$ on the evening side. Even when assuming the extreme case of no wavelength dependence for $u_1$ and $u_2$, the general shape of the spectra still agrees. This suggests that assuming a simple linear wavelength dependence for $q_1$ and $q_2$ does not significantly distort the overall shape of the spectra.

\bibliography{sample631}{}
\bibliographystyle{aasjournal}

\end{document}